\documentclass[notitlepage,eqsecnum,bm,amsmath,preprintnumbers,superscriptaddress,nofootinbib,aps,11pt]{revtex4-1} 

\usepackage{graphicx}
\usepackage{subfigure}
\usepackage{hyperref}
\usepackage{array}
\usepackage{amsmath,bm}

\long\def\del #1 \enddel { }

\usepackage{graphicx}
\usepackage{amsmath}
\usepackage{amssymb}
\usepackage{subfigure}

\usepackage{epsfig}

\usepackage{graphicx}
\usepackage{subfigure}
\usepackage{hyperref}

\usepackage{color}

\newcommand\floor[1]{\lfloor#1\rfloor}

\def\eq#1{(\ref{#1})}

\def\bR{\bar R}
\def\R{\rho}
\def\R{R}
\def\s0#1#2{\mbox{\small{$ \frac{#1}{#2} $}}}
\def\0#1#2{\frac{#1}{#2}}
\def\beq{\begin{equation}}
\def\eeq{\end{equation}}

\def\bea{\arraycolsep .1em \begin{eqnarray}}
\def\eea{\end{eqnarray}}

\usepackage{cancel}
\usepackage{bbold}
\usepackage{appendix}
\usepackage{amsmath}
\usepackage{amssymb}
\usepackage{graphicx}
\usepackage{amssymb}
\usepackage{epstopdf}
\usepackage{hyperref}
\usepackage{subfigure}
\usepackage{epstopdf}
\usepackage{verbatim} 
\usepackage{array}
\usepackage{multirow}
\usepackage{epsfig}
\usepackage{subfigure}
\usepackage{graphicx}
\usepackage{hyperref}
\usepackage{color}
\usepackage{natbib}
\usepackage{xcolor}
\usepackage{colortbl}

\definecolor{Yellow}{rgb}{1,1,0.05}
\definecolor{Gray}{gray}{0.85}
\definecolor{LightGray}{gray}{0.93}
\definecolor{LightGreen}{rgb}{0.88, 1, 0.88}
\definecolor{LightCyan}{rgb}{0.88,1,1}
\definecolor{LightRed}{rgb}{1, 0.85, 0.85}
\definecolor{LightRed}{rgb}{1, 0.88, 0.88}
\definecolor{LightYellow}{rgb}{1, 1, 0.85}
\definecolor{LightBlue}{rgb}{0.87, 0.94, 1}
\definecolor{white}{gray}{1}

\newcolumntype{G}{>{\columncolor{LightGray}}c}

\renewcommand{\thesection}{{\bf \Roman{section}}}

\usepackage{array,mathtools,amssymb,booktabs}
\newcolumntype{C}{>{$}c<{$}}
\AtBeginDocument{
\heavyrulewidth=.16em
\lightrulewidth=.1em
\cmidrulewidth=.03em
\belowrulesep=.4ex
\belowbottomsep=0pt
\aboverulesep=.4ex
\abovetopsep=0pt
\cmidrulesep=\doublerulesep
\cmidrulekern=.5em
\defaultaddspace=.5em
}

\makeatletter
    \def\CT@@do@color{%
      \global\let\CT@do@color\relax
            \@tempdima\wd\z@
            \advance\@tempdima\@tempdimb
            \advance\@tempdima\@tempdimc
    \advance\@tempdimb\tabcolsep
    \advance\@tempdimc\tabcolsep
    \advance\@tempdima2\tabcolsep
            \kern-\@tempdimb
            \leaders\vrule
                    \hskip\@tempdima\@plus  1fill
            \kern-\@tempdimc
            \hskip-\wd\z@ \@plus -1fill }

   \makeatother

\makeatletter
\def\@ssect@ltx#1#2#3#4#5#6[#7]#8{%
  \def\H@svsec{\phantomsection}%
  \@tempskipa #5\relax
  \@ifdim{\@tempskipa>\z@}{%
    \begingroup
      \interlinepenalty \@M
      #6{%
       \@ifundefined{@hangfroms@#1}{\@hang@froms}{\csname @hangfroms@#1\endcsname}%
       {\hskip#3\relax\H@svsec}{#8}%
      }%
      \@@par
    \endgroup
    \@ifundefined{#1smark}{\@gobble}{\csname #1smark\endcsname}{#7}%
  }{%
    \def\@svsechd{%
      #6{%
       \@ifundefined{@runin@tos@#1}{\@runin@tos}{\csname @runin@tos@#1\endcsname}%
       {\hskip#3\relax\H@svsec}{#8}%
      }%
      \@ifundefined{#1smark}{\@gobble}{\csname #1smark\endcsname}{#7}%
      \addcontentsline{toc}{#1}{\protect\numberline{}#8}%
    }%
  }%
  \@xsect{#5}%
}%
\makeatother
\usepackage{setspace}

\begin{document}
\title{Asymptotic safety of quantum gravity beyond Ricci scalars} 
\author{Kevin~Falls}
\affiliation{Institut f\"{u}r Theoretische Physik, University of Heidelberg, 69120 Heidelberg, Germany}
\author{Callum~R.~King}
\author{Daniel F.~Litim}
\author{Kostas~Nikolakopoulos}
\affiliation{Department of Physics and Astronomy, University of Sussex, Brighton, BN1 9QH, U.K.}
\author{Christoph~Rahmede}

\affiliation{Department of Physics and Astronomy, University of Sussex, Brighton, BN1 9QH, U.K.}

\affiliation{Rome International Centre for Material Science Superstripes RICMASS, via dei Sabelli 119A,
00185 Roma, Italy}

\begin{abstract}
We investigate the asymptotic safety conjecture for quantum gravity 
including
curvature invariants beyond Ricci scalars.
Our strategy is put to work for families of gravitational actions which depend on functions of the Ricci scalar, the Ricci tensor, and   products thereof.  
Combining functional renormalisation with high order polynomial approximations and full numerical integration 
we derive the renormalisation group flow for all couplings and analyse their fixed points, scaling exponents, and the fixed point effective action as a function of the background Ricci curvature.  The theory is characterised by three relevant couplings. Higher-dimensional couplings 
show near-Gaussian scaling with increasing canonical mass dimension. We find that Ricci tensor invariants  stabilise the UV fixed point and lead to a rapid convergence of 
polynomial approximations.
We apply our results to models for cosmology and establish that the gravitational fixed point admits inflationary solutions. We also compare findings with those from $f(R)$-type theories in the same approximation and pin-point the key new effects due to Ricci tensor interactions.  
Implications for the asymptotic safety conjecture of gravity are indicated.
\end{abstract}
\maketitle

\begin{spacing}{.94}
\tableofcontents
\end{spacing}

\section{\bf Introduction}
\label{sec:IntroductionRM}

It is widely acknowledged that for quantum field theories to be fundamental and predictive up to highest energies, their short-distance behaviour should be controlled by an ultraviolet (UV)  fixed point under the renormalisation group  \cite{Wilson:1971bg,Wilson:1971dh}. In particle physics, it is known since long that ultraviolet fixed points can be free, such as in asymptotic freedom of non-abelian gauge theories 
\cite{Gross:1973id,Politzer:1973fx}. Ultraviolet fixed points can also be interacting, such as in asymptotic safety \cite{Weinberg:1980gg}. 
In the absence of gravity, general theorems for asymptotic safety \cite{Bond:2016dvk} and  proofs of existence in four dimensional theories  are available, covering simple \cite{Litim:2014uca,Bond:2017tbw}, semi-simple  \cite{Bond:2017lnq}, supersymmetric gauge theories \cite{Bond:2017suy}, and Standard Model extensions \cite{Bond:2017wut}. 
It has also been speculated that a quantum theory for gravity with matter may exist with the help of an interacting UV fixed point \cite{Niedermaier:2006ns,Litim:2006dx,Niedermaier:2006wt,Percacci:2007sz,Litim:2008tt,Litim:2011cp,
Reuter:2012id}. 

In gravity, and much unlike in non-Abelian gauge theories, proofs or theorems for asymptotic safety  are presently not at hand. This is largely due to the dimensional nature of Newton's coupling  whose canonical mass dimension must be compensated by large anomalous dimensions and strong quantum effects.
Still, at strong coupling, the asymptotic safety conjecture can be tested systematically with the help of a  bootstrap search strategy \cite{Falls:2013bv}. By now, an extensive body of  evidence for asymptotic safety has grown over the years using the renormalisation group within increasingly sophisticated approximations \cite{
Reuter:1996cp,
Falkenberg:1996bq,
Souma:1999at,
Lauscher:2001ya,
Lauscher:2002sq,
Litim:2003vp,
Bonanno:2004sy,
Fischer:2006fz, 
Fischer:2006at,Codello:2006in,Codello:2007bd,
Machado:2007ea,Codello:2008vh,
Niedermaier:2009zz,
Benedetti:2009rx,
Eichhorn:2009ah,Manrique:2009uh,
Niedermaier:2010zz,Eichhorn:2010tb,Groh:2010ta,Manrique:2010am,
Niedermaier:2011zz,Manrique:2011jc,
Litim:2012vz,
Benedetti:2012dx,
Donkin:2012ud,
Christiansen:2012rx,
Dietz:2012ic,
 Hindmarsh:2012rc,
  Falls:2013bv,
  Benedetti:2013jk,
  Dietz:2013sba,
  Codello:2013fpa,
Ohta:2013uca,
Christiansen:2014raa,Becker:2014qya,
Falls:2014zba,
Padilla:2014yea,
  Falls:2014tra,
Saltas:2014cta,
Falls:2015qga,Falls:2015cta,
Christiansen:2015rva,Gies:2015tca,
Benedetti:2015zsw,
 Eichhorn:2015bna,Ohta:2015efa,Ohta:2015fcu,
 Gies:2016con,Falls:2016wsa,Falls:2016msz,Christiansen:2016sjn,
 Biemans:2016rvp,Pagani:2016dof,Denz:2016qks,
Falls:2017cze,Hamada:2017rvn,
Houthoff:2017oam, Gonzalez-Martin:2017gza,Knorr:2017fus,Knorr:2017mhu,
Becker:2017tcx,
Christiansen:2017cxa,
Falls:2017cze,Christiansen:2017bsy}.
Invariably, the fluctuations of the metric field are found to be strong, strong  enough for gravity to become anti-screening 
\cite{Litim:2006dx,Niedermaier:2006ns,Niedermaier:2006wt}.  
Canonical power counting continues to be a ``good'' ordering principle \cite{Weinberg:1980gg,Falls:2013bv}:
canonically relevant couplings remain relevant at interacting fixed points, canonically irrelevant couplings remain irrelevant,
and the spectrum becomes 
near-Gaussian with increasing mass dimension \cite{Falls:2013bv,Falls:2014tra}. It would then seem that asymptotically safe gravity  becomes ``as Gaussian as it gets''.

Much of the above picture has been established  for $f(R)$ quantum gravity and high-order polynomial approximations thereof. In this work, we extend investigations of the asymptotic safety conjecture towards actions beyond Ricci scalars. We use the method of functional renormalisation to derive renormalisation group flow equation for gravity including at strong coupling.
Our primary goal is to generalise the successful flow equations for $f(R)$ type approximations to much more general classes of actions involving Ricci and Riemann tensor invariants, or products and combinations thereof.  A main challenge for such a  programme is that explicit flow equations for general actions of the form $f(R,R_{\mu\nu},R_{\mu\nu\rho\sigma},\cdots)$ are not available. The main new idea is the observation that for certain subsets of actions, flow equations can nevertheless be derived and analysed. For want of example, we consider the UV fixed point of quantum gravity based on actions of the form
\beq\label{idea}
f(R,R_{\mu\nu})=F(R_{\mu\nu}R^{\mu\nu})+R\cdot Z(R_{\mu\nu}R^{\mu\nu})\,.
\eeq
where $F$ and $Z$ are unspecified functions, ultimately determined by the UV fixed point itself. 
For small curvature, the action reduces to the standard Einstein-Hilbert action. For large curvature, higher order terms modify the action and the dynamics of the theory. The virtue of actions of the form \eq{idea} is that, when Taylor-expanded in curvature invariants, only a single term arises per canonical mass dimension. In other words, evaluating the renormalisation group equations on spaces with constant curvature is  sufficient to find the running for all polynomial couplings of the theory, as well as the  non-perturbative renormalisation group  flow of the functions $F$ and $Z$. By and large, these ideas are similar to what had been done previously for $f(R)$ type actions. Generalising those for actions involving more general curvature invariants will enable us to investigate the impact on the gravitational UV fixed point of curvature invariants beyond Ricci scalars. Unlike $f(R)$ theories, 
our approximations  also incorporate the full kinematics of fourth order gravity alongside higher order curvature invariants. As a first application of our results, we investigate the availability of scaling solutions for cosmology such as inflation, and compare findings with $f(R)$-type theories within the exact same approximation.\\

The outline of the paper is as follows. In Sec.~\ref{sec:RG}, we introduce the main new idea in more detail, and discuss functional renormalisation group techniques and approximations to be used in the sequel. In Sec.~\ref{sec:QAlgorithm} we explain the technical algorithm used to compute the operator traces. Sec.~\ref{SecRmunu} introduces our models and the derivation of the RG flow for all couplings.
In Sec.~\ref{fixedpoints}, we search for fixed points within polynomial approximations and beyond. This is followed by an overview of scaling exponents in Sec.~\ref{scaling}. We also show that our findings satisfy the boostrap test for asymptotic safety, Sec.~\ref{sec:bootstrap}. 
In Sec.~\ref{sec:cosmology}, we apply our findings to quantum cosmology and establish the existence of de Sitter solutions for small Ricci curvature.  In
Sec.~\ref{sec:discussion} results are compared contrasted with those based on $f(R)$ type actions. A summary of findings and an outlook is given in 
Sec.~\ref{sec:ConclusionsRM}. Technicalities are summarised in three appendices.

\section{\bf Renormalisation group}\label{sec:RG}

In this section we will put forward our main new idea, also recalling the general setup for the renormalisation group for gravity that we shall adopt.

\subsection{Main idea and approximations}
\label{subsec:Approximations}

Our starting point is the scale-dependent effective action for gravity $\Gamma_k[\phi, \bar{g}_{\mu\nu}]$ where momentum modes down to the wilsonian RG scale $k$ have been integrated out. Here, $\bar{g}_{\mu\nu}$ denotes the background metric and $\phi$ the set of dynamical fields. In the case of interest $\phi$ includes the metric fluctuation $h_{\mu\nu} \equiv g_{\mu\nu} - \bar{g}_{\mu\nu}$ as well as ghosts and auxiliary fields arising from the functional measure. The effective average action obeys the flow equation \cite{Wetterich:1992yh,Morris:1993qb}
\begin{equation}
k\partial_k\Gamma_k[\phi,\bar{g}_{\mu\nu}]=\frac{1}{2}\rm{STr}\left[\left(\Gamma^{(2)}_k+\mathcal{R}_k\right)^{-1} k\partial_k\mathcal{R}_k\right],\label{FlowEquation}
\end{equation}
where $\Gamma^{(2)}_k$ denotes the second variation of the action with respect to the dynamical fields $\phi$ and $\mathcal{R}_k$ is the regulator function which vanishes for large momenta $p^2 \gg k^2$ while suppressing low momenta $p^2< k^2$ in the path integral. The RG scale $k$ therefore splits modes into high and low momentum and the left hand side of \eq{FlowEquation} gives the change in the effective action as $k$ is varied. Evaluation of the  super-trace on the right hand side of the equation allows one to obtain the non-perturbative beta functions.

To find an approximate solution to \eq{FlowEquation} we will specify an ansatz for  $\Gamma_k[\phi,\bar{g}_{\mu\nu}]$ while making simplifying assumptions in order that the trace may be evaluated in closed form. To this end we write the action as
\beq
\Gamma_k[\phi,\bar{g}_{\mu\nu}] = \bar{\Gamma}_k[g_{\mu\nu}] + \hat{\Gamma}_k[\phi ,\bar{g}_{\mu\nu}]
\eeq
where $\bar{\Gamma}_k[g_{\mu\nu}]$ only depends on the metric fluctuation $h_{\mu\nu}$ through the full dynamical metric $g_{\mu\nu} = \bar{g}_{\mu\nu} + h_{\mu\nu}$ and we demand that $\hat{\Gamma}_k[0 ,\bar{g}_{\mu\nu}]= 0$. Our first simplification will be to specify that $\hat{\Gamma}_k[\phi ,\bar{g}_{\mu\nu}]$ is given by the sum of terms in the bare action $S$ which arise from the functional measure namely the gauge fixing action $S_{\rm gf}$, the ghost action $S_{\rm gh}$ and the auxiliary field action $S_{\rm aux}$ arising from field redefinitions. Since these terms come from the bare action we will neglect their $k$ dependence and therefore they only appear explicitly in the right hand side of the flow equation as part of the hessian operator  $\Gamma^{(2)}_k$. The left hand side of the flow equation will therefore only contain the beta functions of couplings arising from the action  $\bar{\Gamma}_k[g_{\mu\nu}]$ which is a diffeomorphism functional of the metric tensor $g_{\mu\nu}$. It then suffices to evaluate the flow equation at vanishing field $\phi = 0$ in order to obtain the beta functions for the couplings inside $\bar{\Gamma}_k[g_{\mu\nu}]$. This approximation is known as the single field or background field approximation.
In general, we may introduce an ansatz for $\bar{\Gamma}_k[g_{\mu\nu}]$  by writing it as
\beq\label{Ansatz}
\bar\Gamma_k[g_{\mu\nu}]  = \sum_{n,i}\,\bar\lambda_{n,i}\,\int d^4x\, \mathcal{O}_{n,i}[g_{\mu\nu}] 
\eeq
Here, $\mathcal{O}_{n,i}$ are invariants of  mass dimension $2n$ built out of the metric field and derivatives thereof,
with $i$ differentiating terms of the same dimensionality,  and $\lambda_{n,i}$ the corresponding couplings.  

As a second approximation, we choose to evaluate the flow equation \eq{FlowEquation} on a spherical background with metric $\bar{g}_{\mu\nu} $
obeying the relations
\begin{equation}\label{sphere}
\bR_{\mu\nu}=\frac{\bR}{d}  \bar g_{\mu\nu} \,, \qquad 
\bR_{\mu\nu\rho\sigma}=\frac{\bR}{d(d-1)}\left(  \bar  g_{\mu\rho}  \bar  g_{\nu\sigma}-  \bar  g_{\mu\sigma}  \bar  g_{\nu\rho}\right)\,,
\end{equation}
where $\bR$, $\bR_{\mu\nu}$ and $\bR_{\mu\nu\rho\sigma}$ are the Ricci scalar, Ricci curvature, and Riemann curvature tensors respectively.
This choice then limits our ability to distinguish curvature invariants with the same canonical mass dimension. In order to obtain unique flows for the couplings in\eq{Ansatz},
we will therefore include in our approximations for the effective action operators $\mathcal{O}_n$ 
such that for every $n$ the corresponding operator is proportional to the $n$th power of the scalar curvature when evaluated on the sphere
\beq \label{Oinvariant}
\left.\mathcal{O}_n\right|_{\rm sphere}   \propto \sqrt{g} \bR^n \,.
\eeq
Stated differently, the technical choice \eq{sphere} limits our options for \eq{Ansatz} to those where only one representative per canonical mass dimension is retained.  A well-studied possibility is to choose the set $\mathcal{O}_n = \sqrt{g} R^n$ which corresponds to  an $f(R)$-type action,
\beq\label{fofr}
{\bar\Gamma}_k[g_{\mu\nu}]  = \int d^4x  \sqrt{g} 
\, f_k(R)    \,.
\eeq
Given a choice for the operators $\mathcal{O}_n$ we can close the approximation by first identifying $g_{\mu\nu} = \bar{g}_{\mu\nu}$ and then expanding both sides of the flow equation in powers of the scalar curvature $\bar{R} = R$. 

Our main  point is that \eq{fofr}  is only one out of many possible choices: flows for the running of couplings can be found for any set of $\mathcal{O}_n$ with unequal canonical mass dimensions. In this light, we put forward a strategy to derive RG flows for much more gravitational actions by replacing \eq{fofr} with
\beq\label{modelaction}
{\bar\Gamma}_k[g_{\mu\nu}] = \int d^4x \sqrt{g} \left[ F_k(\bar X) + R\cdot Z_k(\bar X)    \right]\,.
\eeq
where $F_k$ and $Z_k$ are  functions of 
\beq \label{X}
\bar X=a\, R^2 +b\, {\rm Ric}^2+c\, {\rm Riem}^2\,,
\eeq
with $a,b,c$ free parameters. By construction, \eq{X}  is a linear combination of the three independent quadratic curvature invariants with mass dimension four, which are the squares of the scalar curvature $R$, the Ricci tensor ${\rm Ric}$ and the Riemann tensor ${\rm Riem}$. We also require that $\bar X$, when evaluated on spheres with constant scalar curvature $r$, gives $\bar X|_{\rm spheres}=r^2/4$. This normalisation condition implies that only two of the three parameters $a,b$ and $c$ are independent. The family of theories described by \eq{modelaction} contain the scale-dependent cosmological constant $\Lambda_k$ and Newton's coupling $G_k$ as the leading order coefficients \beq
F_k(0)=\frac{\Lambda_k}{{8\pi}}\,,\quad Z_k(0)=-\frac{1}{16 \pi G_k}\,,
\eeq
 of the Taylor expansion in $\bar X$. In addition, the model contains all polynomial couplings proportional to the invariants $\bar X^n$ and  $R\cdot\bar X^m$.
The $f(R)$ models \eq{fofr} correspond to the  choice\beq \label{fofR}(a,b,c)=(1,0,0)\,.\eeq
Here, instead, we are interested in the class of models with
\beq \label{Rmunu}
(a,b,c)=(0,1,0)\,.
\eeq
Unlike \eq{fofR}, the model with \eq{Rmunu}  retains Ricci curvature invariants, which are more sensitive to the dynamics of the graviton.  The first few leading terms of the effective action read \cite{Nikolakopoulos:Thesis}
\beq\label{model}
\bar{\Gamma}_k = \int d^4x \sqrt{g} \left( \bar{\lambda}_0 + \bar{\lambda}_1 R  + \bar{\lambda}_{2} R^{\mu\nu}R_{\mu\nu} +  \bar{\lambda}_{3} R (R^{\mu\nu}R_{\mu\nu}) + \bar{\lambda}_4  (R^{\mu\nu}R_{\mu\nu})  (R^{\alpha\beta}R_{\alpha \beta} )+ ...    \right) \,.
\eeq
The action involves new curvature invariants which have not been studied previously within non-perturbative RG for gravity. We stress that our choice \eq{Rmunu} is {\it ad hoc} inasmuch as \eq{fofR}. The main purpose of this study is to establish whether or not the additional dynamics of the metric field is going to affect the UV behaviour in any significant manner. 

In the remainder of this section we derive the generic form of the flow equation in accordance with the above approximations. 
Many of the results of this section 
are commonly used in RG studies of gravity and can be found in  previous works 
\cite{Reuter:1996cp,Lauscher:2002sq,Dou:1997fg,Codello:2007bd,Machado:2007ea}. We generalise and extend these results for the purpose of our study, also noting that they do not depend 
on the choice for the  gravitational action $\bar{\Gamma}_k[g_{\mu\nu}]$. We investigate the RG flow for the model \eq{model} in Sect.~\ref{SecRmunu}.

\subsection{Field decompositions}
\label{subsec:Decompositions}

In order to calculate the trace we will adopt heat kernel methods. To do so we first bring the second variation into a form where 
it is a function of Laplacians and scalar curvature only. 
For this reason we use the transverse traceless decomposition of the original field $h_{\mu\nu}$ 
\cite{York:1973ia} 
\begin{equation}
h_{\mu\nu}=h^T_{\mu\nu}+\bar \nabla_{\mu}\xi_{\nu}+\bar \nabla_{\nu}\xi_{\mu}+\bar \nabla_{\mu}\bar \nabla_{\nu}\sigma-\frac{1}{d}\bar g_{\mu\nu}\bar \nabla^2\sigma+\frac{1}{d}\bar g_{\mu\nu}h\ ,\label{MetricDecomposition}
\end{equation}
which was first introduced in the context of the functional renormalisation group in \cite{Dou:1997fg}.
The various new fields are subject to the constraints, 
\begin{equation}
\bar g^{\mu\nu}h^T_{\mu\nu}=0, \qquad \bar \nabla^{\mu}h^T_{\mu\nu}=0, \qquad \bar \nabla^{\mu}\xi_{\mu}=0, 
\qquad h=\bar g_{\mu\nu}h^{\mu\nu}.
\end{equation}
Here $h=\bar g^{\mu\nu}h_{\mu\nu}$ is the trace of the fluctuation, $h_{\mu\nu}^T$ denotes the transverse-traceless part of 
$h_{\mu\nu}$, $\xi_{\mu}$ is a transverse vector that together with the scalar $\sigma$ gives the longitudinal-traceless part 
of $h_{\mu\nu}$ according to \eqref{MetricDecomposition}. Such a decomposition is advantageous because it leads to partial 
diagonalisation  of the propagator on a spherical background (except between $h$ and $\sigma$) and therefore it becomes a simple task 
to invert it.

Note that after decomposing our original field $h_{\mu\nu}$ into its transverse traceless decomposition it does not receive any 
contributions from the modes of the $\sigma$ field that obey the conformal Killing equation
\begin{equation}
\bar \nabla_{\mu}\bar\nabla_{\nu}\sigma+\bar\nabla_{\nu}\bar\nabla_{\mu}\sigma=\frac{2}{d}\,\bar g_{\mu\nu} \bar\nabla^2\sigma
\end{equation}
and similarly from the modes of the $\xi^{\mu}$ field that obey the Killing equation
\begin{equation}
\bar \nabla_{\mu}\xi_{\nu}+\bar\nabla_{\nu}\xi_{\mu}=0.
\end{equation}
These modes, when evaluated on the sphere, correspond to the lowest two modes of $\sigma$ and the lowest mode of $\xi^{\mu}$ 
respectively and therefore they should be excluded from the trace evaluation. For details of the heat kernel coefficients of the 
constrained fields as well as for the exclusion of lowest modes see appendix \ref{app:HeatKernel}.

\subsection{Gauge fixing and ghosts}
\label{subsec:GaugeFixingPart}

Because of the diffeomorphism invariance of the metric field the effective action has to be supplemented with a gauge fixing term so that only the physical modes of the field are taken into account. It is convenient to choose a gauge fixing term of the form
\begin{equation}
S_{\rm gf}=\frac{1}{2\alpha}\int d^dx\sqrt{\bar g}\,\bar g^{\mu\nu}\mathcal{F}_{\mu}\mathcal{F}_{\nu}\label{GFAction}
\end{equation}
with the gauge fixing condition $\mathcal{F}_{\mu}=0$. Here we will choose $\mathcal{F}_{\mu}$ to be given by 
\begin{equation}
\mathcal{F}_{\mu}=\sqrt{2}\kappa\left(\bar \nabla^{\nu}h_{\mu\nu}-\frac{1+\delta}{d}\bar \nabla_{\mu}h\right),
\end{equation}
with $\kappa=(32\pi G_N)^{-1/2}$ and $h=h^{\mu\nu}\bar g_{\mu\nu}$ being the trace of the fluctuation. 
The gauge fixing condition that was originally used in \cite{Reuter:1996cp} corresponds to $\delta=\frac{d}{2}-1$. Here and from now on bared geometrical quantities, such as $\bar \nabla$ above, mean that they are constructed by the background metric $\bar g_{\mu\nu}$. Then, by substituting $\mathcal{F}_{\mu}$ into the gauge fixing action we get
\begin{equation}
S_{\rm gf}=\frac{\kappa^2}{\alpha}\int d^dx\sqrt{\bar g}\left[\bar \nabla^{\rho}h_{\mu\rho}\bar \nabla^{\lambda}
h^{\mu}_{\lambda}-\left(\frac{1+\delta}{d}\right)^2h\bar \nabla^2h+2\frac{1+\delta}{d}h\bar \nabla_{\mu}\bar \nabla^{\rho}
h^{\mu}_{\rho}\right].\label{GaugeFixingAction}
\end{equation}
For the above gauge fixing, the corresponding ghost action is 
\begin{equation}
S_{\rm gh}=-\sqrt{2}\int d^dx\sqrt{\bar g}\,\bar C^{\mu}M_{\mu\nu}C^{\nu},\label{GhostAction}
\end{equation}
with the Faddeev-Popov operator given by 
\begin{equation}
M_{\mu\nu}=\bar \nabla^{\rho}\bar g_{\mu\nu} \nabla_{\rho}
+\bar \nabla^{\rho}\bar g_{\rho\nu} \nabla_{\mu}-2\frac{1+\delta}{d}\bar \nabla_{\mu} \bar g^{\rho\sigma}
\bar g_{\rho\nu} \nabla_{\sigma}\label{FPOperator}.
\end{equation}
As with the metric fluctuations, it is convenient to decompose the ghost fields into transverse ($C^{T}_{\mu}$ and $\bar C^{T}_{\mu}$) 
and longitudinal ($\eta$ and $\bar\eta$) parts
\begin{equation}
C_{\mu}=C_{\mu}^T+\bar \nabla\eta, \qquad \qquad \bar C_{\mu}=\bar C^T_{\mu}+\bar \nabla_{\mu}\bar\eta. \label{GhostDecomposition}
\end{equation}
In this decomposition, the modes that are unphysical and must be excluded from the trace evaluation are the constant modes of $\eta$ and $\bar\eta$ corresponding to lowest modes of the Laplacian. In addition, it has also been argued  that the lowest mode of $C^T$ and $\bar{C}^T$ and the second lowest modes of  $\eta$ and $\bar\eta$ should be removed to ensure an exact cancellation with the  fluctuations of the field $\xi$ and $\sigma$ from the gauge fixing term (see below) in the gauge $\delta =0$, \cite{Codello:2007bd,Machado:2007ea,Codello:2008vh}. 
Here, we point out an independent motivation for leaving out these additional modes by noting 
that they corresponds to zero modes of the ghost operator. As such, if retained, these modes would lead to a vanishing Fadeev-Popov determinant, which is why we will leave them out in this paper. We come back to this aspect in Sect.~\ref{sec:FlowEquationRM}. More technical details about this construction can be found  in App.~\ref{app:HeatKernel}.

The gauge fixing action \eqref{GaugeFixingAction} is already quadratic in the fields. Now we can substitute the metric decomposition \eqref{MetricDecomposition} into \eqref{GaugeFixingAction}  to express $S_{\rm gf}$ in terms of the metric components as
\begin{equation}
\begin{split}
S_{\rm gf}=\frac{\kappa^2}{\alpha}\int d^dx\sqrt{\bar g}&
\left\{ \xi_{\nu}\left[\bar \Box^2+2\frac{\bar R}{d}\bar \Box+\frac{\bar R^2}{d^2}\right]\xi^{\nu}\right.\\
&\ \ -\sigma\left[\left(\frac{d-1}{d}\right)^2\bar \Box^3+2\frac{\bar R}{d^2}(d-1)\bar \Box^2+\frac{\bar R^2}{d^2}\bar \Box\right]\sigma\\
&\ \ \left.+\frac{2}{d^2}h\left[(d-1)\delta\bar \Box^2+\delta \bar R\,\bar \Box\right]\sigma-\frac{\delta^2}{d^2}h\bar \Box h \right\}.
\end{split}
\end{equation}
It follows that the contributions to the hessians  coming from the gauge fixing action take the form
\begin{eqnarray}
\left(S^{(2)}_{\rm gf}\right)^{\xi\xi}&=& \frac{\kappa^2}{\alpha}\left(\Box^2+2\frac{R}{d} \Box+\frac{ R^2}{d^2}\right)\\
\left(S^{(2)}_{\rm gf}\right)^{\sigma \sigma}&=&\frac{\kappa^2}{\alpha}\left(-\left(\frac{d-1}{d}\right)^2 \Box^3-2\frac{ R}{d^2}(d-1) \Box^2-\frac{ R^2}{d^2} \Box\right)
\\
\left(S^{(2)}_{\rm gf}\right)^{hh}&=&\frac{\kappa^2}{\alpha}\left(-\frac{\delta^2}{d^2} \Box\right)
\end{eqnarray}
\begin{eqnarray} 
\left(S^{(2)}_{\rm gf}\right)^{h\sigma}&=&\frac{\kappa^2}{\alpha}\left(\delta\frac{2}{d^2}(d-1) \Box^2+\delta \frac{2}{d^2} R\, \Box\right)
\end{eqnarray}
where we have dropped the bars for notational simplicity, since after computing the Hessians we set $g_{\mu\nu}=\bar g_{\mu\nu}$ and therefore it remains only one metric field. However, it should be kept in mind that all the geometric quantities are constructed with the background metric.

After substituting the ghost decomposition \eqref{GhostDecomposition} into the ghost action \eqref{GhostAction} we have 
\begin{equation}
S_{\rm gh}=-\sqrt{2}\int d^dx\sqrt{\bar g}\left\{\bar C^{\mu T}M_{\mu\nu}C^{\nu T} +\bar C^{\mu T}M_{\mu\nu}\bar \nabla^{\nu}\eta+\bar \nabla^{\mu}\bar\eta \,M_{\mu\nu} C^{\nu T} +\bar \nabla^{\mu}\bar\eta \, M_{\mu\nu}\bar \nabla^{\nu}\eta\right\}.
\end{equation}
Now we substitute the Faddeev-Popov operator \eqref{FPOperator} into the above equation and we perform the second variation in order to get for the Hessians of each ghost component field
\begin{eqnarray} \label{bare_gh_hessians}
\left(S^{(2)}_{\rm gh}\right)^{\bar C^T C^T }&=&-\sqrt{2}\,\Box-\sqrt{2}\frac{R}{d}\\
  \label{bare_gh_hessians_eta} \left(S^{(2)}_{\rm gh}\right)^{\bar \eta\eta }\,\,\,\,\,\,\,\,&=&\frac{2\sqrt{2}}{d}\left[d-\delta-1\right]\Box^2+\frac{2\sqrt{2}}{d}R\,\Box\,,
\end{eqnarray}
where again we have dropped the bars after setting $g_{\mu\nu}=\bar g_{\mu\nu}$.

\subsection{Auxiliary fields}
\label{subsec:AuxiliaryFields}

The metric decomposition \eqref{MetricDecomposition} is merely a coordinate transformation and as such it induces the Jacobian of the transformation $J_{\rm gr}$. Here we are going to determine this quantity and we will follow the Faddeev-Popov trick so that we transform the contributions from the determinants into contributions from auxiliary fields. We begin by writing the following relation between the original field $h_{\mu\nu}$ and the components of the transverse traceless decomposition 
\begin{equation}
h_{\mu\nu}h^{\mu\nu} = 
h^T_{\mu\nu}h^{T\mu\nu}+\frac{1}{d}hh - 2\,\xi_{\nu}\left[\bar \Box + \frac{\bar R}{d}\right]\xi^{\nu}+\sigma \left[\left(\frac{d-1}{d}\right)\bar \Box^2+\frac{\bar R}{d}\bar \Box\right]\sigma\,,
\end{equation}
which is valid when integrated over $\int d^dx \sqrt{\bar g}$. Then, at the level of the path integral the Jacobian of the field transformation takes the form
\begin{equation}
J_{\rm gr}=\left({\rm det} M_{(0)}\right)^{1/2}\,\left({\rm det} M_{(1T)}\right)^{1/2}
\end{equation}
with the operators $M_{(0)}$ and $M_{(1T)}$ coming from the contributions of the scalar $\sigma$ and the transverse vector $\xi^{\mu}$ respectively and having the form
\begin{eqnarray}
M_{(0)}&=&\left(\frac{d-1}{d}\right)\bar \Box^2+\frac{\bar R}{d}\bar \Box \ ,\\
M_{(1T)}&=&\bar \Box + \frac{\bar R}{d}\,.
\end{eqnarray}
The terms containing the transverse-traceless field $h^{T}_{\mu\nu}$ and the trace field $h$ do not contribute to the transformation Jacobian since they do not involve operators and under the path integral they are simple Gaussian integrals contributing only a constant. Now, we would like to express these determinants as new contributions to the action in terms of auxiliary fields. For this reason, we follow the Faddeev-Popov trick and write them as Gaussian integrals. We start with the determinant of the scalar field
\begin{equation}
\left({\rm det} M_{(0)}\right)^{1/2}=\frac{{\rm det} M_{(0)}}{\left({\rm det} M_{(0)}\right)^{1/2}}=\int \mathcal{D}\lambda\mathcal{D}\bar\lambda\mathcal{D}\omega\cdot\exp\left[-\int d^dx\sqrt{\bar g}\left\{\bar\lambda M_{(0)}\lambda+\omega M_{(0)}\omega\right\}\right]
\end{equation}
where $\lambda$ and $\bar\lambda$ are complex Grassmann fields coming from the denominator of the above expression and $\omega$ is a real field coming from the numerator. Thus the action for the scalar auxiliary fields reads $S_{\rm aux\,(0)}=\int d^dx\sqrt{\bar{g}}\left[\bar\lambda M_0\lambda+ \frac{1}{2}\omega M_0\omega\right]$ and the Hessians for $\bar\lambda$, $\lambda$ and $\omega$ are given, after dropping the bars, by
\begin{equation}  \label{bare_aux0_hessian}
\left(S^{(2)}_{\rm aux\, (0)}\right)^{\bar\lambda\lambda}=\left( S^{(2)}_{\rm aux\, (0)}\right)^{\omega\omega}=\left(\frac{d-1}{d}\right)\Box^2+\frac{R}{d}\Box\,.
\end{equation}
Similarly for the determinant ${\rm det} M_{(1T)}$ can be written in terms of a path integral with the auxiliary action $S_{\rm aux\,(1)}=\int d^dx\sqrt{\bar g}\left\{\bar c^T_{\mu} M_{(1T)}c^{T\mu}+\zeta^T_{\mu} M_{(1T)}\zeta^{T\mu}\right\}$ coming from the contribution of the transverse vector
\bea
\left({\rm det} M_{(1T)}\right)^{1/2}&=&\frac{{\rm det} M_{(1T)}}{\left({\rm det} M_{(1T)}\right)^{1/2}}  \\
&=&\int \mathcal{D}c^T_{\mu}\mathcal{D}\bar c^{T\mu}\mathcal{D}\zeta^T_{\mu}\cdot\exp\left[-\int d^dx\sqrt{\bar g}\left\{\bar c^T_{\mu} M_{(1T)}c^{T\mu}+\zeta^T_{\mu} M_{(1T)}\zeta^{T\mu}\right\}\right] \nonumber
\eea
with $c^T_{\mu}$ and $\bar c^T_{\mu}$ being complex Grassmann transverse vector fields coming from the denominator of the above expression and $\zeta^T_{\mu}$ a real transverse vector field coming from the numerator. The corresponding Hessians, after dropping the bars, are
\begin{equation} \label{bare_aux1_hessian}
\left(S^{(2)}_{\rm aux\, (1)}\right)^{\bar c^Tc^T}=\left(S^{(2)}_{\rm aux\, (1)}\right)^{\zeta^T\zeta^T}=\Box + \frac{R}{d}.
\end{equation}
In the same way that the metric decomposition  \eqref{MetricDecomposition} induces the Jacobian of the transformation we get contributions $J_{\rm gh}$ from the decomposition of the ghost fields \eqref{GhostDecomposition}. Now, the original ghost fields $C_{\mu}$ and $\bar{C}^\nu$ obey the following identity with the components $C^{T\mu}$, $\bar C^{T\mu}$,  $\eta$ and $\bar \eta$
\begin{equation}
\int d^dx\sqrt{\bar g}\, \bar C_{\mu}C^{\mu} =\int d^dx\sqrt{\bar g} \left\{ \bar C^T_{\mu}C^{T\mu}-\bar\eta\bar\Box\eta\right\}.
\end{equation}
The fields $C^T_{\mu}$ and $\bar C^T_{\mu}$ do not involve differential operators and thus they contribute only a constant. Now, at the level of the path integral the Jacobian of the transformation comes only from the scalars $\eta$ and $\bar \eta$ and after performing the Gaussian integral it can be written as
\begin{equation}
J_{\rm gh}=\left(\textnormal{det}\left[-\bar\Box\right]\right)^{-1}\ .
\end{equation}
As before we rewrite this expression in terms of the contribution to the action of auxiliary fields, so that it takes the form 
\begin{equation}
J_{\rm gh}=\int\mathcal{D}\bar s\mathcal{D} s\exp\left[-\int d^dx\sqrt{\bar g}\bar s\left[-\bar\Box\right] s\right],
\end{equation}
where now the fields $\bar s$ and $s$ are complex conjugate scalars with the corresponding auxiliary field action $ S_{\rm aux \,{\rm (gh)}} =  -\int d^dx\sqrt{\bar g}\bar s\left[-\bar\Box\right] s$. The  Hessian for this auxiliary field, after dropping the bars, is given by
\begin{equation} \label{bare_auxgh_hessian}
\left(S^{(2)}_{\rm  aux\, (gh) }  \right)^{\bar s s}=-\Box\,.
\end{equation}
Since we exclude the two lowest modes of the ghosts $\bar{\eta}$ and $\eta$ we must also remove these modes from the spectrum of $\bar{s}$ and $s$.
This completes the introduction of the basic degrees of freedom and auxiliary fields.

\subsection{Gravitational path integral}
We now turn to the introduction of a Wilsonian momentum cutoff and a path integral formulation of the theory.
After gauge fixing and employing the field decompositions we can collect all propagating  fields into a  ``superfield" $\varphi$ with components
\beq \label{superfield}
\varphi_i = \{  
\underbrace{h^T_{\mu\nu},\  \xi_\mu ,\ h,\ \sigma}_{\rm metric \, fluctuations}\ , \ 
\underbrace{C_\mu^T,\ \bar{C}_\nu^T,\  \eta,\  \bar{\eta}}_{\rm ghosts}\ ,      
\underbrace{ \bar{\lambda},\ \lambda,\ \omega}_{\rm scalar\, auxiliary \, fields} ,       
\underbrace{c^T_\mu,\ \bar{c}^T_\nu,\  \zeta^T_\mu}_{\rm vector\, auxiliary \, fields} ,  
\underbrace{ s\ \  ,\ \ \bar{s}}_{\rm ghost \,auxiliary  \,fields}         \}
\eeq
where we indicate the origin of each field. We also recall that $C_\mu^T,\, \bar{C}_\nu^T\,\eta, \, \bar{\eta},\, c^T_\mu,\, \bar{c}^T_\nu,\,\bar{\lambda}$ and $\lambda$ are Grassmann fields. In the conventions put forward here, the (gravitational) partition function can then be written as a functional integral over the set of fields \eq{superfield},
\beq\label{partition}
{Z}_k[J] = \int \mathcal{D}\varphi \, 
\exp\left({-S[\varphi]
-\Delta S_k[\varphi]
-\int d^dx\, \sqrt{\bar g}\,\varphi_i(x)\, J_i(x)}\right)
\eeq
with $J$ denoting external currents and the fields \eq{superfield} the integration variables.  The term $S$ in \eq{partition} denotes the fundamental action for all fields of the theory. The term $\Delta S_k$ introduces a cutoff at momentum scale $k$, in the sense of Wilson. Its task is to regularise the path integral \eq{partition} and to suppress the propagation of momentum modes with momenta $q$ smaller than the RG scale $k$. Technically, this is achieved by taking $\Delta S_k$ to be quadratic in the fluctuation fields,
\beq\label{cutoff}
\Delta S_k[\varphi]=\frac{1}{2}\int d^dx \, \sqrt{\bar g}\,\varphi_i(x)\,
\mathcal{R}^{ij}_k
\,\varphi_j(x)\,.
\eeq
The  cutoff functions $\mathcal{R}^{ij}_k
$, which take the role of momentum-dependent mass terms, are at our disposal.
We demand that $\Delta S_k \to 0$ for $k\to 0$, to ensure that the partition function reduces to that of the full  physical theory in the limit where the momentum cutoff is removed. In the UV limit $k\to \infty$, we also require $\Delta S_k$  to grow large to ensure that fluctuations are switched off.
Coming back to  our  settings in the context of gravity, we have that
\bea\label{S}
S[\varphi, \bar{g}] &=& S_{\rm grav}[h^T_{\mu\nu}, \xi_\mu , h,\sigma;\bar{g} ] + S_{\rm gf}[h^T_{\mu\nu}, \xi_\mu , h,\sigma;\bar{g}]  \nonumber \\&& + S_{\rm gh}[C_\mu,\bar{C}^\nu, \eta, \bar{\eta}, h^T_{\mu\nu}, \xi_\mu , h,\sigma ;\bar{g}] + S_{\rm aux}[\bar{\lambda},\lambda,\omega, c^T_\mu, \bar{c}^T_\nu, \zeta^T_\mu,s ,\bar{s}; \bar{g} ]
\eea
The auxiliary field action denotes the sum 
$S_{\rm aux} =  S_{\rm aux\,(0)}  +  S_{\rm aux\, (1)} +  S_{\rm aux \,{\rm (gh)}}$, which, together with the gauge fixing $S_{\rm gf}$ and ghost action $S_{\rm gh}$,  has been defined in the previous subsection. In turn, the gravitational action $S_{\rm grav}$ has been kept arbitrary for now. It will be specified  in Sect.~\ref{SecRmunu} for the  purposes of the present paper.
Similarily, the cutoff action \eq{cutoff} takes the form
\bea
\Delta S_k[\varphi;\bar{g}] &=& \ \ \, \int d^dx  \sqrt{\bar{g}} 
\left( 
\frac{1}{2} h_{\mu\nu}^T \, \mathcal{R}^{\mu\nu\rho\sigma}_{h^Th^T}[\bar{g}]  \, h_{\rho \sigma}^T 
+  \frac{1}{2} \xi_{\mu} \,  \mathcal{R}^{\mu\nu}_{\xi \xi}[\bar{g}] \, \xi_\nu 
+    \frac{1}{2} h\,   \mathcal{R}_{h h}[\bar{g}]\,  h    
+    \frac{1}{2} \sigma \,  \mathcal{R}_{\sigma  \sigma }[\bar{g}] \, \sigma    \right) \nonumber  \\
&& + \int d^dx \sqrt{\bar{g}} \left( \bar{C}_{\nu}\,  \mathcal{R}_{\bar{C} C}^{\mu\nu}\,  C_{\mu}   
+   \bar{\eta} \, \mathcal{R}_{\bar{\eta} \eta} \, \eta   
+   \bar{\lambda} \, \mathcal{R}_{\bar{\lambda} \lambda}\,  \lambda  
+  \frac{1}{2} \omega \, \mathcal{R}_{\omega \omega} \, \omega       \right)  \nonumber \\ 
 &&+ \int d^dx  \sqrt{\bar{g}} \left( \bar{c}^T_{\nu} \, \mathcal{R}_{\bar{c}^T c^T}^{\mu\nu}\,  c_{\mu}^T   
+  \frac{1}{2} \zeta^T_{\nu}\,  \mathcal{R}_{\zeta^T \zeta^T}^{\mu\nu} \, \zeta_{\mu}^T   
+   \bar{s}\,  \mathcal{R}_{\bar{s} s}\,  s         \right)  \,.
\eea
Notice that the momentum cutoff is ``diagonal" in field space, \eq{superfield}.  Off-diagonal terms are not required and have been omitted for simplicity. Further considerations, including our rationale for the explicit choice of  cutoff and profile functions, are laid out  in Sect.~\ref{subsec:RegulatorSchemes} and~\ref{subsec:TheRegulatorTerm} below.

With the help of the partition function $Z_k[J]$ in \eq{partition}, the gravitational effective action $\Gamma_k$ follows via a Legendre transformation $\Gamma_k[\phi]=\sup_J(-\ln Z_k[J]+\phi \cdot J) +\Delta S_k[\phi]$ \cite{Litim:1998nf,Litim:2011cp}, where $\phi=\langle\varphi\rangle_J$ stands for the expectation value of the quantum field in the presence of the source $J$.\footnote{In a slight abuse of notation we will use the same symbols $(h_{\mu\nu}^T , \xi,\cdots)$  to denote the components of $\phi$ and $\varphi$.} The renormalisation group scale-dependence of $\Gamma_k$,  \eq{FlowEquation}, is then obtained  from the exact  flow of the partition function $\partial_t Z_k=-\langle \partial_t \Delta S_k\rangle_J$  by standard manipulations \cite{Wetterich:1992yh,Morris:1993qb,Reuter:1996cp}.
With the above ingrediences at hand, and following textbook procedures, we are led to a path integral representation for the scale-dependent gravitational effective action 
\beq\label{Gamma}
e^{-\Gamma_k[\phi,\bar{g}]} = \int \mathcal{D} \varphi \, 
\exp\left(
{-S[\varphi;\bar{g}]   - \Delta S_k[\varphi- \phi;\bar{g}]    + \int d^dx \sqrt{\bar g}\sum_i \frac{\delta \Gamma_k[\phi, \bar{g}]}{ \delta \phi_i(x)}  (\varphi_i(x) - \phi_i(x)) }
\right)\,.
\eeq
Notice  that in \eq{Gamma}, the cutoff term \eq{cutoff} is now evaluated around the shifted fields $\varphi\to \varphi-\phi$. Technically, the representation \eq{Gamma}  takes the form of an integro-differential equation for the quantum effective action of gravity. Moreover, it is compatible with the flow equation \eq{FlowEquation}. 
Below, we will primarily be concerned with identifying  gravitational effective actions which display   {\it interacting} UV fixed points $\Gamma_*$, and, as such, qualify as fundamental candidates for a quantum theory of gravity  \cite{Litim:2011cp}.

\subsection{Wilsonian momentum cutoff}
\label{subsec:RegulatorSchemes}

The last ingredient of the flow equation \eq{FlowEquation} consists in the choice of the regulator term $\mathcal{R}_k$. In general, for each component 
field, the second variation of the effective action
$ \frac{1}{\sqrt{\bar{g}} } {\delta^2\Gamma_k}/{\delta \phi_i\delta \phi_j}$,
which we will often denote as 
$\Gamma_k^{(2)\phi_i\phi_j}$,
 will be a function of some differential operator $\Delta$. The regularised inverse propagator is then given by
\begin{equation}
\label{inverse}
\left(\Gamma^{(2)}_k\right)^{\phi_i\phi_j}+\mathcal{R}_k^{\phi_i\phi_j}\,.
\end{equation} 
The general prescription we are going to adapt is that the Wilsonian regulator term
$\mathcal{R}_k^{\phi_i\phi_j}$ for the  inverse propagator   \eq{inverse}
should be chosen in such a manner that it leads to the replacement
\begin{equation}
\Delta\to\Delta+R_k(\Delta)
\end{equation} 
for all component fields (for explicit examples, see Sect.~\ref{subsec:TheRegulatorTerm}). Here, $R_k(\Delta)$ denotes the profile function of the Wilsonian momentum cutoff. Hence, at this stage we are still left with  some freedom for choosing the cutoff 
scheme by taking different definitions for the differential operators $\Delta$, and different choices for the  profile function $R_k$. For the former, one may write the differential operator 
as 
\beq
\Delta=- \bar \nabla^2+\mathbf{E}
\eeq
where $ \bar \nabla^2$ represents the covariant derivative for both diffeomorphism invariance and all other possible gauge symmetries, and 
$\mathbf{E}$ denotes a linear map acting on the field.
Following the classification in \cite{Codello:2008vh} a type I cutoff is defined as $\mathbf{E}=0$ such that the Wilsonian regulator will only depend on $-\nabla^2$ and 
not on any endomorphism $\mathbf{E}$. Type II regulators include a non-vanishing  endomorphism $\mathbf{E}$ which is independent of the RG scale $k$. Finally, type III regulators are those where the endomorphisms  $\mathbf{E}$ additionally depend on the RG momentum scale.

To proceed we also have to make a choice for the profile function. For our calculation we are going to choose an optimised cutoff \cite{Litim:2000ci,Litim:2001up,Litim:2001fd,Litim:2006ag} 
\begin{equation}
R_k(Z)=(k^2-Z)\theta(k^2-Z) \label{OptimisedCutoff}
\end{equation}
where $Z$ denotes a suitably chosen differential operator in real space. Most of the time we will be using $Z=\Delta$ (type I cutoff). In momentum space, $Z$ is taken to be  momentum squared $q^2$.  
A key feature of the optimised cutoff is that it vanishes identically for momenta larger than the RG cutoff scale  $q^2>k^2$. For smaller momenta, it acts like a momentum-dependent mass term. Its scale derivative equally vanishes for momenta larger than the cutoff scale, and is given by
\begin{equation}\label {dOpt}
\partial_t\,R_k(Z)=2 k^2\,\theta(k^2-Z) -(\partial_t Z)\,\theta(k^2-Z) \,.
\end{equation}
The second term is absent provided that $Z$ carries no scale-dependence of its own, that is for type I and type II momentum cutoffs. It has been shown that this regulator leads to particularly good stability and convergence properties of 
the functional flows \cite{Litim:2000ci,Litim:2001fd}.  It also provides algebraic simplifications for the flows which will become 
of great importance when we will be concerned with gravity. Another key benefit of the optimised cutoff in gravity is that the  heat kernel expansions truncates at a finite order, allowing for closed expressions for the flow equations (see Sect.~\ref{sec:TraceAlgorithm}).  For a more detailed analysis of the optimised regulator benefits see 
\cite{Litim:2007jb,Litim:2010tt}. 
Since its introduction, 
the regulator \eqref{OptimisedCutoff} has been very popular in the context of quantum gravity and it was used in the previous studies of $f(R)$ 
gravity \cite{Codello:2007bd,Machado:2007ea}, too.

\section{\bf Trace computation algorithm}\label{sec:QAlgorithm}
\label{sec:TraceAlgorithm}
In this section we give a general algorithm for computing functional traces of a general operators which can be written as a Taylor series in terms of $-\Box=- \bar \nabla^\mu \bar \nabla_\mu$  up to a maximal power $p$,
\begin{equation}\label{Am}
\left(\Gamma^{(2)}_k\right)^{\phi_i\phi_j}=\sum_{m=0}^{p} \mathcal{A}^{\phi_i\phi_j}_m (-\Box)^m\, .
\end{equation}
The coefficient functions $\mathcal{A}^{\phi_i\phi_j}_m$ no longer contain derivative operators. The value of $p$ can take any integer value $1 \leq p \leq \infty $. The case $p = \infty$ can arise if we consider, for example, terms in the action of the form $\int d^dx \sqrt{g} R  \, e^{\Box/M^2} \, R$ for some mass scale $M$. Here we shall consider type I regulators explicitly however the results are easily generalised to the case of type II and III regulators. In particular for a more general regulator one replaces $-\Box \to \Delta= -\Box + \mathbf{E}$ which in turn leads to different values for the coefficients $\mathcal{A}^{\phi_i\phi_j}_m$.

\subsection{Heat kernels}\label{sec:heat}
For the computation of the functional trace in \eqref{FlowEquation} we use the heat kernel techniques 
\cite{Avramidi:2000bm,Gilkey:1995mj} which we will recall in this section. 
The RHS of the flow equation \eqref{FlowEquation} consists of the functional trace (i.e. the sum over 
all indices and the integration over all momenta) over the quantity 
$(\Gamma_k^{(2)}+\mathcal R_k)^{-1}\partial_t \mathcal R_k$. 
In general, the functional trace of a functional $W(\Delta)$ of an operator $\Delta$ is given by the sum 
\begin{equation}
\textnormal{Tr} \,W(\Delta)=\sum_nW(\lambda_n), \label{EigenvaluesSum}
\end{equation} 
where $\lambda_i$ are the eigenvalues of the operator. By introducing the Laplace anti-transform 
$W(\Delta)=\int_0^\infty dt \,e^{- t \Delta} \tilde W(t)$
we can write \eqref{EigenvaluesSum} as
\begin{equation}
\textnormal{Tr}_sW(\Delta)=\int_0^{\infty}dt\,\tilde W(t)\,\textnormal{Tr}_se^{-t\Delta} \label{TraceHeatEquation}
\end{equation}
where  $K(t)=e^{-t\Delta}$ is the heat kernel of the operator $\Delta$. In equation \eqref{TraceHeatEquation} the 
subscript $s$ denotes the spin of the field which the operator $\Delta$ acts on and takes the values $0$ for scalars, $1$ for 
vectors and $2$ for second rank tensors. The trace of the heat kernel has a well known early time ($t\to0$) asymptotic expansion given by \cite{Avramidi:2000bm}
\begin{equation}
\begin{split}
\textnormal{Tr}_se^{-t\Delta}=\frac{1}{(4\pi)^{d/2}}\int d^dx\sqrt{g}&\left[\textnormal{tr}_s\mathbf{b}_0(\Delta)t^{-\frac{d}{2}}+\textnormal{tr}_s\mathbf{b}_2(\Delta)t^{-\frac{d}{2}+1}+\ldots\right.\\
&\left.+\textnormal{tr}_s\mathbf{b}_d(\Delta)+\textnormal{tr}_s\mathbf{b}_{d+2}(\Delta)t+\ldots\right]. \label{HKbexpansion}
\end{split}
\end{equation}
The coefficients $\mathbf{b}_n$ are called the heat kernel coefficients and 
 are sums of curvature invariants and their derivatives. 
The traces of the  coefficients that we will need here evaluated on the spherical background
are given in appendix \ref{app:HeatKernel}. Now we define $B_n=\int d^dx\sqrt{g}\,\textnormal{tr}_s\mathbf{b}_n(\Delta)$ 
and $Q_n[W]=\int_0^{\infty}dt\,t^{-n}\tilde W(t)$, so that we can write equation \eqref{TraceHeatEquation} in the form
\begin{equation}
\textnormal{Tr}W(\Delta)=\frac{1}{(4\pi)^{d/2}}\sum_{n=0}^{\infty}Q_{\frac{d}{2}-n}[W]B_{2n}(\Delta).\label{GeneralQ&B}
\end{equation}
Now the functional trace is expressed as the early time expansion for the heat kernel of the operator $\Delta$ instead 
of the spectral sum \eqref{EigenvaluesSum}. This becomes apparent when we make use of the Mellin transform and we express the functional $Q_n[W]$ in terms of the original function $W(z)$. Then for every $n>0$ 
\begin{equation}
Q_n[W]=\frac{1}{\Gamma(n)}\int_0^{\infty}dz\,z^{n-1}W(z) \label{PositiveQDefinition}
\end{equation}
and similarly for $Q_{-n}$ with $n\geq0\in\mathbb{Z}$
\begin{equation}
Q_{-n}[W]=(-1)^n\left.\frac{d^nW(z)}{dz^n}\right|_{z=0}. \label{NegativeQDefinition}
\end{equation}
In the following, we develop a general algorithm for computing these functionals just with the knowledge of the second variation. 
It will then become clear that the usage of the optimised cutoff  \eqref{OptimisedCutoff} truncates the series and we end up with a closed expression for the flow equation which
is one of the great benefits of the optimised 
cutoff.  Therefore with a finite number 
of heat kernel coefficients we can derive the full flow. That the series truncates beyond some order becomes apparent from
the last formula. As we will see later, with the optimised cutoff, $W$ will be a polynomial function of $z$ with the highest power
determined by the number of derivative operators in the second variation of the action.

As will become clear later, there are some cases where we have to exclude certain modes from the trace computation due to constraints 
that some fields obey. Here we are going to state the general mechanism of how this is taken into account. More details about the 
specific exclusions that we are going to use can be found in App.~\ref{app:HeatKernel}. Since the trace of an arbitrary smooth functional $W(\mathcal{O})$ can be represented as its spectral sum, the general rule for omitting the missing modes from the trace of an operator valued function is
\begin{equation}\label{modeexclusion}
\textnormal{Tr}_s^{'...'}[W(-\nabla^2)]=\textnormal{Tr}_s[W(-\nabla^2)]-\sum_{l=1}^{l=m}D_l(d,s)W(\Lambda_l(d,s))
\end{equation} 
where the $m$ primes on the LHS indicate the first $m$ modes to be subtracted, $l$ labels the eigenvalues $\Lambda_l(d,s)$ 
in the spectrum of the operator $-\nabla^2$,  $D_l(d,s)$ gives the multiplicity of each eigenvalue, $d$ is the spacetime dimension and 
$s=0,1,2$ for scalars, vectors and tensor respectively. 
For the operator $-\nabla^2$ acting on scalars, transverse vectors and transverse-traceless symmetric tensors 
$D_l(d,s)$ and $\Lambda_l(d,s)$ are given in Table \ref{Multiplicities} of App.~\ref{app:HeatKernel}. 

In what follows we present an algorithm for computing the trace having as input the general form of the second variation. 
We then determine the form that the regulator term should have, we move on by implementing the optimised cutoff and finally we 
evaluate the integrals for the functionals $Q_n[W]$ defined in \eqref{PositiveQDefinition} and \eqref{NegativeQDefinition}. 
We also evaluate these functionals for the case that we have off-diagonal terms in the second variation with the mixing of 
two components. Finally, we compute the two lowest excluded modes for a scalar field and the lowest excluded mode for a vector field.

\subsection{Wilsonian momentum cutoff}
\label{subsec:TheRegulatorTerm}

We will now discuss the specifics of the regulator term $\mathcal{R}_k^{\phi_i\phi_j}$, which needs to be added to \eq{Am}. For convenience, we introduce the short-hand notation 
\begin{equation}
\label{G}
G_{\phi_i\phi_j}=\left(\Gamma^{(2)}_k\right)^{\phi_i\phi_j}+\mathcal{R}_k^{\phi_i\phi_j}
\end{equation} 
to denote the ${\phi_i\phi_j}$-component of the inverse regularised propagator. We will drop the $\phi_i\phi_j$ indices when it is clear from the context.  The coefficients $\mathcal{A}_m$ \eq{Am}  depend on the momentum scale $k$ through the couplings of the theory. 
For our purposes it is enough to determine the regulator for \emph{type I cutoff}. 
However, the process described here has a straightforward generalisation to the other types of cutoffs described in \cite{Codello:2008vh}. 
For \emph{type I cutoffs}  we define the Wilsonian regulator (see Sec.~\ref{subsec:RegulatorSchemes}) by demanding that the addition of the 
regulator term $\mathcal{R}_k^{\phi_i\phi_j}$ to the inverse propagator of the corresponding field leads to the replacement of 
the operator $-\Box$ by $-\Box+R_k(-\Box)$, where $R_k(-\Box)$ is a suitably chosen profile function characterising the Wilsonian momentum cutoff for the field modes. Using \eq{G} together with \eq{Am} and the requirement on the regulator, the inverse regularised propagator takes the form
\bea \label{D}
G 
&=&\sum_{m=0}^{p} \mathcal{A}_m (-\Box+R_k)^m \eea
Solving \eq{D} with \eq{G}  for $\mathcal{R}_k$, we find
\begin{equation}
\mathcal{R}_k=\sum_{m=1}^{p} \mathcal{A}_m \left[(-\Box+R_k)^m-
 (-\Box)^m\right]\label{GeneralRegulator}\,.
\end{equation}
It is readily confirmed that $R_k\to 0$ entails ${\cal R}_k\to 0$, as it must, alongside $ {\cal R}_k\to \infty$ for $R_k\to \infty$. Now that we have an explicit form for the regulator term we can proceed to determine its scale derivative, which is an essential component of the flow equation \eqref{FlowEquation}. For \emph{type I cutoff} the operator $-\Box$ does not contain any couplings and thus it does not carry any RG scale-dependence. Therefore the $\partial_t$ derivative of the regulator reads
\begin{equation}
\partial_t\mathcal{R}_k
=C_1\,\partial_tR_k+C_2\label{partialR}
\end{equation}
where 
\begin{eqnarray}
\label{C1} C_1&=&\sum_{m=1}^{p} m\mathcal{A}_m (-\Box+R_k)^{m-1}\\
\label{C2} C_2&=&\sum_{m=1}^{p} (\partial_t\mathcal{A}_m )\left[(-\Box+R_k)^m-
(-\Box)^m\right] \, .
\end{eqnarray}
In this paper, we will use one and the same shape function $R_k$ for all fields. This choice can be omitted in more general settings. 
\subsection{Diagonal pieces} 
\label{subsec:FRGDiagonal}
Now we have all the ingredients that we need in order to evaluate the trace of the flow equation. For this purpose we split the RG flow into two parts according to \eqref{partialR}. Using \eq{C1}, \eq{C2} the equation for the diagonal part reads
\begin{equation}
\frac{1}{2}\textnormal{Tr}\left(
\,G^{-1}\partial_t\mathcal{R}_k\right)
=\frac{1}{2}\textnormal{Tr}(G^{-1}\,C_1\partial_tR_k)+\frac{1}{2}\textnormal{Tr}(G^{-1}\,C_2)\label{TraceForTheDiagonalPiece}
\end{equation}
where $G$ is defined in \eq{D} in terms of the coefficients $\mathcal{A}_m$ and the profile function $R_k$.
With the optimised cutoff \eqref{OptimisedCutoff} and its
scale derivative \eq{dOpt}
we can evaluate the functionals \eqref{PositiveQDefinition} and \eqref{NegativeQDefinition} for the above expressions. We note that for a type III cutoff there will be an extra term in \eq{dOpt} proportional to $\partial_t \mathbf{E}$.
One observes that both the above integrals in \eqref{TraceForTheDiagonalPiece} will have as a common factor 
the step function $\theta(k^2-z)$ and therefore we can change the integral limits from $\int_0^{\infty}$ to $\int_0^{k^2}$ 
and replace the step functions by $1$ everywhere.  
After substituting $-\Box=z$, we then have for the coefficients $C_1$, $C_2$ and $ G$ in \eq{TraceForTheDiagonalPiece} the following results,
\begin{equation}\label{c1c2G_opt}
\begin{array}{rcl}
C_1&=&
\displaystyle
\sum_{m=1}^{p} m\mathcal{A}_m \,k^{2m-2} \\[1ex]
C_2(z)&=&
\displaystyle
\sum_{m=1}^{p} (\partial_t\mathcal{A}_m) \,(k^{2m}-z^m) \\[1ex]
G&=&
\displaystyle
\sum_{m=0}^{p} \mathcal{A}_m \,k^{2m}\ .
\end{array}
\end{equation}
Both $G$ and $C_1$ have become independent of $z$ for the optimised cutoff. Now we turn our attention to the evaluation of the integrals $Q_n[W]$ for positive $n$, \eq{PositiveQDefinition}. 
The integrand $W(z)$  in \eqref{PositiveQDefinition}  can be split in two parts according to \eqref{TraceForTheDiagonalPiece}. Defining 
$W_1=(C_1\partial_tR_k)/(2G)$ and $W_2=C_2/(2G)$
we evaluate the corresponding integrals $I_n^i=Q_n[W_i]$ separately. 
Also these two integrals contain step functions which are treated as above.
Then we have
\begin{eqnarray}
I_n^1
&=&
\frac{k^{2n}}{\Gamma(n)}\frac{1}{n}\frac{\sum_{m=1}^{p} m\mathcal{A}_m k^{2m}}{\sum_{m=0}^{p} \mathcal{A}_m k^{2m}} 
\label{I1}\\
\label{I2}
I_n^2&=&\frac12 \frac{k^{2n}}{\Gamma(n)}\frac{1}{n}\frac{\sum_{m=1}^{p}\frac{m}{m+n}\partial_t\mathcal{A}_mk^{2m}}{\sum_{m=0}^{p} \mathcal{A}_m k^{2m}}.
\end{eqnarray}
By summing the two contributions \eq{I1} and \eq{I2}  we get an expression for the coefficients $Q_n$ of the heat kernel expansion for positive $n$, 
\begin{equation}
\label{Qn+}
Q_n=I_n^1+I_n^2
\, ,\qquad (n>0)\,.
\end{equation}
For negative integer $n$, 
the $z$-dependence of $W(z)$ in the $Q$-functionals \eqref{NegativeQDefinition} comes only from the $\partial_t\mathcal{R}_k^{\phi_i\phi_j}$ term. It is then enough to compute 
the derivative of this term. In the previous sections we have disregarded the explicit form of the $\theta(k^2-z)$-functions,
whose effect, as overall factors, amounts to a change of integration limits. 
Here, we use the properties of the step function as introduced above, and additionally exploit 
its idempotence, and that its derivative
$d\theta(k^2-z)/dz=-\delta(k^2-z)$ will not contribute at $z=0$ and $k^2\neq 0$.
Altogether, we find
\begin{equation}
\frac{d^n\partial_t\mathcal{R}_k^{\phi_i\phi_j}}{d z^n}
=
-\sum_{m=n}^{p}\frac{m!\,z^{m-n}}{(m-n)!}(\partial_t\mathcal{A}_m)\,.
\end{equation}
After evaluating this expression at $z=0$, the only term that survives is the one with $m=n$. Therefore, we conclude that the coefficients $Q_{-n}$ are given by 
\begin{equation}\label{Qn-}
Q_{-n}=\frac{1}{2}\frac{(-1)^{n+1}\,n!\,\cdot\left(\partial_t\mathcal{A}_n\right) }{\sum_{m=0}^{p} \mathcal{A}_m \,k^{2m}}\, , \qquad (n\ge0)\,.
\end{equation}
The results \eq{Qn+} and \eq{Qn-} will be needed for the subsequent derivation of the RG flows of couplings.

\subsection{Non-diagonal pieces}
\label{subsec:FRGNonDiagonal}
When there are off-diagonal components in the second variation \eq{Am},
the inversion of \eq{Am} and \eq{G} become more difficult. 
For the case where a mixing arises between two field components $\phi_1$ and $\phi_2$, the inverse of the corresponding submatrix $G_{\phi_i\phi_j}$ involving the fields $\phi_1$ and $\phi_2$  is given by
\begin{equation}
\left(G_{\phi_i\phi_j}
\right)^{-1}=
\frac{1}{\textnormal{Det}(G_{\phi_i\phi_j})}
\cdot\begin{pmatrix}
\ \ G_{\phi_2\phi_2} & -G_{\phi_1\phi_2} \\
-G_{\phi_2\phi_1} & \ \ G_{\phi_1\phi_1}
\end{pmatrix}  \,.
\end{equation}
Performing the matrix multiplication in the field space $\phi_1$ and $\phi_2$, the trace takes the form
\begin{equation}
\frac{1}{2}\textnormal{Tr}\,
\left(G_{\phi_i\phi_j}\right)^{-1}\,\partial_t\mathcal{R}_k^{\phi_i\phi_j}
=\frac{1}{2}\textnormal{Tr}
\frac{{\cal M}}{\textnormal{Det}(G_{\phi_i\phi_j})}
\,,
\end{equation}
where
\beq
{\cal M}=
G_{\phi_2\phi_2}\partial_t\mathcal{R}_k^{\phi_1\phi_1} 
-G_{\phi_1\phi_2} \partial_t\mathcal{R}_k^{\phi_2\phi_1}
-G_{\phi_2\phi_1} \partial_t\mathcal{R}_k^{\phi_1\phi_2}
+ G_{\phi_1\phi_1}\partial_t\mathcal{R}_k^{\phi_2\phi_2}
\eeq
The expressions for $G_{\phi_i\phi_j}$ \eq{D}, for the regulator \eqref{GeneralRegulator} and the coefficients of the expansion \eqref{partialR} remain the same as given previously, except that we restore the indices $\phi_i\phi_j$ in all explicit expressions involving the momentum cutoff. 
For the remaining step we   assume that the second variation $G_{\phi_i\phi_j}$ is symmetric. Adopting the optimised momentum cutoff \eq{OptimisedCutoff}, $G$ becomes  $z$-independent and no longer contributes to integration over $z$, see \eq{c1c2G_opt}. The only remaining $z-$integration is the one over $\partial_t\mathcal{R}_k$, which is known from the previous section. Putting everything together, 
we find that the coefficients $Q_n$ for the non-diagonal piece for positive and negative index are given by
\begin{eqnarray}
Q_n&=&\frac{1}{2\,\Gamma(n)}\cdot\frac{
G_{\phi_2\phi_2}\mathcal{I}^{\phi_1\phi_1}
-2G_{\phi_1\phi_2}\mathcal{I}^{\phi_1\phi_2}
+G_{\phi_1\phi_1}\mathcal{I}^{\phi_2\phi_2}}{G_{\phi_1\phi_1}G_{\phi_2\phi_2}-(G_{\phi_1\phi_2})^2}\,,\quad (n>0)\\
Q_{-n}&=&(-1)^{n+1}\,\frac{n!}{2}
\frac{
G_{\phi_2\phi_2}\mathcal{K}^{\phi_1\phi_1}
-2G_{\phi_1\phi_2}\mathcal{K}^{\phi_1\phi_2}
+G_{\phi_1\phi_1}\mathcal{K}^{\phi_2\phi_2}}{G_{\phi_1\phi_1}G_{\phi_2\phi_2}-(G_{\phi_1\phi_2})^2}\,,
\quad (n\ge 0)\,,
\end{eqnarray}
respectively. Here, we have used the expressions
\begin{eqnarray}
\mathcal{I}^{\phi_i\phi_j}&=&\frac{k^{2n}}{n}\left[2\sum_{m=1}^{p} m\mathcal{A}^{\phi_i\phi_j}_m k^{2m}+\sum_{m=1}^{p}\frac{m}{m+n}(\partial_t\mathcal{A}^{\phi_i\phi_j}_m) k^{2m}\right]\\
\mathcal{K}^{\phi_i\phi_j}&=&\partial_t\mathcal{A}^{\phi_i\phi_j}_n
\end{eqnarray}
and $G_{\phi_i\phi_j}$ is given as in \eq{c1c2G_opt}.

\subsection{Excluded modes}
\label{subsec:ExcludedModes}

As explained in Sec.~\ref{sec:heat} we often encounter the trace of a function where some eigenmodes of the 
operator $-\Box$ 
have to be excluded. For example we saw in  Sec.~\ref{subsec:Decompositions} that after performing the decomposition of 
the fluctuation $h_{\mu\nu}$ into its components through the transverse traceless decomposition, the lowest mode of
$\xi^{\mu}$ and the two lowest modes of $\sigma$ have to be excluded. 
In order to incorporate this fact into the evaluation of the traces we make use of  \eq{modeexclusion}.
For our calculation we will need the form of $\textnormal{Tr}_{(0)}^{'}$, $\textnormal{Tr}_{(0)}^{''}$and
$\textnormal{Tr}_{(1T)}^{'}$. We start with the case where the lowest mode of a scalar has to be excluded. 
According to Tab.~\ref{Multiplicities} in App.~\ref{app:HeatKernel}, for the lowest mode of a scalar field we have
\begin{equation}
\textnormal{Tr}_{(0)}^{'}[W(-\Box)]=\textnormal{Tr}_{(0)}[W(-\Box)]-W(0).
\end{equation}
The part $W(0)$ of the above equation which has to be excluded from the scalar trace is simply evaluated by setting $z=0$ in the 
expression for $W(z)$. Then we denote the lowest excluded mode of a scalar as $X^{'}_{(0)}$ and we have
\begin{equation}
X^{'}_{(0)}=W(0)=\frac{\sum_{m=1}^{p}\partial_t\mathcal{A}_mk^{2m}}{\sum_{m=0}^{p}\mathcal{A}_mk^{2m}}+\frac{\sum_{m=1}^{p}m\mathcal{A}_mk^{2m}}{\sum_{m=0}^{p}\mathcal{A}_mk^{2m}}\ .
\end{equation}
For the exclusion of the two lowest modes of a scalar field we have according to Tab.~\ref{Multiplicities} 
\begin{equation}
\textnormal{Tr}_{(0)}^{''}[W(-\Box)]=\textnormal{Tr}_{(0)}[W(-\Box)]-W(0)-(d+1)\,W\left(\frac{1}{d-1}R\right)\ .
\end{equation}
In general, the result will involve theta functions coming from the choice of the profile function and more precisely from 
\eqref{OptimisedCutoff}
and \eqref{dOpt}. For the specific expression under consideration the theta functions are of 
the form $\theta(k^2-\frac{R}{d-1})$. Since in the following we are going to focus on an expansion in small $\frac{R}{k^2}$ we 
can evaluate these theta functions in the approximation $\frac{R}{k^2}\ll 1$ in order to get for the exclusion of the two lowest 
scalar modes
\begin{equation}
\begin{split}
X^{''}_{(0)}=X^{'}_{(0)}+(d+1)\frac{1}{\sum_{m=0}^{p}\mathcal{A}_mk^{2m}}&\left[2\sum_{m=1}^{p}m\mathcal{A}_mk^{2m}+\sum_{m=1}^{p}\partial_t\mathcal{A}_m\left(k^{2m}-\frac{R^m}{(d-1)^m}\right)\right].
\end{split}
\end{equation}
Finally we have to determine the exclusion for the lowest mode of a transverse vector. Again, by reading off the multiplicities and the eigenvalues from Tab.~\ref{Multiplicities}, we find
\begin{equation}
\textnormal{Tr}_{(1T)}^{'}[W(-\Box)]=\textnormal{Tr}_{(1T)}[W(-\Box)]-\frac{d(d+1)}{2}W\left(\frac{R}{d}\right)
\end{equation}
As before we evaluate the theta functions coming from the profile function for the approximation  $\frac{R}{k^2}\ll 1$ and we have for the lowest exclusion mode of the vector
\begin{equation}
X^{'}_{(1T)}=\frac{d(d+1)}{2}\frac{1}{\sum_{m=0}^{p}\mathcal{A}_mk^{2m}}\left[2\sum_{m=1}^{p}m\mathcal{A}_mk^{2m}+\sum_{m=1}^{p}\partial_t\mathcal{A}_m\left(k^{2m}-\frac{R^m}{d^m}\right)\right]\,.
\end{equation}
This completes the collection of the technical tools that we need to calculate the renormalisation group flow for certain general classes of gravitational actions projected 
on the sphere. In the next section we will apply this formalism for a concrete model of quantum gravity involving Ricci tensor invariants.

\section{\bf Ricci tensor expansion}
\label{SecRmunu}

In this Section, we derive RG flows for theories of gravity whose actions depend on certain Ricci tensor invariants, generalising the $f(R)$ type actions towards more complicated curvature invariants. 
This will give us new information on gravitational renormalisation group flows complementary to existing results in the $f(R)$-approximation.
A new qualitative feature is that now the propagator of the tensor modes will also contain higher-derivative contributions in contrast to the $f(R)$-approximation.

\subsection{Hessian}

Here we are going to use the methods developed in the previous Section to derive the Hessians for all the fields that contribute to our 
ansatz. As explained in Sect.~\ref{subsec:Approximations} the effective average action takes the form
\begin{equation}
\Gamma_k[g,\bar g, c, \bar c]=\bar \Gamma_k[g]+S_{\rm gf}[h_{\mu\nu};\bar g]+S_{\rm gh}[h_{\mu\nu},C_\mu,\bar C_\nu;\bar g]  + S_{\rm aux}. \label{AnsatzRM}
\end{equation}
with $S_{\rm gf}[h_{\mu\nu};\bar g]$, $S_{\rm gh}[h_{\mu\nu},C_{\mu},\bar C_{\nu};\bar g]$ and $S_{\rm aux}$ taking the forms specified in Sect.~\ref{sec:RG}.
Consequently, in our approximation, the Hessians for the ghost and auxiliary fields are  equal to the ones obtained from the "bare" action and given by \eq{bare_gh_hessians},   \eq{bare_gh_hessians_eta}, \eq{bare_aux0_hessian}, \eq{bare_aux1_hessian}  and \eq{bare_auxgh_hessian}, meaning that $\Gamma^{(2)}_{\phi^i \phi^j} = S^{(2)}_{\phi^i \phi^j}$ for these fields. The missing element is the computation of the second variation for the gravitational part  $\bar \Gamma_k^{(2)}[g]+S_{\rm gf}^{(2)}[h;\bar g]$ since we are yet to fix $\bar \Gamma_k[g]$, which we termed $S_{\rm grav}$ in \eq{S}. To that end, we assume that the gravitational effective average action is formed of two unspecified functions of the square of the Ricci tensor  $F_k(R_{\mu\nu}R^{\mu\nu})$ and $Z_k(R_{\mu\nu}R^{\mu\nu})$, with the latter one additionally multiplied by the Ricci scalar
\begin{equation}\label{ansatzeaa}
\bar\Gamma_k[g]=\int d^dx
\sqrt{g}\left(F_k(R_{\mu\nu}R^{\mu\nu})+R\,Z_k(R_{\mu\nu}R^{\mu\nu})\right)\,.
\end{equation}
Once the equation is evaluated on a spherical background and expanded in power of the curvature we then have exactly one invariant $\mathcal{O}_n \propto R^n$ (see \eq{Oinvariant}) for all integer values of $n$. The ansatz \eq{ansatzeaa} provides a natural extension to preceding work on the effective average action in gravity including more complicated tensor structures with non-trivial dynamics.
In order to compute the Hessians and to evaluate the flow equation we need to know $\Gamma^{(2)}_k$. 

To proceed we expand the
 gravitational action $\bar\Gamma_k[g]$ as
\begin{equation}
\bar\Gamma_k[\bar g+\bar h;\bar g]=\bar \Gamma_k[\bar g; \bar g]+ \mathcal{O}(h)+\bar \Gamma_k^{\rm{quad}}[\bar g + h_{\mu\nu} ;\bar g]+\mathcal{O}(h^3).
\end{equation} 
to extract the part quadratic in $h_{\mu\nu}$ which takes the form 
\begin{eqnarray}
\bar\Gamma^{\rm{quad}}_k&=&\int d^d x\left(\delta^{(2)}(\sqrt{g})\left[F_k+R\,Z_k\right]+2\delta(\sqrt{g})\delta(R_{\mu\nu}R^{\mu\nu})\left[F_k'+R\,Z_k'\right]\right.\nonumber\\
&&\left.+\sqrt{g}(\delta(R_{\mu\nu}R^{\mu\nu}))^2\left[F_k''+R\,Z_k''\right]+\sqrt{g}\delta^{(2)}(R_{\mu\nu}R^{\mu\nu})\left[F_k'+R\,Z_k'\right]\right.\nonumber\\
&&\left.+\sqrt{g}\delta^{(2)}(R)\,Z_k+2\delta(\sqrt{g})\delta(R)\,Z_k+\sqrt{g}\delta(R)\delta(R_{\mu\nu}R^{\mu\nu})Z_k'\right)\ ,\label{QuadraticPartFR}
\end{eqnarray}
where all the geometric quantities of the above equation are constructed from the background metric with $\delta g_{\mu\nu} = h_{\mu\nu}$. From now on we drop the bar notation since we identify the metric with the background. Moreover it is understood from now on that $F_k\to F_k(R_{\mu\nu}R^{\mu\nu})$ as well as $Z_k\to Z_k(R_{\mu\nu}R^{\mu\nu})$ and that the primes denote derivatives with respect to the argument. Using the expressions in App.~\ref{app:Variations} we find that the quadratic part takes the form
\begin{equation}
\begin{split}
\bar\Gamma^{\rm{quad}}_k=&\int d^d x \sqrt{g}\,\left\{  h_{\mu\nu}\left\{  \frac{1}{2}(F_k'+R\,Z_k')\nabla^4
+\left[\frac{d-3}{d-1}\frac{R}{d}(F_k'+R\,Z_k')+\frac{1}{2}Z_k\right] \nabla^2
\right.\right. \\
&\qquad\quad\left.\left.+2\frac{d^2-3d+3}{(d-1)^2}\frac{R^2}{d^2}(F_k'+R\,Z_k')-\frac{1}{2}F_k\right\}h^{\mu\nu}  
\right.\\
&\left.+
\,h \left\{  \left[\frac{1}{2}(F_k'+R\,Z_k')+4\frac{R}{d}\left(\frac{R}{d}F_k''
+Z_k'+\frac{R^2}{d}Z_k''\right)\right]\nabla^4
\right.\right.\\
&\qquad\quad\left.\left.+
\left[-\frac{d-5}{2(d-1)}\frac{R}{d}(F_k'+R\,Z_k')\nabla^2+8\frac{R^2}{d^2}\left(\frac{R}{d}F_k''+Z_k'
+\frac{R^2}{d}Z_k''\right)-\frac{1}{2}Z_k\right]\nabla^2
\right.\right.\\
&\qquad\quad\left.\left.
-2\frac{(d-2)^2}{(d-1)^2}\frac{R^2}{d^2}(F_k'+R\,Z_k')+4\frac{R^3}{d^3}\left(\frac{R}{d}F_k''+Z_k'
+\frac{R^2}{d}Z_k''\right)
\right.\right.\\
&\qquad\quad\left.\left.
+\frac{1}{4}(F_k+R\,Z_k)-\frac{d-2}{d-1}\frac{R}{d}\,Z_k\right\}h
\right.\\
&+\left.(\nabla^{\mu}\nabla^{\nu}h_{\mu\nu}) \left\{ \left[-(F_k'+R\,Z_k')-8\frac{R}{d}\left(\frac{R}{d}F_k''+Z_k'
+\frac{R^2}{d}Z_k''\right)\right]\nabla^2
\right.\right.\\
&\qquad\quad+\left.\left.
2\frac{R}{d}(F_k'+R\,Z_k')-8\frac{R^2}{d^2}\left(\frac{R}{d}F_k''+Z_k'+\frac{R^2}{d}Z_k''\right)
+Z_k\right\}h
\right.\\
&+\left.
(\nabla_{\alpha}h^{\alpha\beta})\left\{[F_k'+R\,Z_k']\nabla^2+3\frac{R}{d}(F_k'+R\,Z_k')+Z_k\right\}
(\nabla^{\mu}h_{\mu\beta})
\right.\\
&+\left. 
(\nabla^{\mu}\nabla^{\nu}h_{\mu\nu})\left\{F_k'+R\,Z_k'+4\frac{R}{d}\left(\frac{R}{d}F_k''
+Z_k'+\frac{R^2}{d}Z_k''\right)\right\}(\nabla^{\alpha}\nabla^{\beta}h_{\alpha\beta})
\right\} \label{SecondVariation}
\end{split}
\end{equation}
 We then use the metric field decomposition as defined in Sect.~\ref{subsec:Decompositions} to find the Hessians in terms of each component field. These are given in App.~\ref{app:Variations}. In Tab.~\ref{FRmunuSummary} we summarise the contribution from each individual component field after adding the contributions from the gauge fixing part, the ghost part and the auxiliary fields.


\begin{center}
\addtolength{\tabcolsep}{2pt}
\begin{table*}[ht]
\scalebox{0.75}{
\setlength{\extrarowheight}{10pt}
\normalsize
\begin{tabular}{c c} 
\toprule 
\rowcolor{Yellow}
$\bf fields\  \bm{\phi_i\phi_j}$
& 
\bf matrix element $\bm{\left(\Gamma^{(2)}_k\right)^{\phi_i\phi_j}}$
\\ 
\toprule
&\\
{\multirow{-2}{*}{\bm{$h^{T\mu\nu}\,h^T_{\mu\nu}$} }}
&{\multirow{-2}{*}{\bm{$\left(\frac{1}{2}\Box^2+\frac{(d-3)R}{d(d-1)}\Box+\frac{2(d^2-3d+3)R^2}{d^2(d-1)^2}\right)(F_k'+R\,Z_k')
-\frac{1}{2}(F_k+R\,Z_k)
+\left(\frac{1}{2}\Box
+\frac{d-2}{d(d-1)}R\right)Z_k
$}}}
\\ \midrule 
\rowcolor{LightGray}&\\
\rowcolor{LightGray}{\multirow{-2}{*}{\bm{$\xi^{\mu}\,\xi_{\mu}$}}} & 
{\multirow{-2}{*}{ 
\bm{$\left[\frac{1}{\alpha}(\Box+\frac{R}{d})-\frac{4}{d^2}R^2(F_k'+R\,Z_k')+F_k+R\,Z_k-\frac{2}{d}R\,Z_k\right](\Box+\frac{R}{d})$}}}
\\ \midrule 
{\multirow{4}{*}{ $\bm{\sigma\,\sigma}$}} 
&{\multirow{2}{*}{\bm{$\frac{d-1}{2d}\left[-\frac{2(d-1)}{\alpha d} (\Box+\frac{R}{d-1})
+(\Box^2-\frac{2(d-4)R}{d^2}\Box+\frac{4R^2}{d^2}) (F_k'+R\,Z_k')
-(F_k+R\,Z_k)
\right.$}}}\\&\\
&{\multirow{2}{*}{\bm{$\left.
 +\frac{8(d-1)R}{d^2}\Box(\Box+\frac{R}{d-1})\left(\frac{R}{d}F_k''+Z_k'+\frac{R^2}{d}Z_k''\right)
\right] (\Box+\frac{R}{d-1})\Box
+\frac{R}{2 d}(\Box+\frac{2R}{d})\Box Z_k$}}}
\\&
\\ \midrule 
\rowcolor{LightGray}
& 
\\
\rowcolor{LightGray}&{\multirow{-2}{*}{\bm{$-\frac{\delta^2}{\alpha}\frac{1}{d^2}\Box
+\left(\frac{d-1}{2d}\Box^2-\frac{(d^2-10d+8)R}{2d^3}\Box-2\frac{(d-3)R^2}{d^3}\right)(F_k'+R\,Z_k')$}}}\\
\rowcolor{LightGray}&
\\
\rowcolor{LightGray}
{\multirow{-4}{*}{\bm{$h\,h$}}} &
{\multirow{-2}{*}{\bm{$+\frac{4(d-1)^2R}{d^3}(\Box+\frac{R}{d-1})^2\left(\frac{R}{d}F_k''+Z_k'+\frac{R^2}{d}Z_k''\right)
+\frac{d-2}{4d}(F_k+R\,Z_k)-\frac{(d-1)(d-2)}{2d^2}(\Box+\frac{2R}{d-1})\,Z_k
$}}}
\\\midrule 
{\multirow{2}{*}{ $\bm{h\,\sigma$} }}
&
{\multirow{2}{*}{\bm{$\frac{d-1}{d}\left[\frac{2\delta}{d\alpha}
-(\Box-\frac{2(d-4)R}{d^2})(F_k'+R\,Z_k')
 -\frac{8(d-1)R}{d^2}(\Box+\frac{R}{d-1})\left(\frac{R}{d}F_k''+Z_k'+\frac{R^2}{d}Z_k''\right)
+\frac{d-2}{d} Z_k\right]
(\Box+\frac{R}{d-1})\Box
$}}}
\\&
\\
\bottomrule
\end{tabular}}
\caption{Summary of second variations of the gravitational action, suitably decomposed into independent component fields.}\label{FRmunuSummary}
\end{table*} 
\end{center}

\begin{center}
\addtolength{\tabcolsep}{2pt}
\begin{table*}[h!]
\scalebox{1}{
\setlength{\extrarowheight}{10pt}
\normalsize
\begin{tabular}{c c} 
\toprule 
\rowcolor{Yellow}
{\bf fields}\ $\bm{\phi_i\phi_j}$
& 
{\bf matrix element $\bm{\left(\Gamma^{(2)}_k\right)^{\phi_i\phi_j}}$}
\\ 
\toprule
\rowcolor{LightGray}
{\bm{$\bar{\bm{C}}^{\bm T}_{\bm \mu}\,\bm{C^{T,\mu} }$}}
&
{\bm{$-\sqrt{2}\left(\Box+\frac{R}{d}\right)$}} \\
\midrule
{$\bar{\bm \eta}\,\bm \eta$}
&
\bm{$\frac{2\sqrt{2}}{d}\left[d-\delta-1\right]\left(\Box+\frac{R}{d-\delta-1}\right)\Box$}
\\
\midrule 
\rowcolor{LightGray}
$\bar{\bm\lambda}\,\bm\lambda$& 
$\bm{\frac{d-1}{d}\left(\Box+\frac{R}{d-1}\right)\Box}$
\\ \midrule
$\bm{\omega\,\omega}$
&
$\bm{\frac{d-1}{d}\left(\Box+\frac{R}{d-1}\right)\Box}$
\\
\midrule
\rowcolor{LightGray}
$\bar{\bm c}^{\bm T}_{\bm \mu}\,\bm{c^{T\mu}}$& 
\bm{$\Box+\frac{R}{d}$}
\\ \midrule
$\bm{\zeta^T_{\mu}\,\zeta^{T\mu}}$&
$\Box+\frac{R}{d}$
\\ \midrule 
\rowcolor{LightGray} 
$\bar{\bm s}\, \bm s$&
\bm{$-\Box$}
\\
\bottomrule
\end{tabular}}
\caption{Summary of second variations (continued) of the gravitational action, suitably decomposed into independent component fields.}\label{FRmunuSummary2}
\end{table*} 
\end{center}

\subsection{Gravitational flow equation}
\label{sec:FlowEquationRM}

Having evaluated the second variation for our ansatz we are ready to compute the flow equation using the machinery developed in 
Sect.~\ref{sec:TraceAlgorithm}. We need to introduce dimensionless variables for the two functions that make up the effective 
average action. Note that after taking the second variation and before computing the trace we evaluate all the expressions 
on the spherical background metric. Thus the arguments of the functions $F_k$ and $Z_k$ become $\frac{R^2}{d}$ and we will mostly drop
the arguments. Then we define
\begin{eqnarray}\label{deff}
f\left(\frac{\R^2}{d}\right)
&=&\frac{1}{16\pi}k^{-d}F_k\left(\frac{\bR^2}{d}\right)\\
\label{defz}
 z\left(\frac{\R^2}{d}\right)
 &=&\frac{1}{16\pi}k^{-d+2}Z_k\left(\frac{\bR^2}{d}\right)\,.
\end{eqnarray}
Here, and in order to avoid a proliferation of symbols, we denote the dimensionful Ricci scalar as $\bR$, and the dimensionless Ricci scalar curvature as $R$, with
\beq\label{rho}
R= \frac{\bR}{k^2}\,.
\eeq 
Hence, from now on, $R$ denotes the Ricci scalar in units of the RG scale $k$. 
For the RG scale derivatives of the functions $F_k$ and $Z_k$ we have 
\begin{eqnarray}
\partial_tF_k^{(n)}&=&16\pi k^{d-4n}\left( (d-4\,n) f^{(n)}-4\frac{\R^2}{d}f^{(n+1)}+\partial_tf^{(n)}\right)\\
\partial_tZ_k^{(n)}&=&16\pi k^{d-4n-2}\left( (d-4\,n-2) z^{(n)}-4\frac{\R^2}{d}z^{(n+1)}+\partial_tz^{(n)}\right).
\end{eqnarray}
For the dimensionless functions $f$ and $z$ the arguments become $\frac{\R^2}{d}$. Moreover, the notation $f^{(n)}$ denotes the $n$th derivative of $f$  with respect to its argument. Gathering together all the above we can compute the LHS of the flow equation for the effective average action given by \eqref{AnsatzRM}. Then we have
\begin{equation}\label{LHS}
\partial_t\bar\Gamma_k=\frac{24\pi}{\R^2}\left[4f+2\R\, z-\R^2\left(f'+\R\, z'\right)+\partial_tf+\R\,\partial_tz\right]\,.
\end{equation} 
The RHS of the equation is given by the sum of the traces for the individual components after using the algorithm described in Sec.~\ref{sec:TraceAlgorithm} and subtracting the excluded modes, as indicated. Then, in terms of the components we find
\bea
\partial_t\Gamma[\bar g,\bar g]&=&\frac{1}{2}\textnormal{Tr}_{(2T)}\left[\frac{\partial_t\mathcal{R}_k^{h^Th^T}}{\Gamma^{(2)}_{h^Th^T} + \mathcal{R}_k^{h^Th^T} }\right]+\frac{1}{2}\textnormal{Tr}^{'}_{(1T)}\left[\frac{\partial_t\mathcal{R}_k^{\xi\xi}}{\Gamma^{(2)}_{\xi\xi} + \mathcal{R}_k^{\xi\xi}}\right]+\frac{1}{2}\textnormal{Tr}^{''}_{(0)}\left[\frac{\partial_t\mathcal{R}_k^{\sigma\sigma}}{\Gamma^{(2)}_{\sigma\sigma} + \mathcal{R}_k^{\sigma\sigma} }\right]   \nonumber\\ && +\frac{1}{2}\textnormal{Tr}_{(0)}\left[\frac{\partial_t\mathcal{R}_k^{hh}}{\Gamma^{(2)}_{hh} + \mathcal{R}_k^{hh} }\right]
+\textnormal{Tr}^{''}_{(0)}\left[\frac{\partial_t\mathcal{R}_k^{\sigma h}}{\Gamma^{(2)}_{\sigma h} + \mathcal{R}_k^{\sigma h} }\right]
 -\textnormal{Tr}_{(0)}^{''}\left[\frac{\partial_t\mathcal{R}_k^{\bar\lambda\lambda}}{\Gamma^{(2)}_{\bar \lambda\lambda}   +   \mathcal{R}^{\bar \lambda \lambda} }\right]
\nonumber \\ &&
+\frac{1}{2}\textnormal{Tr}_{(0)}^{''}\left[\frac{\partial_t\mathcal{R}_k^{\omega\omega}}{\Gamma^{(2)}_{\omega\omega} +   \mathcal{R}^{\omega \omega }   }\right]   -\textnormal{Tr}_{(1T)}^{'}\left[\frac{\partial_t\mathcal{R}_k^{\bar c^Tc^{T}}}{\Gamma^{(2)}_{\bar c^Tc^{T}}   + \mathcal{R}^{\bar c^Tc^{T}}   }\right]+\frac{1}{2}\textnormal{Tr}_{(1T)}^{'}\left[\frac{\partial_t\mathcal{R}_k^{\zeta^T\zeta^{T}}}{\Gamma^{(2)}_{\zeta^T\zeta^{T}}   + \mathcal{R}^{\zeta^T \zeta^T}    }\right]  \nonumber \\ &&
    -\textnormal{Tr}_{(1T)}^{'}\left[\frac{\partial_t\mathcal{R}_k^{\bar C^TC^{T}}}{\Gamma^{(2)}_{\bar C^TC^{T}}   + \mathcal{R}^{\bar C^TC^{T}}   }\right]   -\textnormal{Tr}_{(0)}^{''}\left[\frac{\partial_t\mathcal{R}_k^{\bar\eta\eta}}{\Gamma^{(2)}_{\bar\eta\eta}  + \mathcal{R}^{\bar \eta \eta}   }\right] +   \textnormal{Tr}_{(0)}^{''}\left[\frac{\partial_t\mathcal{R}_k^{\bar ss}}{\Gamma^{(2)}_{\bar ss}    + \mathcal{R}^{\bar s s}   }\right]\label{OperatorTraceFR}
\eea
where the Hessians for each field component are given in Tab.~\ref{FRmunuSummary} and~\ref{FRmunuSummary2}. The primes at the traces denote the number of 
lowest modes to be excluded as described in Sect.~\ref{sec:QAlgorithm} and the traces are to be computed using the algorithm 
for the $Q$-functionals and the relevant $\mathbf{b}_n$ coefficients listed in App.~\ref{app:HeatKernel}. Moreover, the auxiliary fields have inherited the primes from the fields from which they originate. We note that the prescription to additionally leave out  the lowest two modes from the ghosts as detailed in Sect.~\ref{subsec:GaugeFixingPart} has lead to an additional prime on each of the three traces in the last line of \eq{OperatorTraceFR}. 

In order to simplify the computation of the traces we fix the gauge and focus on a specific spacetime dimension $d=4$. For the gauge parameters, we choose
\begin{eqnarray}\label{gauge}
\alpha&\to&0\\
\nonumber
\delta&=&0\ .
\end{eqnarray}
This choice of gauge results in three simplifications of the flow equation. Firstly, since we take the Landau gauge  limit $\alpha\to0$ the gauge fixing terms diverge. 
This entails that the Landau gauge is a fixed point under the renormalisation group flow \cite{Litim:1998qi}. Secondly,
since terms proportional to $\frac{1}{\alpha}$ are also included in the regulator, only the terms proportional to $\frac{1}{\alpha}$ survive in the limit $\alpha\to
0$. As a result the non-diagonal term $\sigma h$ vanishes since it has no dependence on $\alpha$ (for $\delta=0$) while the denominator which involves the components $hh$ and $\sigma \sigma$ becomes very large. Finally, the limit \eq{gauge} entails  that the physical and the gauge degrees of freedom totally decouple. The gravitational degrees of freedom are encoded in $h^Th^T$ and $hh$, while the gauge degrees of freedom are contained in $\xi\xi$ and $\sigma\sigma$.
Together with several cancellations which occur in the gauge \eq{gauge} the flow equation \eq{OperatorTraceFR} simplifies substantially and takes the final form
\bea
\partial_t\Gamma[\bar g,\bar g]&=&\frac{1}{2}\textnormal{Tr}_{(2T)}\left[\frac{\partial_t\mathcal{R}_k^{h^Th^T}}{\Gamma^{(2)}_{h^Th^T} + \mathcal{R}_k^{h^Th^T} }\right]  +\frac{1}{2}\textnormal{Tr}_{(0)}\left[\frac{\partial_t\mathcal{R}_k^{hh}}{\Gamma^{(2)}_{hh} + \mathcal{R}_k^{hh} }\right]
\nonumber \\ && - \frac{1}{2} \textnormal{Tr}_{(0)}^{''}\left[\frac{\partial_t R_k }{  - \Box - \frac{R}{3}   +   R_k }\right]  - \frac{1}{2} \textnormal{Tr}_{(1T)}^{'}\left[\frac{\partial_k R_k }{ -\Box - \frac{R}{4}   + R_k   }\right] \,.
\label{OperatorTraceFR_delta0}
\eea
After performing the trace computation on the RHS of \eq{OperatorTraceFR_delta0}  and combining with the corresponding LHS given in \eq{LHS}, we obtain the explicit flow for the two functions $f$ and $z$ introduced in \eq{deff} and \eq{defz} as 
\begin{equation}
\partial_t f+ 4 f-\R^2\,f'+\R\left(\partial_t z+2  z-\R^2  z' \right)=I[f,z](\R) \,.\label{FlowEquationRM}
\end{equation}
Here, the expression on the RHS encodes the contributions from fluctuations and it splits in several parts as
\begin{equation}
\begin{split}
I[f,z](\R)=&I_0[f,z](\R)+\partial_tz\,I_1[f,z](\R)+\partial_tf'\,I_2[f,z](\R)+\partial_tz'\,I_3[f,z](\R)\\
&+\partial_tf''\,I_4[f,z](\R)+\partial_tz''\,I_5[f,z](\R)\ .\label{TheFlowEquation}
\end{split}
\end{equation}
All terms appearing in \eqref{TheFlowEquation} arise due to quantum fluctuations, and are summarised explicitly in App.~\ref{AppFlow}. Here, we only point out the following features. Firstly, at a fixed point, all terms proportional to the functions $I_i$ with $i\neq 0$ no longer contribute, because they are proportional to the flows of $f$ and $z$, which both vanish. Therefore, fixed points are solely determined through the function $I_0[f,z]$. Secondly, we observe that all terms in \eqref{TheFlowEquation}  are of homogeneity degree zero in $f$ and $z$, meaning that 
\beq\label{homo}
I[a\cdot f,a\cdot z]=I[f,z]
\eeq
for any $a\neq 0$. For this reason, the RHS becomes negligible in the limit $1/f\to 0$ and $1/z\to 0$, corresponding to the classical limit where quantum fluctuations are absent. Thirdly, the scaling exponents of the theory are sensitive to small deviations form the fixed point. To determine the scaling exponents, we also need to know the functions  $I_i$, $i=1,\cdots 5$, in addition to the fixed point solution itself. The fixed point and the scaling exponents are determined in Sect.~\ref{fixedpoints} and Sect.~\ref{scaling}, respectively.

\section{\bf Interacting fixed points}\label{fixedpoints}

In this section, we turn to a detailed analysis of fixed points of the gravitational flow \eq{FlowEquationRM} and \eq{TheFlowEquation}
for gravitational actions of the form \eqref{AnsatzRM}, \eq{ansatzeaa}, particularly  in view of the asymptotic safety conjecture. Interacting ultraviolet fixed points of the gravitational effective action $\Gamma_*$ allow for a fundamental definition of quantum gravity, \eq{S}. We discuss the classical and quantum fixed points, using polynomial expansions and numerical integration techniques.

\subsection{Classical fixed points}
We begin by discussing the classical fixed points in the absence of fluctuations, following \cite{Falls:2014tra}. In particular, the RG flow \eq{FlowEquationRM} simplifies and becomes
\begin{equation}
\partial_t f+ 4 f-\R^2\,f'+\R\left(\partial_t z+2  z-\R^2  z' \right)=0 
\label{classical}
\end{equation}
Most notably, we observe that while the functions $f$ and $z$ mix on the quantum level, the mixing is absent  in the classical limit. Consequently, the RG flow \eq{classical} can be solved analytically by introducing the new function
\begin{equation}
\label{fbar}
\bar f(\R)= f({\R^2}/4)+\R\cdot z(\R^2/4)\,,
\end{equation}
in terms of which \eq{classical} becomes
\begin{equation}
\left(\partial_t+ 4 -2 \R\,\partial_\R\right)\,\bar f=0 \,.
\label{classical2}
\end{equation}
Alternatively, we can write down a flow for the inverse,
\begin{equation}
\left(\partial_t- 4 -2 \R\,\partial_\R\right)\,(\bar f^{-1})=0 \,.
\label{classical3}
\end{equation}
Solutions and fixed points of \eq{classical2} and of \eq{classical3} have been analysed in \cite{Falls:2014tra}. There, it has been shown that theories described by \eq{classical2}  develop a Gaussian fixed point $\bar f_*=0$ and an infinite Gaussian fixed point $1/\bar f_*=0$.
The most general solution of \eq{classical2} is given by
\beq\label{FPclassical}
\bar f(\R,t)=\R^2 \cdot H\left(\R e^{2t}\right)
\eeq
for arbitrary function $H(x)$ which is solely determined by the boundary conditions at some reference energy scale $t=0$.

\begin{figure}[t]
\centering
\begin{center}
\includegraphics[width=.8\hsize]{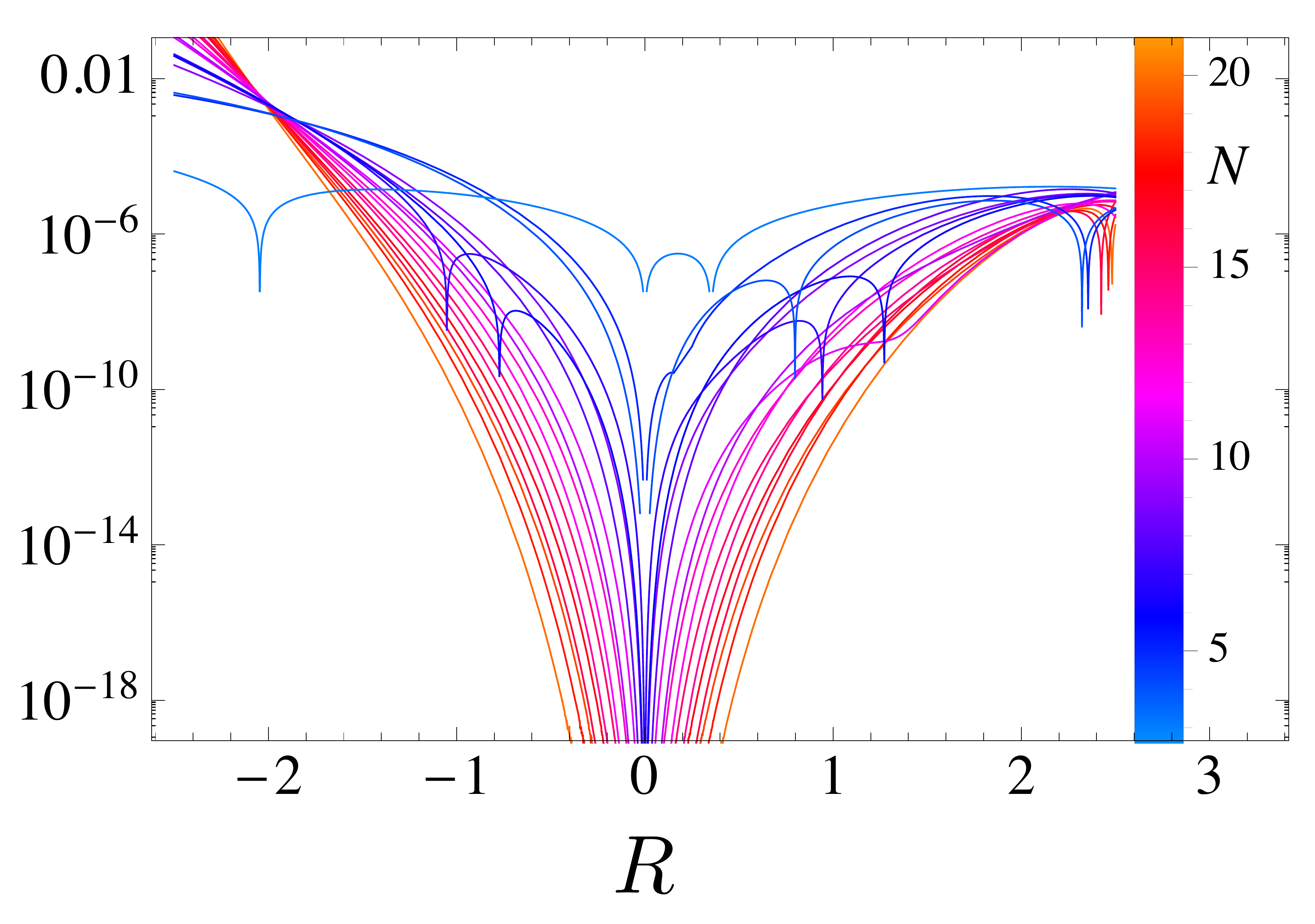}
\caption{\label{pAcc} The accuracy of the fixed point solution, showing the absolute value of \eq{FPequation} for different orders in the polynomial approximation $N$ with colour-coding detailed in the inset. We observe that the accuracy in field space continues to grow with the approximation order $N$ as long as the scalar curvature remains within the radius of convergence of the polynomial expansion \eq{Rb}.}
\end{center}
\end{figure}

\subsection{Quantum fixed points}

Next we turn our attention to quantum fixed points in the presence of fluctuations.
At an interacting fixed point of the theory we observe $\partial_t f = 0 = \partial_t z$ and the RG flow reduces to   
\begin{equation}
\label{FPequation}
4 f_*+2\R\,  z_*-\R^2\left( f'_*+\R\,  z' _*\right)
-I_0[f_*,z_*](\R) =0\,.
\end{equation}
We then proceed to expand \eq{FPequation} in powers of the dimensionless curvature $\R$.
To this end we approximate the functions $f$ and $z$ polynomially as
\begin{eqnarray}
\label{f}
f(X)&=&
\sum_{n=0}^{m_1}
\lambda_{2n}  \,X^n   \\
\label{z}
z(X)&=&
\sum_{n=0} ^{m_2}
\lambda_{2n + 1}  \,X^n  \,.
\end{eqnarray}
where the couplings $\lambda_n$ are dimensionless. We recall that the argument $X$ of the functions $f=f(X)$ and $z=z(X)$  is given by $X=\R^2/4$. It arises from evaluating the square of the Ricci tensor $R_{\mu\nu}R^{\mu\nu}$ on spheres with constant scalar curvature $R=\R\, k^2$, all in units of the RG scale $k$. 
The overall normalisation of the functions $f$ and $z$ is chosen such that the couplings $\lambda_0$ and $\lambda_1$ relate to the dimensionless Newton coupling $g=G_k\,k^2$ and the dimensionless cosmological constant $\lambda=\Lambda_k/k^2$ as 
\begin{eqnarray}
\lambda&=&-\lambda_0/(2\lambda_1)\\
g&=&-1/\lambda_1\,.
\end{eqnarray}
 Classically, in four dimensions, the couplings $\lambda_0$ and $\lambda_1$ are relevant, $\lambda_2$ is marginal, and the couplings $\lambda_n$ for $n\ge 3$ are irrelevant. Their $\beta$-functions 
\begin{equation}
\beta_n(\lambda_i)=\partial_t\lambda_n
\end{equation}
for all couplings are obtained by suitable projection from  our basic RG flow \eq{FlowEquationRM} by solving the linear system of equations
\begin{equation}\label{beta}
\beta_n=U_n+V_{nm}\,\beta_m\,.
\end{equation}
The functions $U_n$ and $V_{nm}$ are known expressions of the couplings, defined implicitly via \eq{FlowEquationRM}. Below, we consider polynomial approximations of the action by retaining invariants with an increasing canonical mass dimensions. At approximation order $N$, we retain $N\ge 2$ independent curvature-invariants corresponding  to the upper summation limits
$m_1=\floor{(N-1)/2}$ and $m_2=\floor{(N-2)/2}$ in \eq{f} and \eq{z}, together with 
 $N = 2+m_1 + m_2$.

\begin{center}
\addtolength{\tabcolsep}{2pt}
\setlength{\extrarowheight}{2pt}
\begin{table*}[t]
\scalebox{0.9}{
\normalsize
\begin{tabular}{G|c G  c  G  c  G} 
\toprule 
\rowcolor{Yellow}
$\ \ \bm{N}\ \ $&
$\bm{\lambda_0}$&
$\bm{\lambda_1}$&
$\bm{\lambda_2}$&
$\bm{\lambda_3}$&
$\bm{\lambda_4}$&
$\bm{10^2\times \lambda_5}$
\\ 
\midrule 
 2 & 0.262694\ \, & $-$1.01608\  \ \,\,&\cellcolor{white}  & \cellcolor{white}  &\cellcolor{white}   & \cellcolor{white}  \\
 3 & 0.232891\ \, & $-$0.681762\ \, & 0.230963\ \, &\cellcolor{white}   &\cellcolor{white}   &\cellcolor{white}   \\
 4 & 0.3190549 & $-$0.8405815 & 0.1866115 & $-$0.5842019 &  &\cellcolor{white}   \\
 5 & 0.3247020 & $-$0.8312001 & 0.1912809 & $-$0.6271953 & $-$0.2189009 &\cellcolor{white}   \\
 6 & 0.325324\ \, & $-$0.830164\ \, & 0.19186\ \,\ \, & $-$0.631297\ \, & $-$0.245865\ \, & $-$3.52258\ \,\ \, \\
 7 & 0.3250134 & $-$0.8306768 & 0.1915745 & $-$0.6292062 & $-$0.2295505 & $-$0.4578745 \\
 8 & 0.3250091 & $-$0.8306836 & 0.1915709 & $-$0.6291752 & $-$0.2292068 & $-$0.3635126 \\
 9 & 0.3250101 & $-$0.8306819 & 0.1915719 & $-$0.6291810 & $-$0.2292164 & $-$0.3544319 \\
 10 & 0.3250066 & $-$0.8306878 & 0.1915686 & $-$0.6291580 & $-$0.2290678 & $-$0.3362079 \\
 11 & 0.3250068 & $-$0.8306875 & 0.1915687 & $-$0.6291590 & $-$0.2290763 & $-$0.3379264 \\
 12 & 0.3250068 & $-$0.8306876 & 0.1915687 & $-$0.6291589 & $-$0.2290738 & $-$0.3369824 \\
 13 & 0.3250067 & $-$0.8306877 & 0.1915686 & $-$0.6291584 & $-$0.2290714 & $-$0.3368553 \\
 14 & 0.3250067 & $-$0.8306876 & 0.1915687 & $-$0.6291586 & $-$0.2290729 & $-$0.3371316 \\
 15 & 0.3250067 & $-$0.8306876 & 0.1915687 & $-$0.6291586 & $-$0.2290727 & $-$0.3370663 \\
 16 & 0.3250067 & $-$0.8306876 & 0.1915687 & $-$0.6291586 & $-$0.2290726 & $-$0.3370645 \\
 17 & 0.3250067 & $-$0.8306876 & 0.1915687 & $-$0.6291586 & $-$0.2290727 & $-$0.3370860 \\
 18 & 0.3250067 & $-$0.8306876 & 0.1915687 & $-$0.6291586 & $-$0.2290727 & $-$0.3370778 \\
 19 & 0.3250067 & $-$0.8306876 & 0.1915687 & $-$0.6291586 & $-$0.2290727 & $-$0.3370774 \\
 20 & 0.3250067 & $-$0.8306876 & 0.1915687 & $-$0.6291586 & $-$0.2290727 & $-$0.3370799 \\
 21 & 0.3250067 & $-$0.8306876 & 0.1915687 & $-$0.6291586 & $-$0.2290727 & $-$0.3370791
 \\
 \midrule 
\rowcolor{Yellow}$\bm{N}$&
$\bm{\lambda_6}$&
$\bm{10^2\times\lambda_7}$&
$\bm{10^2\times\lambda_8}$&
$\bm{10^2\times\lambda_9}$&
$\bm{10^2\times\lambda_{10}}$&
$\bm{10^2\times \lambda_{11}}$\\
\midrule 
7 & 0.1059614 &\cellcolor{white}   &  &\cellcolor{white}   &  &\cellcolor{white}   \\
8 & 0.1114891 & 0.5933014 &  & \cellcolor{white}  &  &\cellcolor{white}   \\
 9 & 0.1126503 & 0.8182385 &\cellcolor{LightRed}  \ \ 0.7619809 &\cellcolor{white}   &  &\cellcolor{white}   \\
 10 & 0.1125038 & 0.6088539 & $-$0.7102757 & $-$1.465145 &  &\cellcolor{white}   \\
 11 & 0.1124355 & 0.6067056 & $-$0.6708707 & $-$1.375498 & 0.2991940 &\cellcolor{white}   \\
 12 & 0.1125046 & 0.6163190 & $-$0.6542211 & $-$1.406077 & 0.0194100
 & $-$0.2675564\ \, \\
 13 & 0.1124839 & 0.6096092 & $-$0.6883268 & $-$1.427912 & 0.0912924
& $-$0.04964086 \\
 14 & 0.1124753 & 0.6099047 & $-$0.6790793 & $-$1.412076 & 0.1306336 & $-$0.06997614 \\
 15 & 0.1124795 & 0.6104102 & $-$0.6786530 & $-$1.414547 & 0.1134571 & $-$0.08342165 \\
 16 & 0.1124782 & 0.6100440 & $-$0.6803688 & $-$1.415459 & 0.1181812 & $-$0.07172904 \\
 17 & 0.1124776 & 0.6100764 & $-$0.6796060 & $-$1.414205 & 0.1211160 & $-$0.07361072 \\
 18 & 0.1124781 & 0.6101417 & $-$0.6795417 & $-$1.414507 & 0.1189456 & $-$0.07536190 \\
 19 & 0.1124780 & 0.6101137 & $-$0.6796811 & $-$1.414592 & 0.1192647 & $-$0.07445450 \\
 20 & 0.1124779 & 0.6101143 & $-$0.6796066 & $-$1.414452 & 0.1196498 & $-$0.07457352 \\
 21 & 0.1124780 & 0.6101227 & $-$0.6795918 & $-$1.414479 & 0.1194069 & $-$0.07480735 \\
 \bottomrule 
 \end{tabular}  
}
\caption{\label{FixedPointsScaled} Summary of the fixed point couplings $\lambda_n(N)$ for $n=0,...,11$ and for selected approximation orders $N$. Overall, we observe a fast convergence. Of all couplings, only $\lambda_8(N)$ (red-shaded) starts off with the wrong sign at first appearance ($N=9)$.}
\end{table*} 
\end{center}

\begin{figure}[t]
\centering
\begin{center}
\includegraphics[width=.8\hsize]{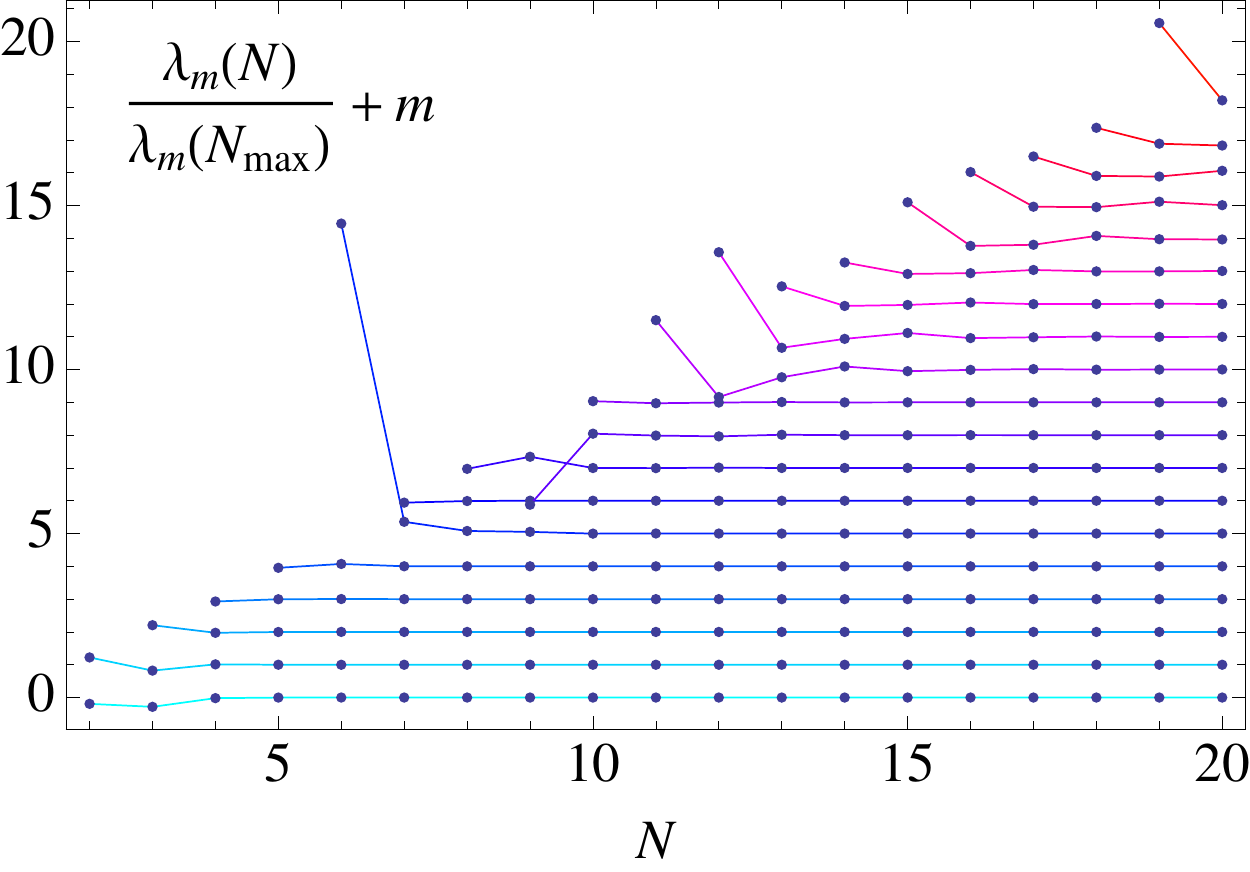}
\caption{\label{Lambdas} Shown are the first 19 fixed point couplings $\lambda_m$ $
(m=0,\cdots,18$ from bottom to top) with increasing approximation order $N$, normalised to their value at the 
highest order in the approximation $N_{\rm max}=21$. The shift by $m$ is added for better display  (see text). We observe a fast convergence with approximation order.}
\end{center}
\end{figure}

\subsection{Fixed point data}\label{FPdata}
We solve the fixed point condition $\beta_i=0$ using \eq{beta} to identify fixed point candidates  at each order of the approximation from $N=2$ to $N=21$ \cite{Nikolakopoulos:Thesis,King:MSc}. The analoguous problem in $f(R)$ type theories of gravity allowed for an exact recursive solution of  all fixed point couplings in terms of the two lowest ones \cite{Falls:2013bv,Falls:2014tra}.

More generally, exact recursive solutions are possible provided that the RG flows $\beta_i$ depend only linearly on the coupling $\lambda_n$ with the highest index $n$.  In the present case, this is not the case as the dependence on the highest coupling comes out quadratic. This new feature arises  owing to the fluctuations of the Ricci tensor as opposed to those of the Ricci scalar. For this reason, we adopt a different strategy to solve for fixed points. We begin with the fixed point in the Einstein-Hilbert approximation $(N=2)$, where a unique interacting fixed point is found by solving $\beta_0=0=\beta_1$ for $(\lambda^*_0,\lambda^*_1)$. As boundary conditions, we impose $\beta_i=0$ and $\lambda_i=0$ for all $i>1$. Provided a fixed point is found, we then increase the approximation order and use the previously-found values $(\lambda^*_0,\lambda^*_1)$ as starting values to search for interacting fixed points in the couplings $(\lambda^*_0,\lambda^*_1,\lambda^*_2)$ in its vicinity. This is repeated from $N=2$ up to $N=21$. We find that the procedure converges rapidly. At each and every order, we observe an UV fixed point in accordance with the asymptotic safety conjecture. 
Starting from order $N=3$ we also find spurious fixed points which either disappear in the next order or show inconsistencies in the number of relevant eigenvalues. These spurious fixed points are discarded.

In Fig.~\ref{pAcc}, we plot the decadic logarithm of the fixed point condition in field space, \eq{FPequation}, for all approximation orders $N$. With increasing $N$, we observe that the accuracy of the fixed point solution and the domain of validity increases for all scalar curvatures within the radius of convergence $\R_c$. The latter can be estimated from  Fig.~\ref{pAcc} as the absolute value of the scalar curvature beyond which the accuracy in the fixed point solution starts decreasing with increasing $N$. We find 
\begin{equation}\label{Rb}
\R_b\approx 2.006\,.
\end{equation}
\begin{figure}[t]
\centering
\begin{center}
\includegraphics[width=.8\hsize]{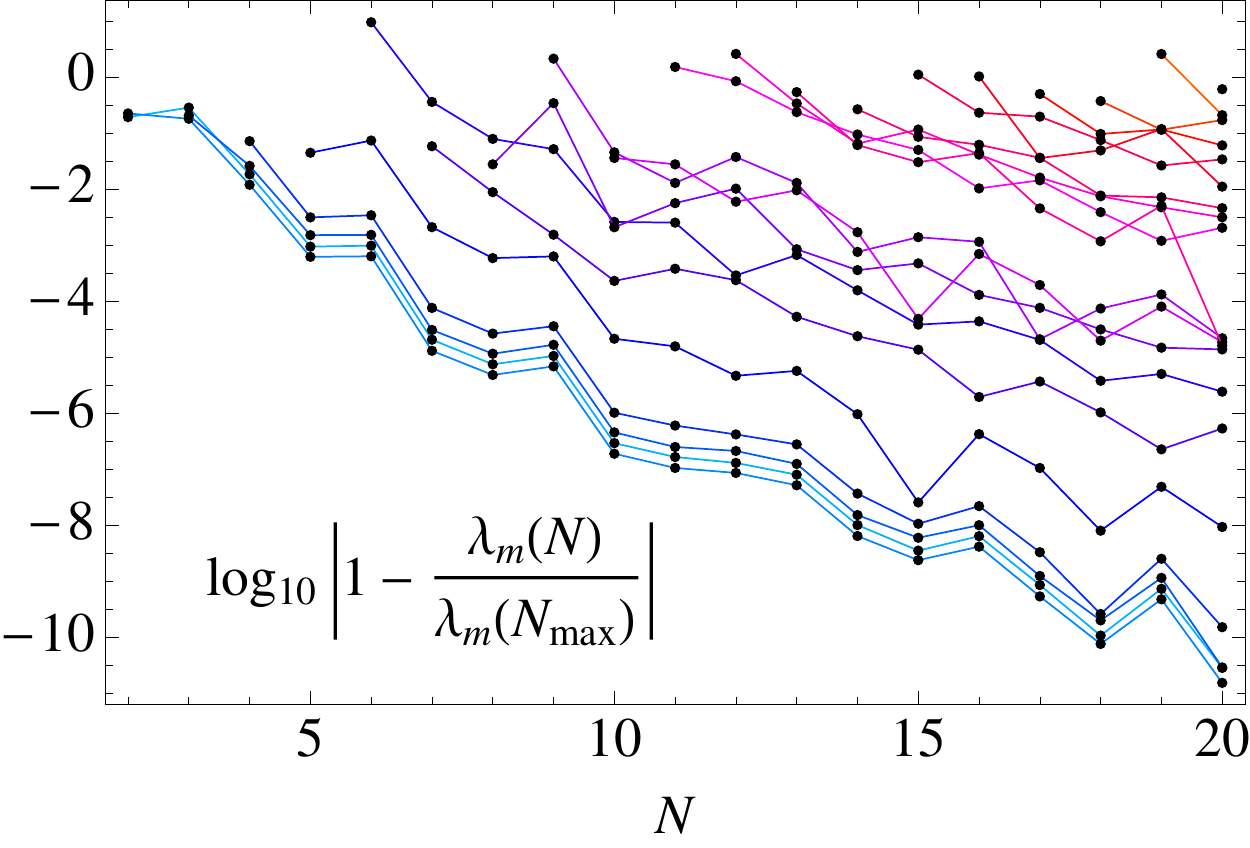}
\caption{\label{pLambdaConvergence} The rate of convergence of the first few fixed point couplings $\lambda_m$ $
(m=0,\cdots,18$ from left to right) with increasing approximation order $N$, normalised to their value at the 
highest order in the approximation $N_{\rm max}=21$. The accuracy in the couplings increases roughly by a decimal place for every $N\to N+2$.}
\end{center}
\end{figure}
The estimate \eq{Rb}  agrees with the location of the first singularity along the real axis of the fixed point equation, discussed in \eq{r0} below. As such, our findings indicate that the polynomial approximation exhausts the maximum radius of convergence set by a pole in the real field axis of the RG flow itself. Poles could have arisen in the complex field plane \cite{Falls:2016wsa,Morris:1994ki,Litim:2002cf,Litim:2016hlb,Juttner:2017cpr,Marchais:2017jqc}, but they do not in the case at hand.

Our numerical results for the fixed points are shown in Tabs.~\ref{FixedPointsScaled} and \ref{tconverge}, and in Fig.~\ref{Lambdas}. 
From the data in Tab.~\ref{FixedPointsScaled} we notice that the convergence of the first few couplings is fast. Also, all couplings arise with the correct sign already at their first appearance, except $\lambda_8$, see Tab.~\ref{FixedPointsScaled}.   At order $N=3$, 
the inclusion of the dimensionless coupling $\lambda_2$ proportional to $R_{\mu\nu}R^{\mu\nu}$ is expected to have a strong effect. 
It does lead to a significant shift on the couplings $\lambda_0$ and $\lambda_1$, with $\lambda_1$ changing by about $25\%$. From approximation order $N=4$ onwards, however, the inclusion of higher dimensional operators rapidly stabilises the fixed point values. 
We estimate the asymptotic values of the first few couplings by taking the average over the four best approximation orders to find
\begin{equation}\label{errors}
\begin{array}{rll}
\langle\lambda_0\rangle=&\ \ \,0.3250067185&\pm 2 \cdot 10^{-10}\\
\langle\lambda_1\rangle=&-0.8306876250&\pm 3\cdot 10^{-10}\\
\langle\lambda_2\rangle=&\ \ \, 0.1915686687&\pm 2\cdot 10^{-10}\\
\langle\lambda_3\rangle=&-0.6291586052&\pm 9\cdot 10^{-10}\\
\langle\lambda_4\rangle=&-0.229072717&\pm 6\cdot 10^{-9}\\
\langle\lambda_5\rangle=&-0.00337078&\pm 2\cdot 10^{-8}\\
\langle\lambda_6\rangle=&\ \ \,0.11247802&\pm 8\cdot 10^{-8}\\
\langle\lambda_7\rangle=&\ \ \,0.0061012&\pm  2\cdot 10^{-7}\,.
\end{array}
\end{equation}
The indicated error corresponds to a standard deviation in the data from the mentioned average. The fast convergence can also be read off from Fig.~\ref{Lambdas}, showing that the accuracy in the couplings increase by roughly a decimal place for approximation order $N\to N+2$. 
 This result is quite the opposite of what has been observed previously in $f(R)$ models of gravity. There, the coupling proportional 
to the square of the Ricci scalar leads to a strong impact on the fixed point coordinate. Also, the $R^2$ coupling itself showed 
a slow convergence with increasing approximation order. 

A brief remark on accuracy and error estimates is in order. The error estimates in \eq{errors} refer to the accuracy with which the relevant fixed point equations have been solved. In principle, results at low polynomial orders cannot be taken for granted due to the non-perturbative nature of the graviton \cite{Falls:2013bv}. However, the high numerical accuracy and the fast convergence make sure that the fixed point is a reliable approximation of the full, non-perturbative, fixed point of \eq{FlowEquationRM}, \eq{TheFlowEquation}. On the other hand, our error estimate in \eq{errors} does not account for proper systematic errors. For strategies to estimate the latter using functional renormalisation we refer to \cite{Litim:2010tt}.

\begin{center}
\addtolength{\tabcolsep}{2pt}
\begin{table*}[t]
\setlength{\extrarowheight}{.5pt}
\normalsize
\begin{tabular}{ G l G  l   G  l G } 
\toprule
\rowcolor{Yellow}
& \multicolumn{3}{c}{\bf fixed point couplings}  &\multicolumn{3}{c}{\bf scaling exponents}\\ 
\rowcolor{Yellow}
\multirow{-2}{*}{${}\quad \bm N\quad$}& ${}\quad \bm{\lambda}_*$ & ${}\quad \bm g_*$ & ${}\quad\bm g_*\cdot\bm \lambda_*$  & ${}\quad \bm\theta'$ & ${}\quad \bm\theta_2$  & ${}\quad \bm\theta'_3$  \\
\midrule 
2 & 0.129268 & 0.984172 & 0.127222 & 2.38241 &  &  \\
 3 & 0.170801 & 1.46679\, \ & 0.250528 & 1.62684 & 21.233 &  \\
 4 & 0.18978224 & 1.189653 & 0.2257749 & 2.45042 & 1.1075 & $-$8.27329 \\
 5 & 0.19532118 & 1.203080 & 0.2349870 & 2.41473 & 0.99548 & $-$5.31158 \\
 6 & 0.19594 & 1.20458\, \ & 0.236025 & 2.40791 & 0.97504 & $-$5.05665 \\
 7 & 0.19563165 & 1.203838 & 0.2355087 & 2.40905 & 0.98114 & $-$5.00970 \\
 8 & 0.19562751 & 1.203828 & 0.2355018 & 2.40904 & 0.98111 & $-$5.00233 \\
 9 & 0.19562851 & 1.203830 & 0.2355035 & 2.40902 & 0.98103 & $-$5.00244 \\
 10 & 0.19562502 & 1.203822 & 0.2354976 & 2.40903 & 0.98118 & $-$5.00295 \\
 11 & 0.19562516 & 1.203822 & 0.2354979 & 2.40903 & 0.98117 & $-$5.00273 \\
 12 & 0.19562515 & 1.203822 & 0.2354979 & 2.40903 & 0.98117 & $-$5.00278 \\
 13 & 0.19562509 & 1.203822 & 0.2354978 & 2.40903 & 0.98117 & $-$5.00281 \\
 14 & 0.19562511 & 1.203822 & 0.2354978 & 2.40903 & 0.98117 & $-$5.00280 \\
 21 & 0.19562511 & 1.203822 & 0.2354978 & 2.40903 & 0.98117 & $-$5.00280 \\
 \bottomrule
\end{tabular}  
\vskip.1cm
\normalsize
\caption{\label{tconverge}
The fixed point values for the dimensionless Newton coupling $g_*$, the dimensionless cosmological constant $\lambda_*$,  the universal product $\lambda\cdot g$, 
and the first four scaling exponents to various orders in the expansion. Within the accuracy of the displayed digits, the values no longer vary between approximation orders $N=14$ and $N=21$.}
\end{table*} 
\end{center}

\subsection{Universal coupling ratios}
Fixed point couplings are non-universal. Still, some universal quantities of interest are given by specific products of couplings which remain invariant under global re-scalings of the metric field
 \begin{equation}\label{ell}
 g_{\mu\nu}\to \ell\, g_{\mu\nu}\,.
 \end{equation}
 Under \eq{ell}, the couplings scale as
\begin{equation}
\lambda_n\to\ell^{4-2n}\,\lambda_n\,.
\end{equation}
Note that the classically dimensionless coupling $\lambda_2$ remains invariant under the rescaling \eq{ell}. All other couplings scale non-trivially. Consequently, various products of couplings can be formed which stay invariant under \eq{ell}. 
Such invariants may serve as a measure for the relative strength of the gravitational interactions \cite{Kawai:1989yh}.  
For couplings including up to $\lambda_3R R_{\mu\nu}R^{\mu\nu}$ we may construct three independent invariants with values
\begin{equation}\label{lambdaP}
\begin{array}{rcl}
\langle\lambda_0/\lambda_1^2\rangle\ =&\ \ \, 0.4709956080&\pm\ 5\cdot 10^{-10}\\
\langle\lambda_0\lambda_3^2\rangle\ =&\ \ \,0.1286508384&\pm\ 4\cdot 10^{-10}\\
\langle\lambda_1\lambda_3 \rangle\ = &\ \ \,0.5226342675&\pm\ 6\cdot 10^{-10}\ .
\end{array}
\end{equation}
Tab.~\ref{tconverge} show the convergence of couplings, universal coupling ratios, and the leading scaling exponents. We postpone a detailed discussion of universality until Sect.~\ref{scaling}.

\begin{figure*}[t]
\begin{center}
\includegraphics[width=.85\hsize]{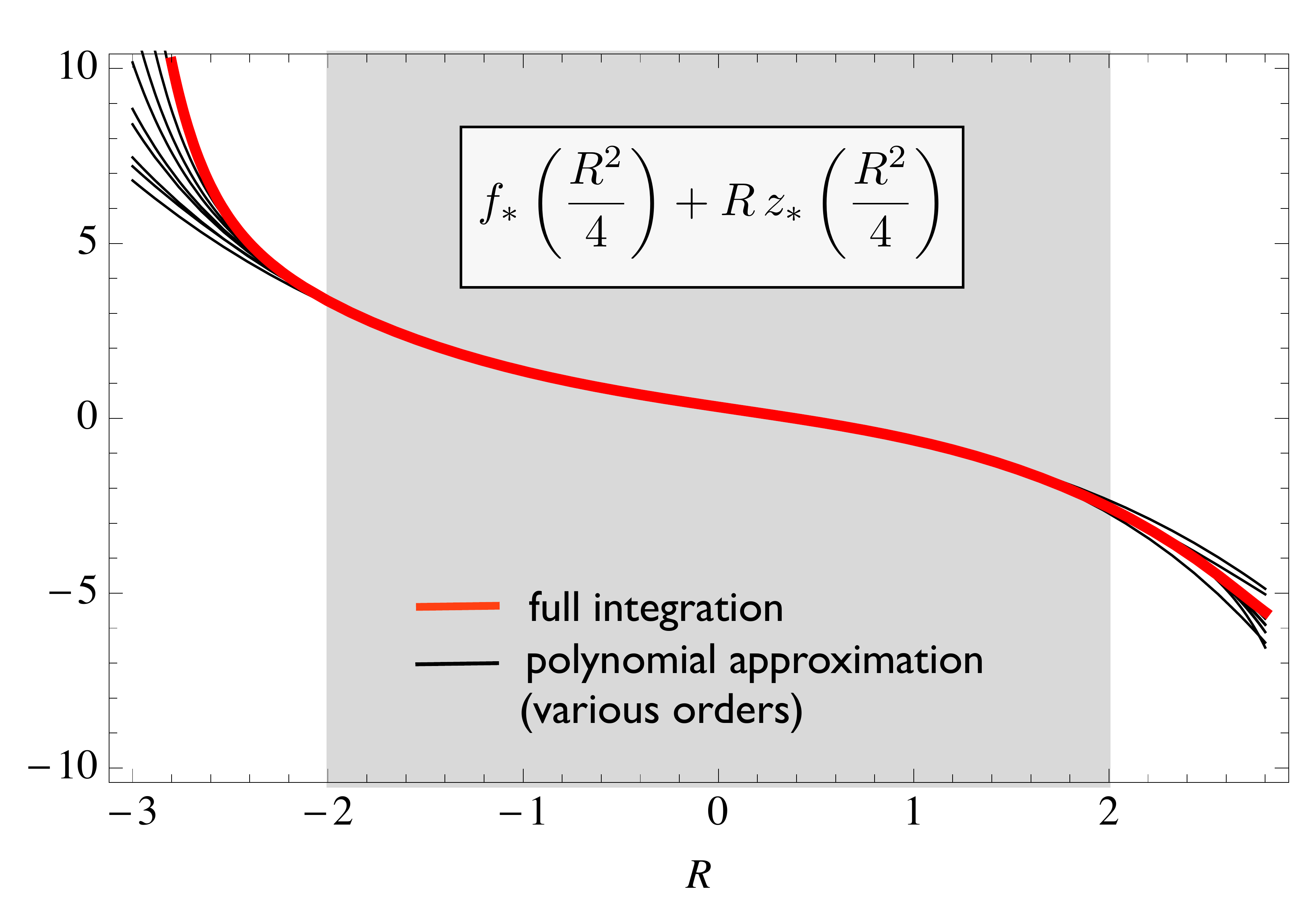}
\caption{\label{Numfeff} Shown is the non-perturbative ``fixed point functional''  $f_*(R^2/4)+R\,z_*(R^2/4)$ given in \eq{f}, \eq{z} as a function of the dimensionless scalar Ricci curvature $R$ and the polynomial approximations from $N=3$ up to $N=19$ (in steps of 2, thin black lines). The highest approximation order (thick red line) also coincides with the numerical integration of \eq{FPeven}, \eq{FPodd}. Notice that the radius of convergence, indicated by the shaded area,  is maximal, and given by $\R_p$  in \eq{r0}.}
\end{center}
\end{figure*}

\subsection{Beyond polynomial expansions}

In order to go beyond the polynomial approximations we can integrate the fixed point condition numerically. In a first step, we consider even and odd parts of the fixed point condition in $\R$ \eq{FPequation}, leading to
\bea
\label{FPeven}
\left(4 f-\R^2f'
\right)=I_0[f,z](\R)+I_0[f,z](-\R)\\
\label{FPodd}
\R \left(2\,  z-\R^2 \,  z' 
\right)=I_0[f,z](\R)-I_0[f,z](-\R)
\eea
These two equations can then be taken as differential equations for the functions $f(x)$ and $z(x)$ which depend on $\R$ through the variable $x= \frac{\R^2}{4}$. Taking the initial conditions from our polynomial solutions we can solve these two equations to give $f$ and $z$. Since the equations depend on up to the third derivatives of these functions we must give three initial conditions for both $f$ and $z$. Resolving them for the highest derivatives, the coupled differential equations take the form
\bea
\label{dF}
\frac{df''}{d\R}=H_f[f,z;\R]\\
\label{dZ}
\frac{dz''}{d\R}=H_z[f,z;\R]
\eea
where the functions $H_f$ and $H_z$ depend on $f$ and $z$ up to including their second derivatives (the explicit expressions are rather lenghty and not given here). 
From the structure of the fixed point equation \eq{I0} we observe that both $f'''$ and $z'''$ are proportional to the polynomial \eq{P0S3}, which vanishes at
\begin{eqnarray}
\R_0\ \ &=&\ \ 0\nonumber\\
\R_p\ \ &=&\ \ 2.00648\label{r0}\\
\R_n\ \ &=&-9.99855\,.\nonumber
\end{eqnarray}
Consequently, at the points \eq{r0}, the third derivatives $f'''$ and $z'''$ in \eq{dF}, \eq{dZ}  are ill-defined.\footnote{The same pattern was observed in $f(R)$-type theories of gravity where the factor \eq{P0S3} appears in front of $f'''(\R)$.}
For the numerical integration, we therefore will have to chose initial conditions away from \eq{r0}. 
Initial conditions are provided through the polynomial fixed point in its domain of validity. This  limits the integration of the fixed point equations \eq{FPeven}, \eq{FPodd} to the regime $0<x<1.00649$. We fix initial conditions at a sufficently small value for $x$, say $x=1/100$. We have checked that results do not depend on this choice.
 In Fig.~\ref{Numfeff} we plot the function $f+ \R z$ as a function of $\R$ and compare it to the polynomial approximations from orders $N=3$ to $N=19$. We observe that the polynomial solutions are in good agreement with the numerical solution all the way to $\R = \pm2.00648$. 
The existence of the formal singularity at $\R_p =2.00648$ and the fact that our polynomial approximation agrees with the numerical solution up to this point indicates that the radius of convergence for the polynomial approximation is maximal: there are no other convergence-limiting singularities in the complex plane of scalar Ricci curvature with $|\R_{\rm sing.}|<\R_p$.

\begin{figure}[t]
\centering
\begin{center}
\includegraphics[width=.8\hsize]{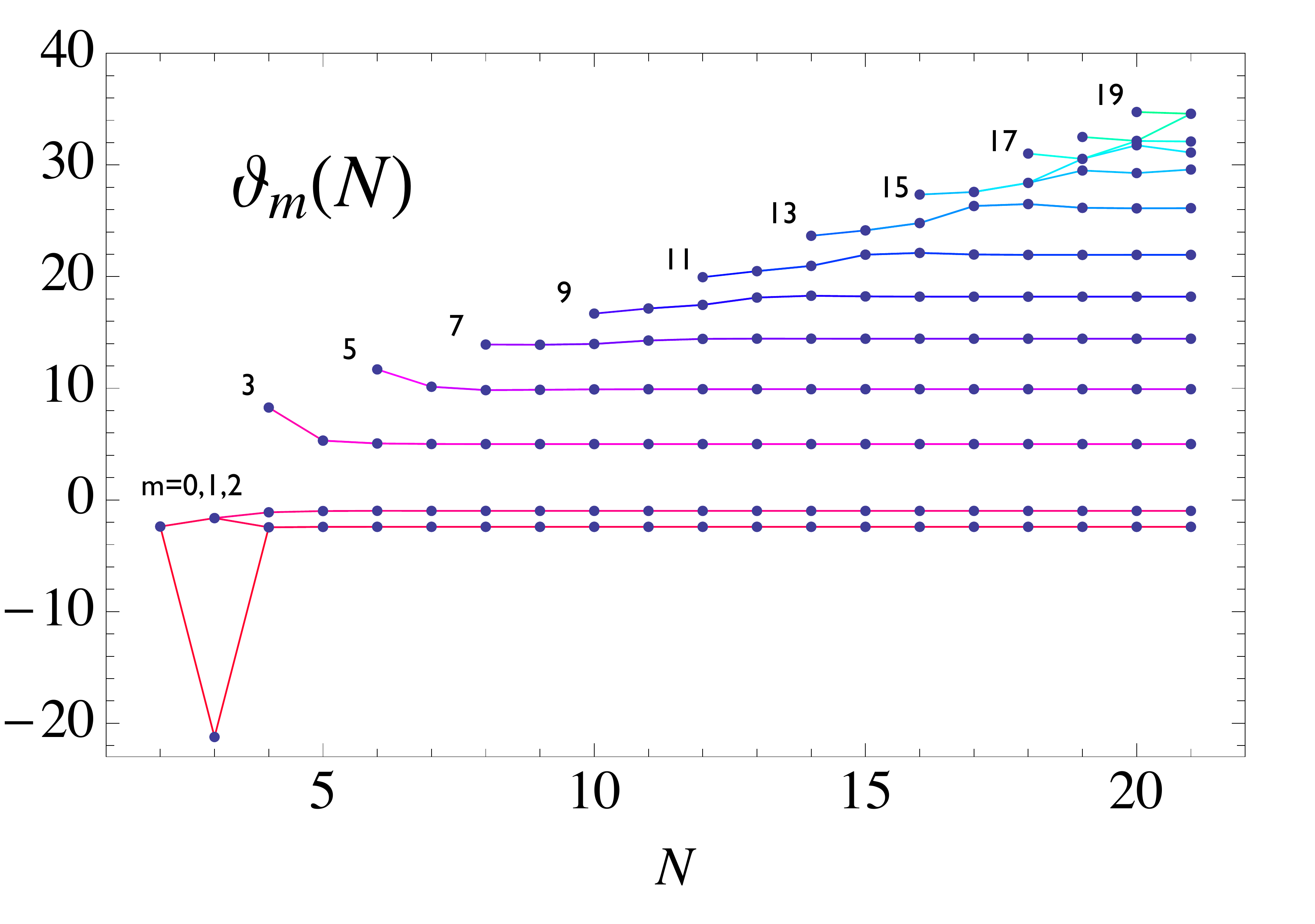}
\caption{\label{pThetas} The universal eigenvalues $\vartheta_m(N)$ (real part if complex) for specific values of $m$ as functions of the order of the polynomial approximation $N$ (see main text).}
\end{center}
\end{figure}

\begin{figure}[t]
\centering
\begin{center}
\includegraphics[width=.8\hsize]{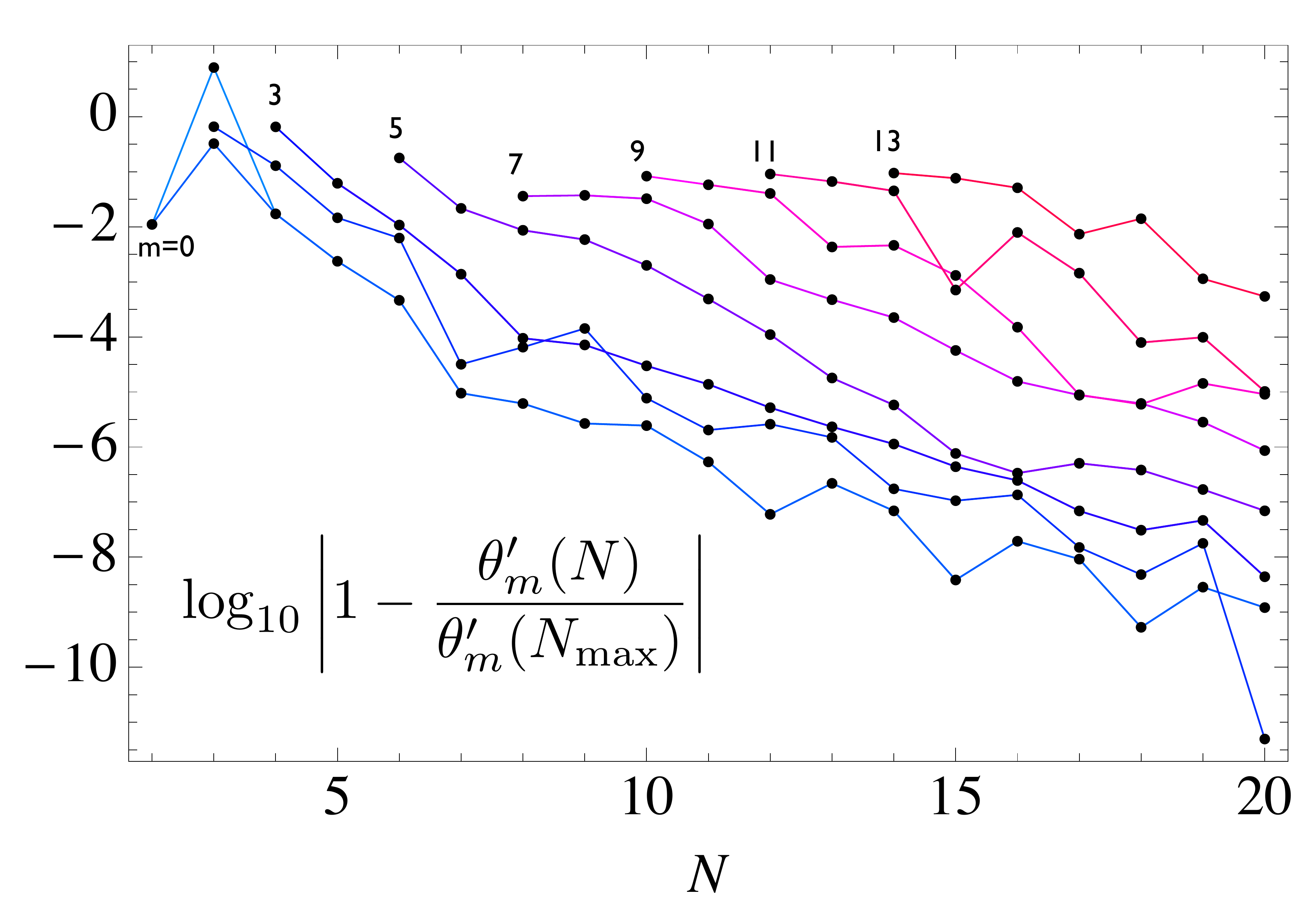}
\caption{\label{pThetaConvergence} The rate of convergence of the real part of universal eigenvalues $\theta'_m$ ($m=0,2,3,5,\cdots 13)$
with increasing approximation order $N$ towards their best values at $N_{\rm max}=21 $ (see main text).}
\end{center}
\end{figure}

\section{\bf Universality and scaling exponents}\label{scaling}
In this section, we turn to the computation of universal scaling exponents $\{\vartheta_n\}$ characterising the UV fixed point. 

\subsection{Scaling exponents}

Unlike fixed point values which are non-universal, scaling exponents are universal and, in principle, mesurable in an experiment. They are deduced as the eigenvalues of the stability matrix at the fixed point $M_{ij}=\partial\beta_j/(\partial\lambda_j)|_*$ , which in turn is obtained from \eq{beta}. At each order $N$ in the approximation we retain $N$ operators in the effective action. We therefore find a set of $N$ eigenvalues 
\beq\label{thetas}
\{\vartheta_n(N), n=0,\cdots,N-1\}\,,
\eeq
which we order according to the magnitude of their real parts, ${\rm Re}\,\vartheta_n\le {\rm Re}\,\vartheta_{n+1}$. Negative (positive) eigenvalues correspond to relevant (irrelevant) operators in the UV. It is expected that a UV fixed point displays, at best, a finite number of relevant operators in order to ensure predictivity of the theory.

At each order of the approximation 
from $N=2$ to $N=21$ we find that there are always three UV attractive directions, corresponding to three negative eigenvalues, 
exactly as was the case for the $f(R)$ approximation up to order $N=35$ \cite{Falls:2013bv,Falls:2014tra}. 
Some scaling exponents are complex conjugate pairs. It is then convenient to denote their real and imaginary parts of the eigenvalues \eq{thetas}  as $\theta' =- {\rm Re}\,\vartheta$ and $\theta'' = \pm {\rm Im}\,\vartheta$.  Specifically, the three leading relevant exponents $\{\vartheta_0,\vartheta_1,\vartheta_2\}$ are a complex conjugate pair together with a real eigenvalue, which we write as $\vartheta_{0}=-\theta'_0+ i\theta''_0$, $\vartheta_{1}=-\theta'_0- i\theta''_0$ and $\vartheta_2=-\theta'_2$. All irrelevant eigenvalues $\vartheta_n$ for $n\ge 3$ settle as complex conjugate pairs,  which we write pairwise as $\vartheta_{n}=-\theta'_n+i \theta''_n$ and $\vartheta_{n+1}=-\theta'_n-i\theta''_n$ ($n\ge 3$ and odd).
Using these conventions, in Fig.~\ref{pThetas} we show the real part of the eigenvalues $\vartheta_n$ as functions of the polynomial approximation order. In  Fig.~\ref{pThetaConvergence}, we show  the fast convergence of $\{\theta'_0,\theta'_2,\theta'_3,\theta'_5,\cdots,\theta'_{15}\}$, again as functions of the  polynomial approximation order.

The values of the critical exponents can be found in 
Tab.~\ref{tconverge}. It is noteworthy that the values are quite stable from order to order, except for the first occurrence of the perturbatively marginal coupling $\lambda_2$ with eigenvalue $\theta_2$ at order $N=3$. The reason for this has been given in \cite{Falls:2014tra}: the invariant $\int\sqrt{g}R_{\mu\nu} R^{\mu\nu}$  is classically marginal meaning that the RG flow for the corresponding coupling does not have a term linear in the coupling. Consequently, higher order interactions are needed to help the classically marginal coupling to develop a fixed point of its own. In other words, the operators with irrelevant scaling dimension are necessary to help the relevant and marginal invariants to develop stable fixed points. This structural pattern is generic for fixed points of quantum fields theories within polynomial or vertex expansions.

We estimate the asymptotic values of scaling exponents by taking an average over the best four  approximation orders. We find
\begin{equation}
\label{theta*}
\begin{array}{rll}
\langle\theta'_0\rangle\ =&\ \ \,2.409028030&\pm\ 5\cdot  10^{-9}\\
\langle\theta'_2 \rangle\ =&\ \ \, 0.98117249 &\pm\ 1 \cdot 10^{-8}\\
\langle\theta'_3 \rangle\ =& -5.0028024 &\pm\ 2\cdot  10^{-7}\\
\langle\theta'_5 \rangle\ =& -9.921004&\pm\ 2\cdot  10^{-6}\\
\end{array}
\end{equation}
as real parts of the first few exponents. Except for the eigenvalue $\theta_2$, all others relate to complex conjugate pairs and appear with the degeneracy two. For the corresponding imaginary parts, we have
\begin{equation}
\label{Imtheta*}
\begin{array}{rll}
\langle\theta''_0\rangle\ =&\ \ \,2.297772857&\pm\ 5\cdot  10^{-9}\\
\langle\theta''_2\rangle\ =&\ \ \,0\\
\langle\theta''_3 \rangle\ =& \ \ \, 3.8264682 &\pm\ 2\cdot  10^{-7}\\
\langle\theta''_5 \rangle\ =& \ \ \, 3.870582 &\pm\ 2\cdot  10^{-6}\,.
\end{array}
\end{equation}
These encouraging results indicate that the number of negative eigenvalues is not affected by the inclusion of more complicated tensor structures.

\begin{figure}[t]
\centering
\begin{center}
\includegraphics[width=.8\hsize]{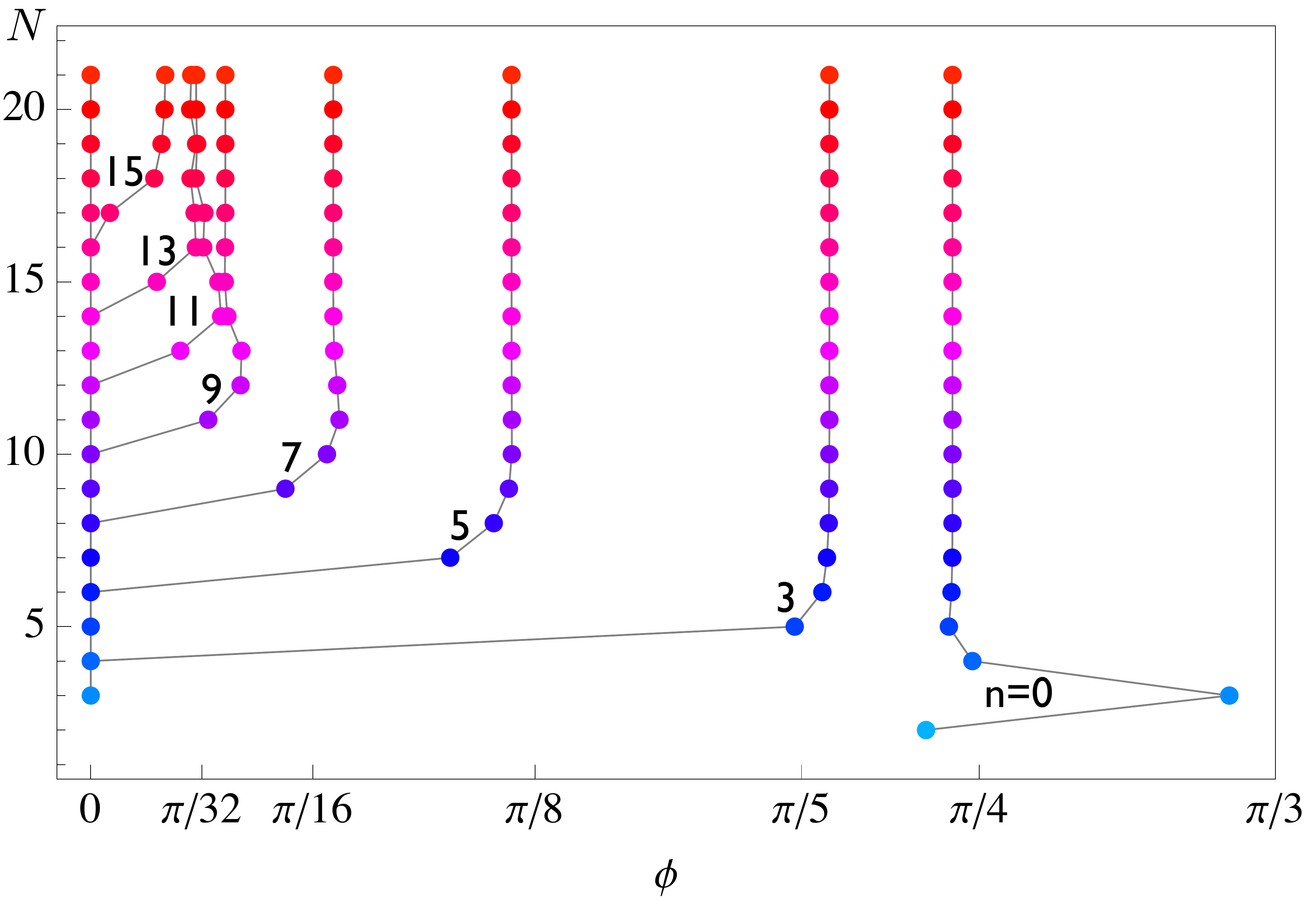}
\caption{\label{pAngles} Tomography of the angles $\phi_n$ defined in \eq{angles}, 
as functions of the approximation order $N$. The non-trivial angles $\phi_n$  are connected by lines ($n=0,3,5,7,9,11,13,15$ from right to left, respectively). With increasing $N$, we observe a rapid convergence. With increasing $n$, we observe that the imaginary parts are increasingly small compared to the real parts.}
\end{center}
\end{figure}

\subsection{Degeneracy}
We briefly comment on the complex eigenvalues.
Complex conjugate pairs of eigenvalues are a consequence of interactions and indicate a degeneracy amongst the scaling of eigenoperators.  Technically, they arise due to large off-diagonal entires in the stability matrix. The  degeneracy indicates in that two linearly  independent eigenperturbations decay with the exact same power law (associated to the real part of the eigenvalue), the sole difference being a relative phase (associated to the imaginary part of the eigenvalue)  \cite{Falls:2014tra}. If complex conjugate exponents persist in the full physical theory, it would be important to fully understand the dynamical origin. However, the degeneracy may very well be an artifact of our inability to retain all gravitational couplings. If so,  we expect that the degeneracy is  lifted, or becoming less pronounced, provided additional gravitational couplings are retained in the effective action beyond those studied here. 
In the present case, we find that imaginary parts (if they arise) are small. In particular, we find that the ratio of real to imaginary part ${\rm Im}\vartheta_n/{\rm Re} \vartheta_n$ and the angles
\begin{equation}\label{angles}
\phi_n=\arctan\frac{{\rm Im}\,\vartheta_n}{{\rm Re} \,\vartheta_n} \,,
\end{equation}
if nonzero, becomes increasingly small with increasing $n$. Our results for the angles \eq{angles} are shown in Fig.~\ref{pAngles}. In view of the discussion given above, small imaginary parts imply that the degeneracy amongst pairs of eigenperturbations is mild, and increasingly milder for increasingly higher order invariants. We conclude that our findings are consistent with the picture given above. 

\begin{figure}[t]
\centering
\begin{center}
\unitlength0.001\hsize
\begin{picture}(900,550)
\put(10,0){\includegraphics[width=.8\hsize]{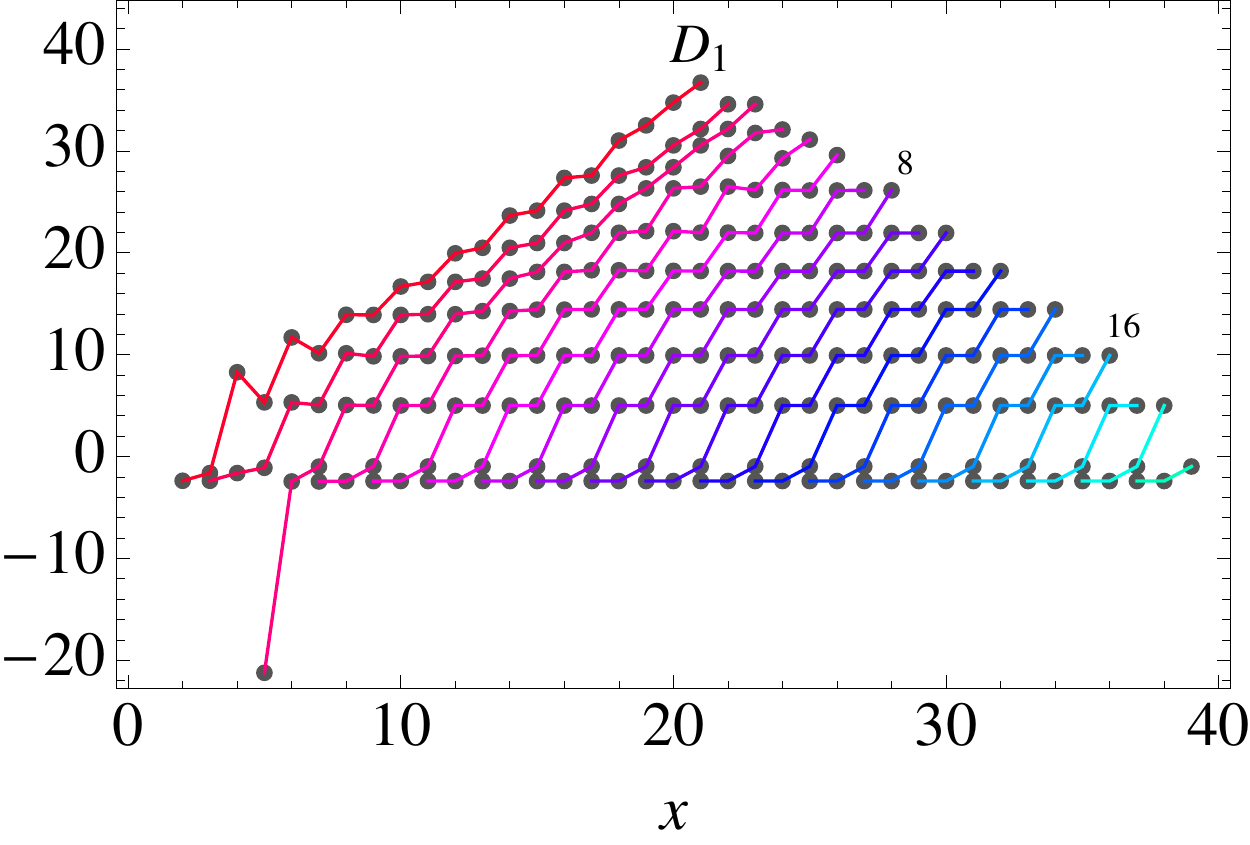}}
\end{picture}
\caption{\label{pBootstrap} The bootstrap test for asymptotic safety: from left to right, each line $D_i$ shows the $i^{\rm th}$ largest eigenvalues from all approximation orders $N\ge i$, connected by a line, and $x(N)=N+1+i$. At fixed $i$, and with increasing approximation order $N$ corresponding to increasing $x$, we observe that all curves $D_i$ consistently grow with $N$. This establishes that invariants with a larger canonical mass dimension lead systematically to larger scaling exponents  (see main text).}
\end{center}
\end{figure}

\subsection{Gap}
We now turn to the gap in the eigenvalue spectrum, following \cite{Falls:2014tra}. Here, our results establish three relevant eigendirections corresponding primarily to the cosmological constant $\int\sqrt{g}$, the Ricci scalar $\int\sqrt{g}R$ and the square of the Ricci tensor $\int\sqrt{g}R_{\mu\nu} R^{\mu\nu}$. The smallest 
negative eigenvalue is of the order  $\approx -1$. In turn, the most relevant of the irrelevant eigendirections, primarily 
induced by the invariant $\int\sqrt{g}R R_{\mu\nu} R^{\mu\nu}$, has the 
eigenvalue $\approx -5.00$. Classically, its Gaussian eigenvalue reads $-2$. The large numerical value means that fluctuations have made
the operator less relevant. Denoting  the difference between the smallest relevant eigenvalue and 
the smallest irrelevant eigenvalue (or their real parts if complex) as the ``gap'' $\Delta$, we then find from \eq{theta*} that the gap in the 
eigenvalue spectrum is given by 
\begin{equation}
\Delta=\theta'_2-\theta'_3\approx 5.98\,.
\end{equation}
Notice that the gap is roughly of the same size as the first irrelevant eigenvalue. Furthermore, the gap is only slightly larger than the gap $\Delta\approx 5.5$ observed in the $f(R)$ approximation \cite{Falls:2013bv,Falls:2014tra}. 
This comparison seems to suggest that the invariant $\int \sqrt{g}\,R\,R_{\mu\nu}R^{\mu\nu}$ is a less relevant operator than $\int \sqrt{g}R^3$.

\begin{figure}[t]
\centering
\begin{center}
\includegraphics[width=.8\hsize]{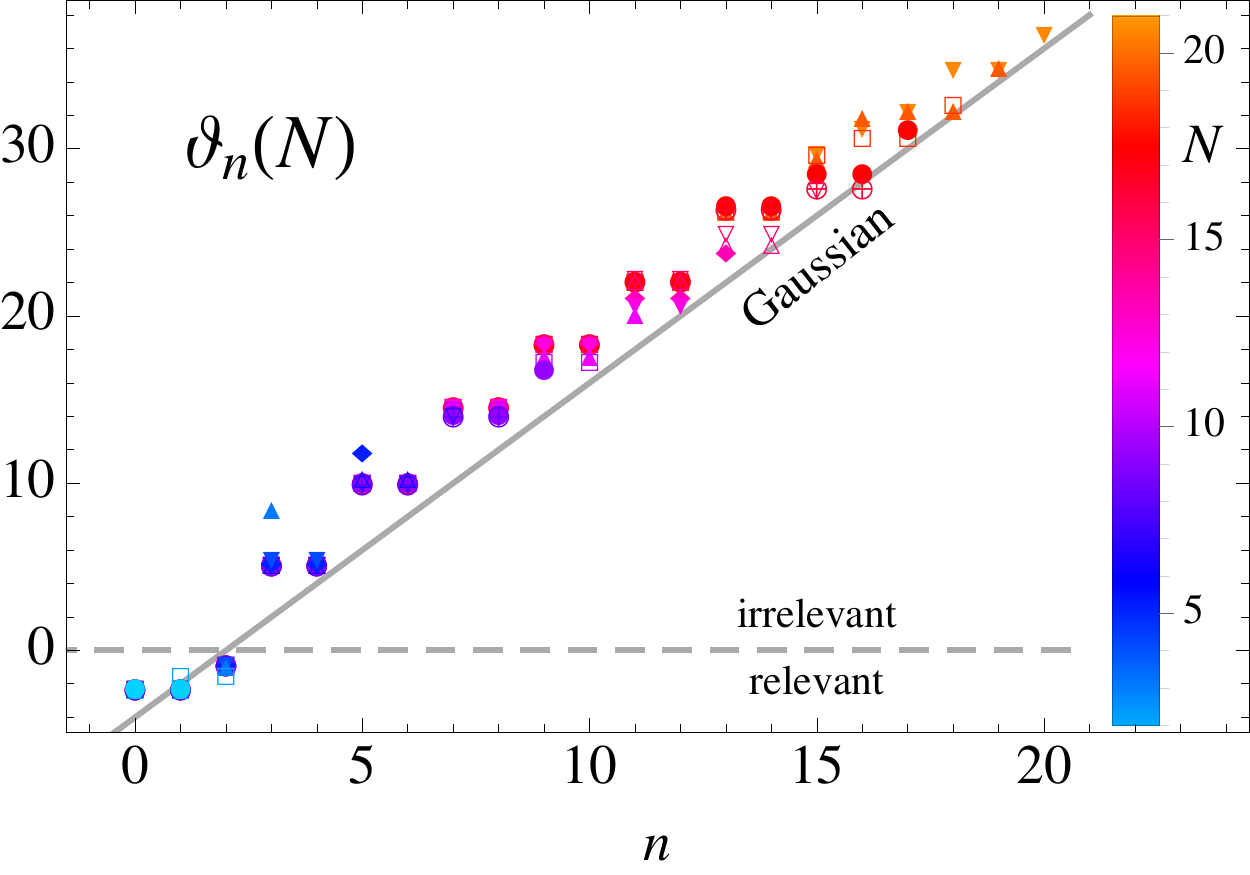}
\caption{\label{pRicci21} The universal scaling exponents $\vartheta_n(N)$ (real parts if complex) are shown for all approximation orders $N$, in comparison with classical exponents (straight line). We observe three relevant eigenvalues. All other eigenvalues are increasingly irrelevant, approaching Gaussian values from above.}
\end{center}
\end{figure}

\section{\bf Bootstrap and near-Gaussianity}\label{sec:bootstrap}

In this section, we apply the bootstrap test for asymptotic safety \cite{Falls:2013bv} and  investigate the large order behaviour of scaling exponents.

\subsection{Bootstrap hypothesis}
A key challenge in the reliable determination of asymptoticaly safe UV fixed points consists in the fact that the set of relevant, marginal, and irrelevant operators at the fixed point are not known beforehand. This implies that the set of universal eigenvalues
\beq
\{\vartheta_n\}
\eeq
at the interacting fixed point  is not known beforehand. This is in stark contrast to theories with asymptotic freedom, where the set of positive, vanishing, or negative eigenvalues is known a priori to coincide with the Gaussian exponents. Furthermore, fixed point searches in interacting theories generically require approximations, and even if stable fixed points are established, one still has to acertain that invariants neglected thus far are not spoiling the result. In contrast, in asymptotically free theories the finite  set of relevant and marginal operators is known beforehand allowing for systematic approximations.

For this reason, in \cite{Falls:2013bv}, a bootstrap strategy has been put forward to help circumnavigate the lack of a priori information about interacting fixed points, applicable for UV and IR fixed points alike. As our working hypothesis we will assume that canonical power counting continues to be a good guiding principle at an interacting fixed point, or, in other words, we assume that
\begin{equation}\label{hypo}
\bullet 
\begin{array}{rl}
&{\rm 
the\ relevancy\ of\ invariants\ at\ an\ interacting\ fixed\ point\ continues}\\ &{\rm to\ be\ governed\ by\ the\ invariant's\ canonical\ mass\ dimension}\,.
\end{array}
\end{equation}
The motivation for this hypothesis is that all known interacting fixed points in quantum field theory share this property. The value of a working hypothesis consists in the fact that it can be tested and falsified, systematically, within given approximations. If successful, it establishes the existence of (asymtpotically safe) fixed points under the hypothesis that canonical power counting is applicable.

\subsection{Testing asymptotic safety}
To establish that our results conform with the bootstrap hypothesis, we display in Fig.~\ref{pBootstrap} the ordered scaling exponents per approximation level. The quantities $D_n$ displayed there are defined as follows,
\begin{equation}\label{Dn}
D_n(N)=\vartheta_{N-n}(N)\,,
\end{equation}
and the scaling exponents $\vartheta_{n}$ refer to \eq{thetas}. The meaning of \eq{Dn} is that $D_n(N)$ gives the $n^{\rm th}$ largest eigenvalue at approximation order $N$. In Fig.~\ref{pBootstrap}, the line $D_1$ as a function of the approximation order $N$ thus shows how the largest eigenvalue increases with increasing approximation order provided $N$ has become sufficiently large. Similarly, we observe that all lines $D_i$ grow with increasing approximation order. This result establishes very clearly that the addition of higher order invariants $N\to N+1$ systematically adds a new eigenvalue to the spectrum which is larger, and thus less relevant, than those already present. We conclude that the asymptotically safe fixed point detected here is fully consistent with our working hypothesis.

 \subsection{Large order behaviour}
  In \cite{Falls:2013bv,Falls:2014tra} is has been observed for $f(R)$ type theories of gravity that higher order powers in the Ricci scalar lead to scaling exponents which become increasingly Gaussian. The result as such is quite intriguing, suggesting that the quantum dynamics of gravity makes the theory ``as Gaussian as it gets'': the scaling of a few invariants with a low mass dimension receives strong corrections with scaling exponents deviating strongly from classical values. On the other hand, the scaling of invariants with large canonical mass dimension receive only very mild corrections leading to near-Gaussian exponents.  Exact Gaussian scaling is clearly not achievable owing to the perturbative non-renormalisability of the theory. Here, we want to check whether this picture persists in case the higher order interactions are substantially different from those investigated in \cite{Falls:2013bv,Falls:2014tra}. 

To that end, 
we display the entire eigenvalue spectrum \eq{thetas} for all approximation orders in Fig.~\ref{pRicci21}. The colour coding (from blue to red) indicates growing approximation order $N$. We make two observations. Firstly, three negative eigenvalues are neatly visible for all approximation orders. They differ by order unity from their classical values. Secondly, with increasing $n$, we observe that the eigenvalues approach Gaussian values from above. Both of these features agree with the picture detected in $f(R)$ type approximations \cite{Falls:2013bv,Falls:2014tra}. The near-Gaussian behaviour thus appears to be a feature of quantum gravity rather than a feature of or limitation to specific technical approximations.

\begin{figure}[t]
\centering
\begin{center}
\includegraphics[width=.8\hsize]{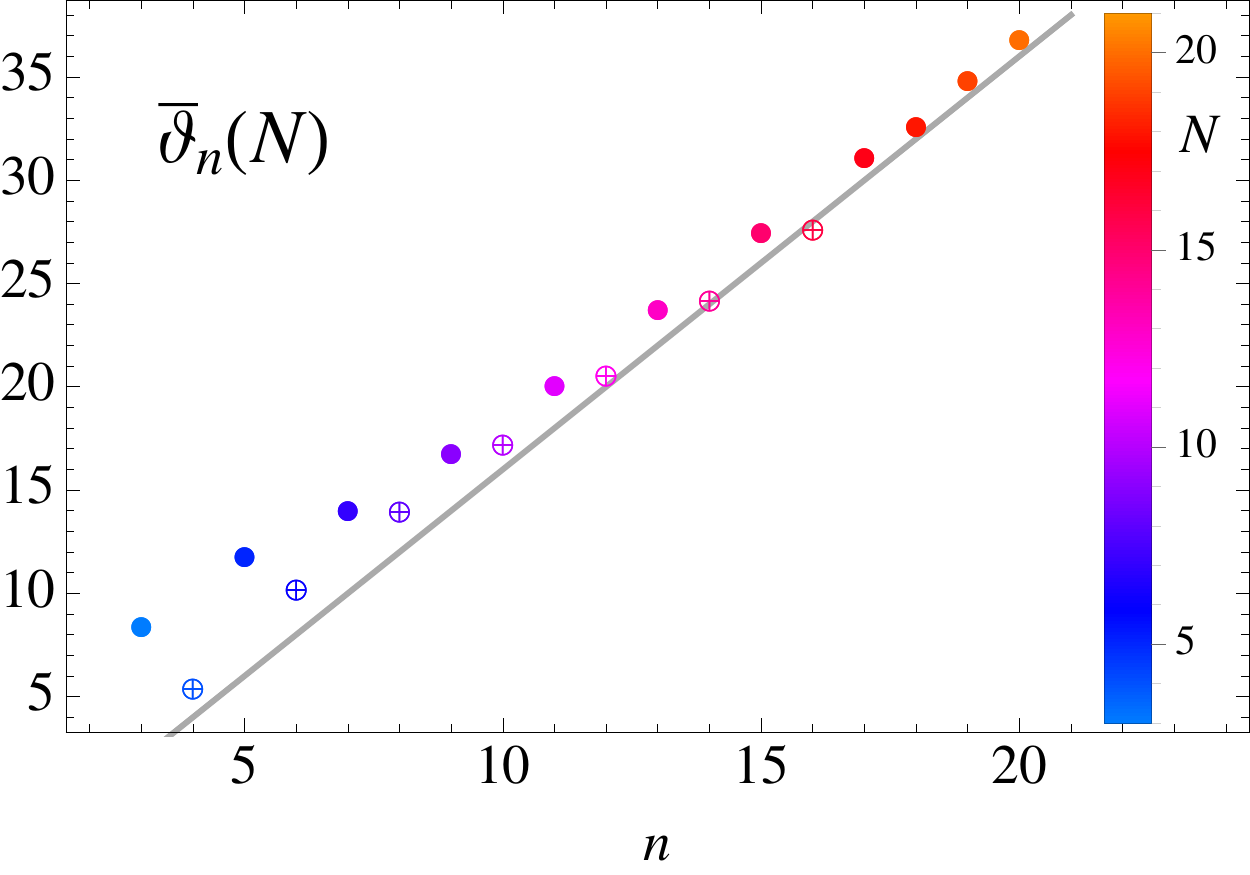}
\caption{\label{pMaxColor} Origin of near-Gaussian scaling exponents. Shown are the scaling exponents $\bar\vartheta_m$, which, for each approximation order $N$, are given by the largest eigenvalue $\vartheta_{m=N-1}(N)$ (full dot) or its real part if complex (crossed dot). We observe that with increasing $m$ (or $N$) the largest eigenvalue approach Gaussian values.}
\end{center}
\end{figure}

\begin{figure}[t]
\centering
\begin{center}
\includegraphics[width=.8\hsize]{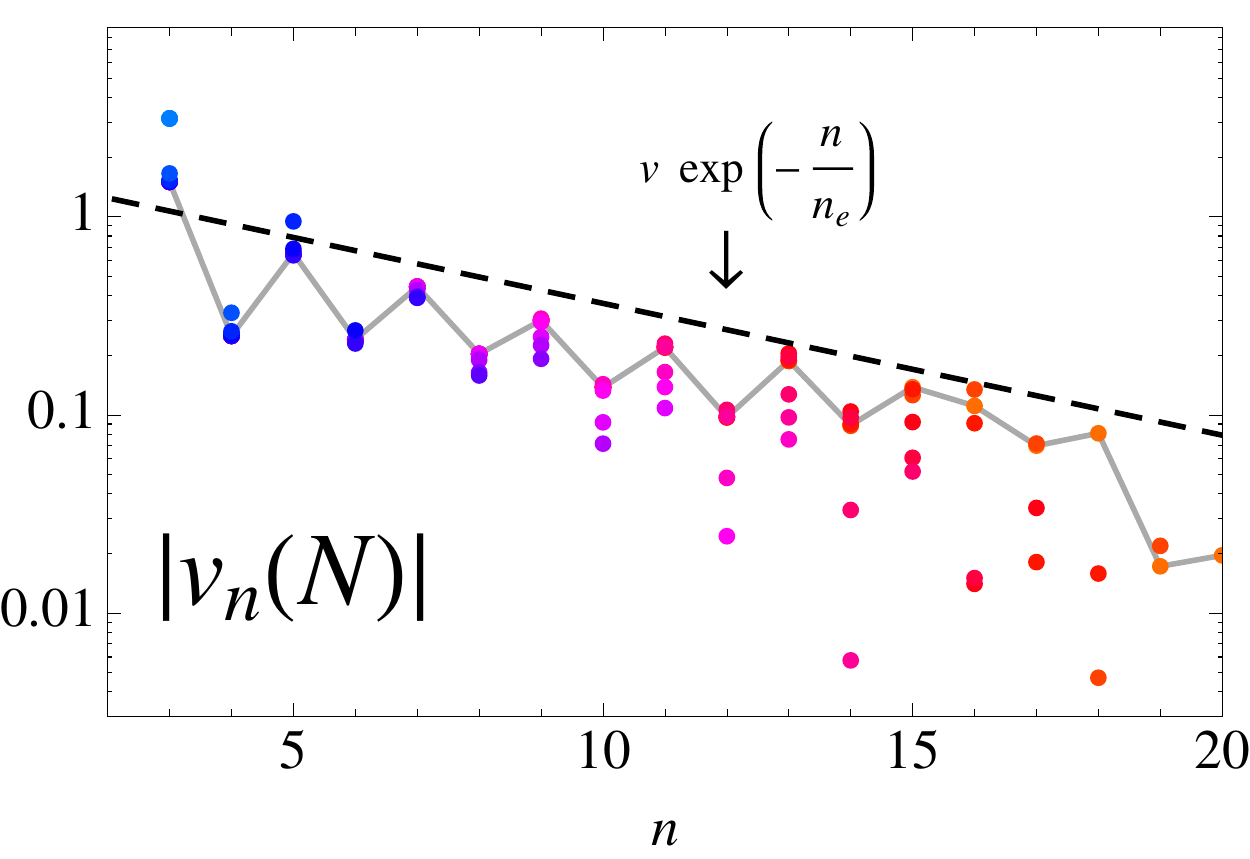}
\caption{\label{pRelative20} The approach towards near-Gaussian scaling: shown is the relative distance $\bar v_m(N)$  \eq{Delta} of  scaling exponents $\vartheta_n$ from Gaussian values $\vartheta_{G,n}$, as a function of the approximation order $N$. The approach to Gaussian values is well-approximated by an exponential (dashed line).}
\end{center}
\end{figure}

 \subsection{Origin of near-Gaussianity}

In Fig.~\ref{pMaxColor}, we further investigate the origin for the near-Gaussian behaviour. Shown  are the scaling exponents $\bar\vartheta_m$, which, for each approximation order $N$, correspond to the largest eigenvalue within the set of eigenvalues \eq{thetas}   or its real part if complex,
  \beq
  \bar\vartheta_m={\rm Re}\,\vartheta_{m=N-1}(N)\,.
  \eeq
We also indicate if the largest exponent is real (full dot) or a complex conjugate pair (crossed dot). We observe that with increasing $m$ (or $N$) the largest eigenvalue rapidly approaches Gaussian values. This pattern is qualitatively similar to the pattern observed for $f(R)$ type approximations \cite{Falls:2014tra}. As opposed to the $f(R)$ models, the imaginary part of high order exponents comes out smaller. For this reason, the approach to near Gaussianity is largely insensitive to wehther the largest exponent has a small imaginary component, or not.

In Fig.~\ref{pRelative20} we investigate the rate at which near-Gaussian behaviour is achieved in the scaling exponents. Shown is a semi-logarithmic plot for  the relative shift of the eigenvalues away from Gaussian values, introducing
\begin{equation}\label{Delta}
{\rm v}_n(N)=1-\frac{{\rm Re}\,\vartheta_n(N)}{\vartheta_{{\rm G},n}}\,.
\end{equation}
The colour-coding of the data shows the trend that $|{\rm v}_n(N)|$ decreases with increasing $N$. 
Based on the data up to $1/N_{\rm max}\approx 0.05$,
we conclude that \eq{Delta} resides in the  10--20\% range, 
\beq\label{asym2}
|{\rm v}_n(N)|<0.005-2\,,
\eeq 
decreasing with increasing $n$. In addition, we estimate the asymptotic behaviour of \eq{Delta} by taking the average values  for each $n$ over all approximation orders $N$. These are indicated in Fig.~\ref{pRelative20} by the dashed black line to guide the eye. The mean values  show a much smoother dependence on $n$, slowly decaying with increasing $n$. Their envelope is  fitted  by a simple exponential, 
\beq\label{asym3}
{\bar {\rm v}_n}\approx v\cdot \exp\left(-\frac{n}{n_e}\right)\,.
\eeq 
All mean values from $n>5$ onwards, and most entries from the high-order data sets, are below the envelope. Quantitatively, we have
\beq\label{asym4}
\begin{array}{rcl}
v&\approx&1.69
\\[.5ex]
n_e&\approx&6.53\,.
\end{array}
\eeq 
The significance of  \eq{asym3} with \eq{asym4} is as follows. The parameter $v$ is a measure for the mean relative variation in \eq{Delta} at low $n$, and the parameter $n_e$ states at which order the relative variation becomes reduced by a factor of $e$. With $N_{\rm max}/n_e\approx 3$, the reduction at $N_{\rm max}$ is by a factor of $20$, consistent with \eq{asym2}.
The important piece of information here is that the data shows a consistent and fast asymptotic decay towards near-Gaussian values. Extrapolation of \eq{asym3}, \eq{asym4}  predicts that 
\begin{equation}\label{vto0}
{\rm v}_n(N)\to 0
\end{equation}
for sufficiently large $n$, and
$1/N\to 0$.
We also note that the apporach towards near-Gaussianity is much faster than in $f(R)$ type theories. In the later case, $v\approx 0.22$ and $n_e\approx 46.7$ has been found \cite{Falls:2014tra}. We may conclude that the dynamics of Ricci tensor fluctuations strengthens the near-Gaussian behaviour of higher order invariants.

As a last comment, we stress that near-Gaussian eigenvalues are not mandatory for the asymptotic safety conjecture to apply \cite{Falls:2013bv,Falls:2014tra}. In fact, even more substantial modifications of eigenvalues up to
\begin{equation}
{\rm v}_n(N)< 1
\end{equation}
at large orders with intrinsically non-Gaussian corrections would still be compatible with the asymptotic safety conjecture. In this light,  the quantum modifications of the high-order eigenvalues at the fixed point of our model are moderate.

\begin{center}
\begin{table}
\normalsize\begin{tabular}{cGcG} 
\toprule
\rowcolor{Yellow}
&\multicolumn{1}{c}{\ \ \bf AdS\ \ }  &\multicolumn{1}{c}{\bf dS}&\multicolumn{1}{c}{\bf dS} 
\\ 
\rowcolor{Yellow}
\multirow{-2}{*}{$\bm N$} & \multicolumn{1}{c}{$\quad\R_-<0\quad$}&\multicolumn{1}{c}{$\quad\R_+>0\quad$}&\multicolumn{1}{c}{$\quad\R_{++}>0\quad$}\\ 
\midrule
2&$-$&0.517073&$-$\\
3&$-$&0.683203&$-$\\
4&$-$2.71391&0.875878&1.83803\\
5&$-$&1.00965&1.33600\\
6&$-$2.91114&$-$&$-$\\
7&$-$2.42360&0.998506&1.46442\\
8&$-$2.47362&0.994904&1.50776\\
9&$-$2.43194&0.993475&1.54196\\
10&$-$2.28713&0.996260&1.45479\\
11&$-$2.27572&0.996105&1.46058\\
12&$-$2.25090&0.996225&1.45297\\
13&$-$2.22012&0.996149&1.46132\\
14&$-$2.19669&0.996174&1.45683\\
15&$-$2.18639&0.996168&1.45842\\
16&$-$2.17282&0.996171&1.45706\\
17&$-$2.16134&0.996169&1.45799\\
18&$-$2.15214&0.996170&1.45749\\
19&$-$2.14661&0.996170&1.45772\\
20&$-$2.13907&0.996170&1.45752\\
21&$-$2.13418&0.996170&1.45762\\
\midrule
\rowcolor{Yellow}
mean value (last four)&$-$2.14300&0.996170&1.45758\\
\rowcolor{Yellow}
standard~deviation&$\pm0.4\%$& $\pm10^{-7}\%$& $\pm10^{-4}\%$\\ 
\bottomrule
\end{tabular}  
\caption{\label{tdeSitter} Shown are the three real anti-de Sitter and de Sitter solutions $\R_-<0<\R_+<\R_{++}$ within the radius of convergence of the underlying expansion and their dependence on the approximation order $N$.}
\end{table} 
\end{center}

\section{\bf From asymptotic safety to cosmology}\label{sec:cosmology}

Cosmology, thanks to the wealth of data available from observation \cite{Ade:2013uln,Ade:2013zuv,Perlmutter:1998np}, offers an important territory to test the asymptotic safety scenario for gravity. Provided that asymptotic safety is realised in nature, it is conceivable that the characteristic quantum gravitational modifications  have impacted during the very early universe,  including its phase of inflationary expansion and the phase of  late time acceleration. A number of studies have explored  these possibilities 
by exploiting characteristics of an asymptotically safe fixed point using renormalisation group improvements of the effective action or of the gravitational equations of motion including those of Friedmann-Robertson-Walker universes \cite{Shapiro:2000dz,Weinberg:2009wa,Reuter:2005kb,Bonanno:2001xi,Bonanno:2001hi,
Bonanno:2009nj,Bonanno:2010bt,Koch:2010nn,Bonanno:2012jy,Frolov:2011ys,Cai:2011kd,
Contillo:2011ag,Hindmarsh:2011hx,Ahn:2011qt,Hindmarsh:2012rc,Copeland:2013vva,
Kaya:2013bga,Saltas:2015vsc, Bonanno:2015fga,Kofinas:2016lcz}.

In this section, we apply our findings to models of cosmology. Our main interest relates to the availability of cosmological scaling solutions such as inflation in the very early universe, or possibly accelerated late time expansion.

\subsection{Stationarity of the quantum effective action}
Given the fast convergence of the polynomial fixed point solution, we now ask the question whether the UV fixed point solves the equation of motion. 
Since we have evaluated the flow equation on the sphere, solutions to the equation of motion correspond to (Euclidean)  de Sitter  spaces which give a stationary point of the action. Specifically, varying the action \eq{ansatzeaa} with respect to the metric and evaluating it on a sphere we obtain
\beq \label{deltaGamma}
\delta \Gamma =\frac{k^4}{16 \pi} \int d^4x \sqrt{g} 
\,E(\R) \,
g^{\mu \nu} \delta g_{\mu \nu}
\eeq
with
\beq \label{E}
E(\R) = \frac1 2 f\left(\frac{\R^2}{4}\right) - \frac{\R^2}{8} f'\left(\frac{\R^2}{4}\right) +  \frac{1}{4} \R\, z\left(\frac{\R^2}{4}\right)- \frac{\R^3}{8} z'\left(\frac{\R^2}{4}\right)
\eeq
The requirement of stationarity, meaning that $\delta \Gamma/\delta g_{\mu \nu}$ vanishes, entails the ``equation of motion''
\begin{equation}\label{deSitter}
E(\R)=0\,.
\end{equation}
Alternatively one obtains the very same condition by first evaluating the entire action on a sphere with constant Ricci curvature, and then searching for  stationary points of the ``potential'' $\Gamma(\R)$. This leads to
\beq\label{potential}
\Gamma(\R) = \frac{24 \pi }{ \R^2} \left[f\left(\frac{\R^2}{4}\right) + \R \cdot z\left(\frac{\R^2}{4}\right)\right]
\eeq
where the prefactor $\frac{24 \pi }{ \R^2}$ originates from the volume of the four-sphere. Requiring stationarity of the potential \eq{potential}, $\partial \Gamma(\R)/\partial\R=0$  with $\R \neq 0$ then leads to \eq{E} with \eq{deSitter}.
Its solutions $\R_{\rm dS}$ identify  stationary points for both (Euclidean) de Sitter and
 anti-de Sitter corresponding to positive and negative $\R_{\rm dS}$ respectively. The set of solutions with positive $\R_{\rm dS}$ are of relevance for cosmological scenarios with inflation and determine the Hubble parameter \cite{Falls:2016wsa}.

 \begin{figure*}[t]
\begin{center}
\includegraphics[width=.8\hsize]{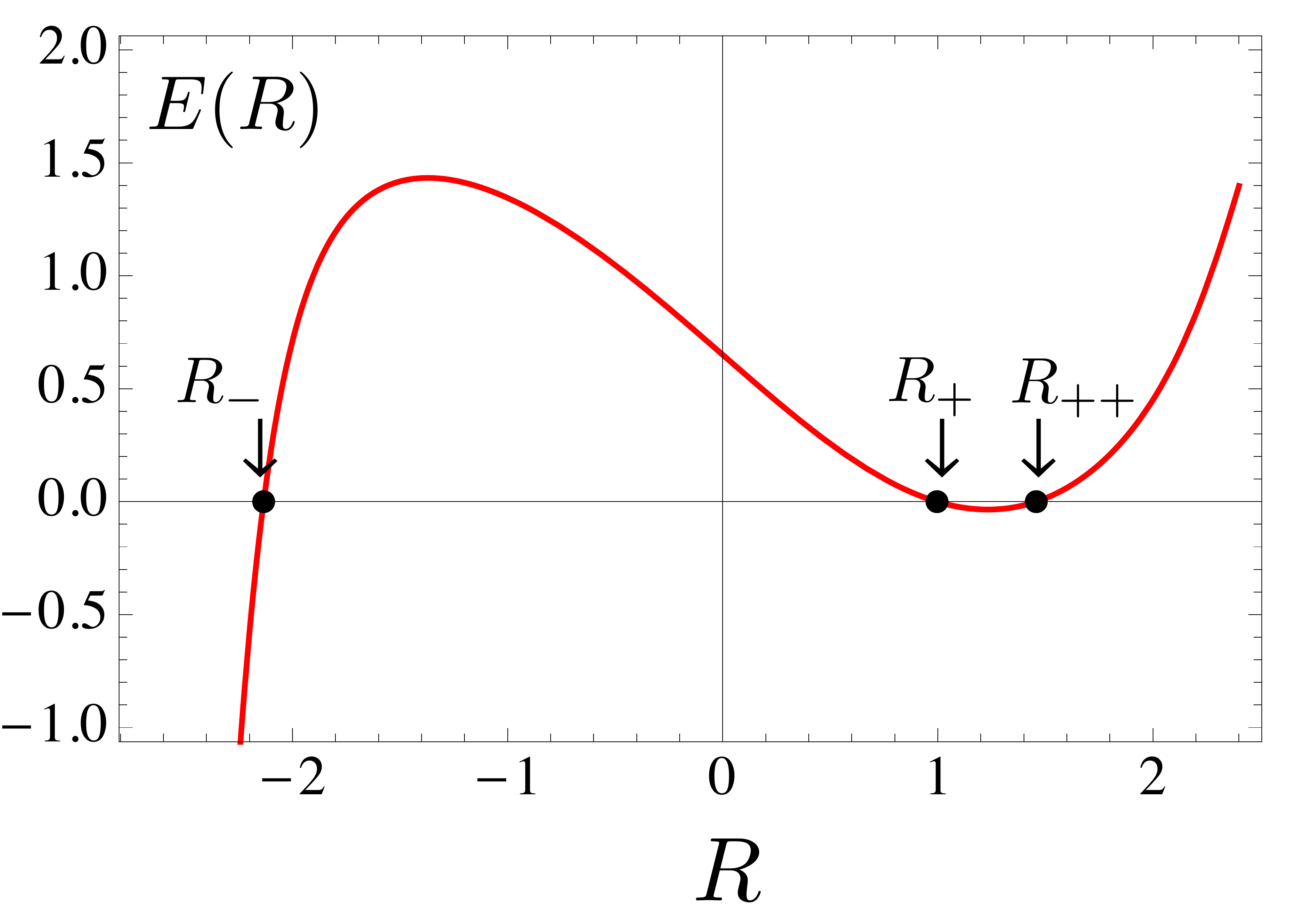}
\caption{\label{pdeSitter} Shown is the equation of motion  \eq{E} including the (anti)-de Sitter solutions, corresponding to zeros of  \eq{deSitter}.
The three solutions $\R_-<0<\R_+<\R_{++}$, indicated with arrows, correspond to two de-Sitter solutions at positive $R$ and an anti-de Sitter solutions at negative $R$. For numerical values, see Tab.~\ref{tdeSitter}.}
\end{center}
\end{figure*}

\begin{figure*}[t]
\begin{center}
\includegraphics[width=.75\hsize]{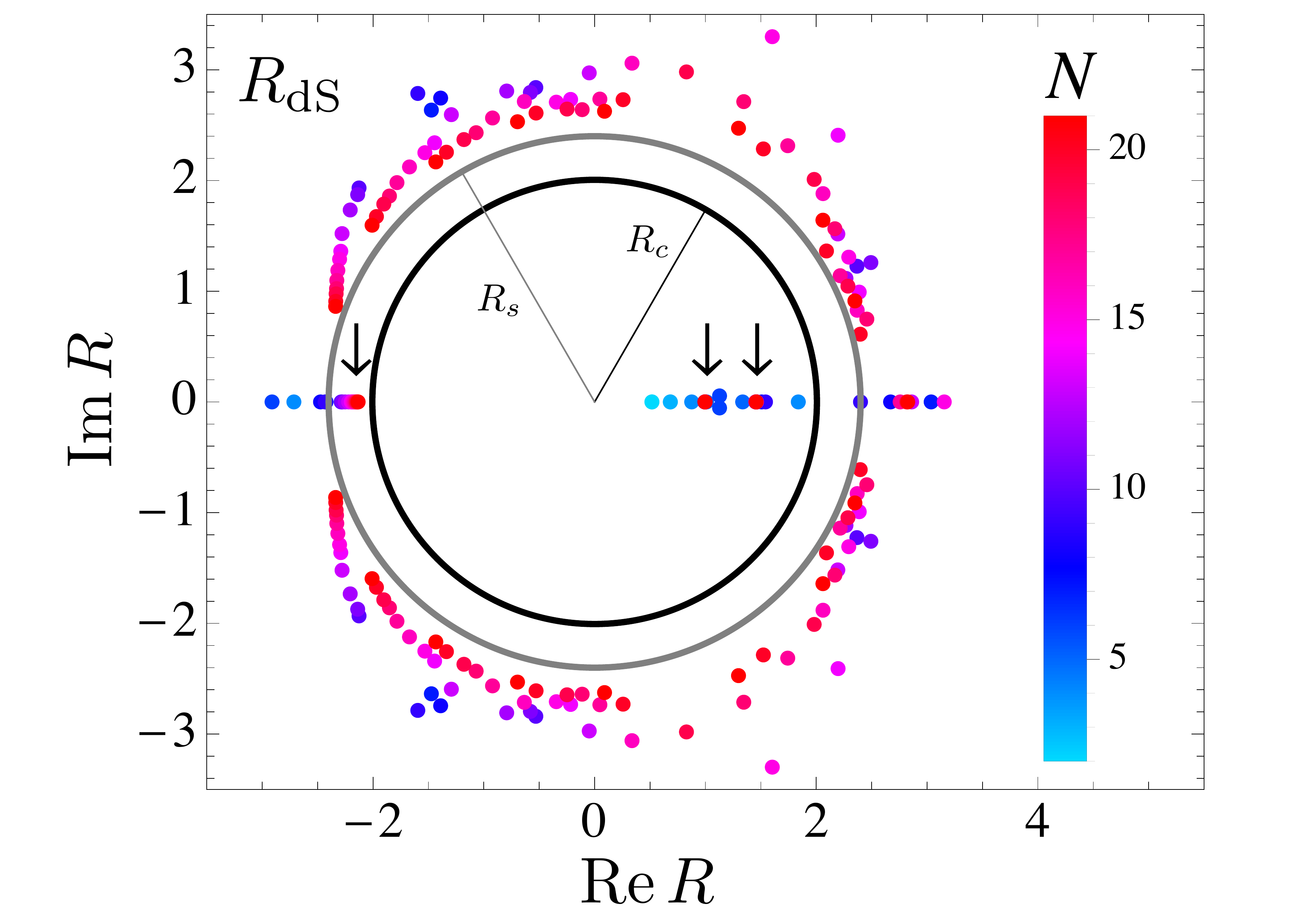}
\caption{\label{deSitterComplex}
Roots of the equations of motion in the complexified field plane. The three arrows indicate the locations of the real (anti-) de Sitter solutions $\R_-<0<\R_+<\R_{++}$, which appear as accumulation points with increasing approximation order $N$. The colour coding of approximations orders is indicated in the inset. The black circles indicate the radius $R_c\approx 2.006$ (inner circle) and $R_s\approx 2.4$ (outer circle).      }
\end{center}
\end{figure*}

   \subsection{De Sitter and anti de Sitter solutions}
  For the polynomial fixed point solutions, the de Sitter conditions  are polynomial as well and can possess several roots as we increase the order $N$.   We can look for 
 solutions of \eq{deSitter} at each order $N$ of the fixed point solution by plotting $\Gamma_*(\R)$ and 
 looking for stationary points. 
  Overall, we observe several real solutions, with
 \beq\label{real}
 \R_-<0<\R_+<\R_{++} \,.
 \eeq
 The positive real solutions $\R_+$ and $\R_{++}$ correspond to a minimum and a maximum of $\Gamma_*$, and the additional   negative root $\R_{-}$ to a minimum. 
  In Tab.~\ref{tdeSitter} we list numerical values for the first negative root $\R_{-}$ and the two smallest 
 positive roots $\R_{+}$ and $\R_{++}$. At orders $N=2$ and $N=3$ in the polynomial approximation there exists only one positive root, $\R_{+}$, proportional to the cosmological constant, because the $R^2$ coupling does not contribute to the equation of motion. At order $N=5$ there is no real negative solution, while at order $N=6$ only the negative real root is present. At order $N=4$ and from order $N=7$ onwards, all three real stationary points \eq{real}  
 are present. Overall, a fast convergence is observed with increasing order of the polynomial approximation. 
 
 The occasional absence of solutions at low orders in Tab.~\ref{tdeSitter} can be linked to the fact that the couplings  have not fully converged to their final values, see Tab.~\ref{FixedPointsScaled}. In fact, if we substitute the more accurate $N_{\rm max}=21$ fixed point values into the expression \eq{E} and then solve \eq{deSitter} at approximation order $N$, we find the roots \eq{real} for all $N$, except for the root $\R_{-}$ which remains absent at $N=5$.

 In Fig.~\ref{pdeSitter} 
we demonstrate  that the full numerical solution shares the same stationary points of the polynomial approximation by plotting the equation \eq{deSitter}. Both the positive $R$ solutions are found in the numerical solution. Notice that the AdS solution is beyond  $\R = - 2.00648$ which is outside of the domain of our numerical solution.

We have also searched for solutions of \eq{deSitter} in the complex field plane, see Fig.~\ref{deSitterComplex}. At each and every order in the polynomial expansion a large number of complex roots are found. Concretely, once $N>3$,  the equation of motion \eq{deSitter} has $N-1$ roots in the complex plane. All of these solutions are shown in  Fig.~\ref{deSitterComplex} for all approximations $N\le N_{\rm max}$. 
We also indicate the radius of convergence set by the singularity, $R_c$,
The radius $R_s$ is estimated from the polynomial expansion of \eq{deSitter} using the same techniques as in \cite{Falls:2016wsa}; the band between $R_c$ and $R_s$ thus serves as an error bar.

 We emphasize that most of the roots are unphysical in that they arise outside the domain of validity where $|R|>R_s$. Moreover, all stable and reliable solutions are on the real axis (highlighted by arrows in Fig.~\ref{deSitterComplex}).  Most importantly, both positive roots are firmly within the domain of validity, while the stable negative root resides at the boundary. 
 
For $f(R)$-type theories within the same set-up, stable solutions in the complex field plane where found, close to the real axis \cite{Falls:2016wsa}.

\section{\bf Comparison}\label{sec:discussion}
We are now in a position to evaluate the impact of Ricci tensor invariants in the effective action. Our results are based on the effective action
  \begin{equation}\label{ansatzfRicci}
\bar\Gamma_k[g]=\int d^dx\,\sqrt{g}\,f_k(R, R_{\mu\nu}R^{\mu\nu})
\end{equation}
where $f_k(R,R_{\mu\nu}R^{\mu\nu})\equiv F_k(R_{\mu\nu}R^{\mu\nu})+R\,Z_k(R_{\mu\nu}R^{\mu\nu})$, see \eq{idea}, \eq{ansatzeaa}. Findings are compared with closely related studies of quantum gravity for an $f(R)$-type action of the form
\begin{equation}\label{ansatzfR}
\bar\Gamma_k[g]=\int d^dx
\sqrt{g}\,f_k(R)\,,
\end{equation}
 see \cite{Falls:2013bv,Falls:2014tra,Falls:2016wsa}. The main point  is that the exact same regularisation schemes and momentum cutoffs have been used in either case. Therefore, qualitative and quantitative differences can now unambiguously be associated to the dynamical differences in the setup \eq{ansatzfRicci} vs \eq{ansatzfR}, that is, the presence (or absence) of Ricci tensors in the effective action. 
 Our results are summarised in Tab.~\ref{tCompare}. A few comments are in order:

 \begin{center}
\begin{table}
\normalsize\begin{tabular}{ccc} 
\toprule
\rowcolor{Yellow}
\bf Fixed point action&$\bm{f(R)}$&\multicolumn{1}{c}{$\bm{f(R, R_{\mu\nu}R^{\mu\nu})}$} 
\\ 
\midrule
\rowcolor{LightGray}
Bootstrap test&Yes\ \ \ \cite{Falls:2013bv,Falls:2014tra}&Yes\\
&3-dimensional\ \ \ \cite{Codello:2007bd,Falls:2013bv,Falls:2014tra}&3-dimensional\\
\multirow{-2}{*}{UV critical surface}
&$\ \ \{\sqrt{g},\sqrt{g}R,\sqrt{g}R^2\}\ \ $
&$\{\sqrt{g},\sqrt{g}R,\sqrt{g}R_{\mu\nu}R^{\mu\nu}\}$\\
\rowcolor{LightGray}
Near-Gaussianity&Yes\ \ \ \cite{Falls:2013bv,Falls:2014tra}&Yes\\
&Moderate\ \ \ \cite{Falls:2016wsa}&Maximal\\
\multirow{-2}{*}{Radius of convergence}&$(R_c/R_{\rm max}\approx 0.41\ldots0.45)$&$(R_c/R_{\rm max}=1)$\\
\rowcolor{LightGray}
Rate of convergence&Slow\ \ \ \cite{Falls:2016wsa}&Fast\\
Recursive relations&Yes\ \ \ \cite{Falls:2014tra}&No\\
\rowcolor{LightGray}
de Sitter solutions&No\ \ \ \cite{Falls:2016wsa}&Yes\\
Anti de Sitter solutions&Yes\ \ \ \cite{Falls:2016wsa}&Yes\\
\bottomrule
\end{tabular}  
\caption{\label{tCompare} Comparison of results from $f(R)$ quantum gravity (middle column) with the findings of this work (right column) in an otherwise identical setup (same gauge fixing, momentum cutoff, heat kernels).}
\end{table} 
\end{center}

\begin{itemize}
\item[a)] 
UV critical surface and scaling. On the basis of either of \eq{ansatzfRicci} and \eq{ansatzfR}, the UV critical surface is three-dimensional, and the UV fixed point satisfies the bootstrap test. Canonical mass dimension continues to offer a ``good'' ordering principle \cite{Falls:2014tra}:  classically relevant couplings receive strong quantum correction and remain relevant at an interacting UV fixed point. Classically irrelevant couplings remain irrelevant, and, moreover, their scaling exponents become increasingly near-Gaussian with increasing canonical mass dimension. The main difference relates to the rate with which exponents approach near-Gaussian values, which is found to be substantially larger as soon as Ricci tensor invariants are retained. Finally, the classically marginal couplings, corresponding to the invariants $\sqrt{g}R^2$ and $\sqrt{g}R_{\mu\nu}R^{\mu\nu}$ respectively, both become relevant in the quantum theory with scaling exponents of order unity.

\item[b)] 
Speed of convergence. Within polynomial expansions of  either of \eq{ansatzfRicci} and \eq{ansatzfR}, couplings at the fixed point and universal scaling exponents show numerical convergence. Quantitatively, for $f(R)$ gravity with \eq{ansatzfR} the rate of convergence is slow, owing to an eigth-fold periodicity pattern hidden underneath the fixed point equation \cite{Falls:2013bv,Falls:2014tra}. It has also been shown that the slow rate of convergence is ultimately  related to the presence of a near-by pole in the complex field plane \cite{Falls:2016wsa}.  Poles in the complex field plane are regular encounters in non-perturbative fixed points \cite{Morris:1994ki,Litim:2002cf,Litim:2016hlb,Juttner:2017cpr,Marchais:2017jqc}. Including Ricci tensor invariants, \eq{ansatzfRicci}, we have observed a substantially faster convergence both for the fixed point couplings as well as for the scaling exponents, see Figs.~\ref{pLambdaConvergence},~\ref{pThetas},and \ref{pThetaConvergence}, and Tab.~\ref{tconverge}. We conclude that Ricci tensor fluctuations have had an impact on the singularity structure of the UV fixed point solution.

\item[c)] 
Radius of convergence. The renormalisation group flow for either of \eq{ansatzfRicci} and \eq{ansatzfR} displays a technical pole at  $R_p\approx 2.00648\cdots$, see \eq{r0}. However, the polynomial radius of convergence for $f(R)$ gravity with \eq{ansatzfR} was found to be much smaller \cite{Falls:2016wsa}, 
$R_c\approx 0.82 \ldots 0.91$.
On the other hand, evaluating the flow for the Ricci tensor action \eq{ansatzfRicci} on spheres with constant curvature, we found that the radius of convergence is maximal,
$R_c=R_p\approx 2.00648$,
and much larger than for $f(R)$ gravity in an otherwise identical setup. The fact that the maximal range in $R$ is exhausted indicates that the fixed point effective action for \eq{ansatzfRicci} does not display convergence-limiting singularities in the complex field plane within $|R|<R_p$. We conclude that Ricci tensor interactions stabilise the theory.

\item[d)] 
Existence of de Sitter solutions. In either case, the fixed point effective actions in the UV, when evaluated on spaces with constant scalar curvature, are functions of the Ricci scalar, $\Gamma_*=\Gamma_*(R)$. For comparison, we have plotted the equations of motion for either of these  in Fig.~\ref{pCompareE}. Results from \eq{ansatzfRicci} are shown with a full line, and those from \eq{ansatzfR} with a dashed one.  
Qualitatively, both equations of motion display the same overall features including an AdS-type zero for negative $R$. For positive $R$, the Ricci tensor contributions push the equation of motion below the zero axis, thereby generating two de Sitter solutions $R_{\rm dS}>0$. Hence, despite the fact that Ricci tensors only offer mild quantititative corrections, these do lead to an imprtant qualitative change: no de Sitter solutions have been found with \eq{ansatzfR} for small Ricci or moderate curvatures.

\item[e)] 
Higher derivative operators. We add a few  comments highlighting structural differences between UV fixed points derived from either \eq{ansatzfRicci} or \eq{ansatzfR}. Differences between the  corresponding  flow  equations \eqref{FlowEquationRM}, \eq{TheFlowEquation}  become evident within a polynomial approximation up to some highest order $N$. Most notably, projecting the flow onto the highest-order coupling and demanding for a fixed point leads in the $f(R)$ case to a linear equation for the highest order coupling. 
Unlike $f(R)$ theories, 
which contain a scalar degree of freedom in addition to the polarisation of the  graviton, the inclusion of terms such as $R_{\mu\nu} R^{\mu\nu}$ leads to a fourth order inverse propagator for the graviton. Consequently, 
as soon as Ricci or Riemann tensor interactions are present, the corresponding equation becomes quadratic.
The reason for the latter is that
the contribution from the transverse 
traceless part $h^{T\mu\nu}$ contains the highest-order coupling itself.
Ultimately, the appearance 
of the highest-order 
coupling in the contribution of $h^{T\mu\nu}$ comes from the fact that its second variation contains a fourth-order differential operator $\Box^2$, a term which is absent in the $f(R)$ case. For further aspects of asymptotic safety in fourth order gravity models, see \cite{Benedetti:2009rx,Niedermaier:2009zz,Niedermaier:2010zz,Christiansen:2016sjn,Becker:2017tcx}.

\begin{figure*}[t]
\begin{center}
\includegraphics[width=.78\hsize]{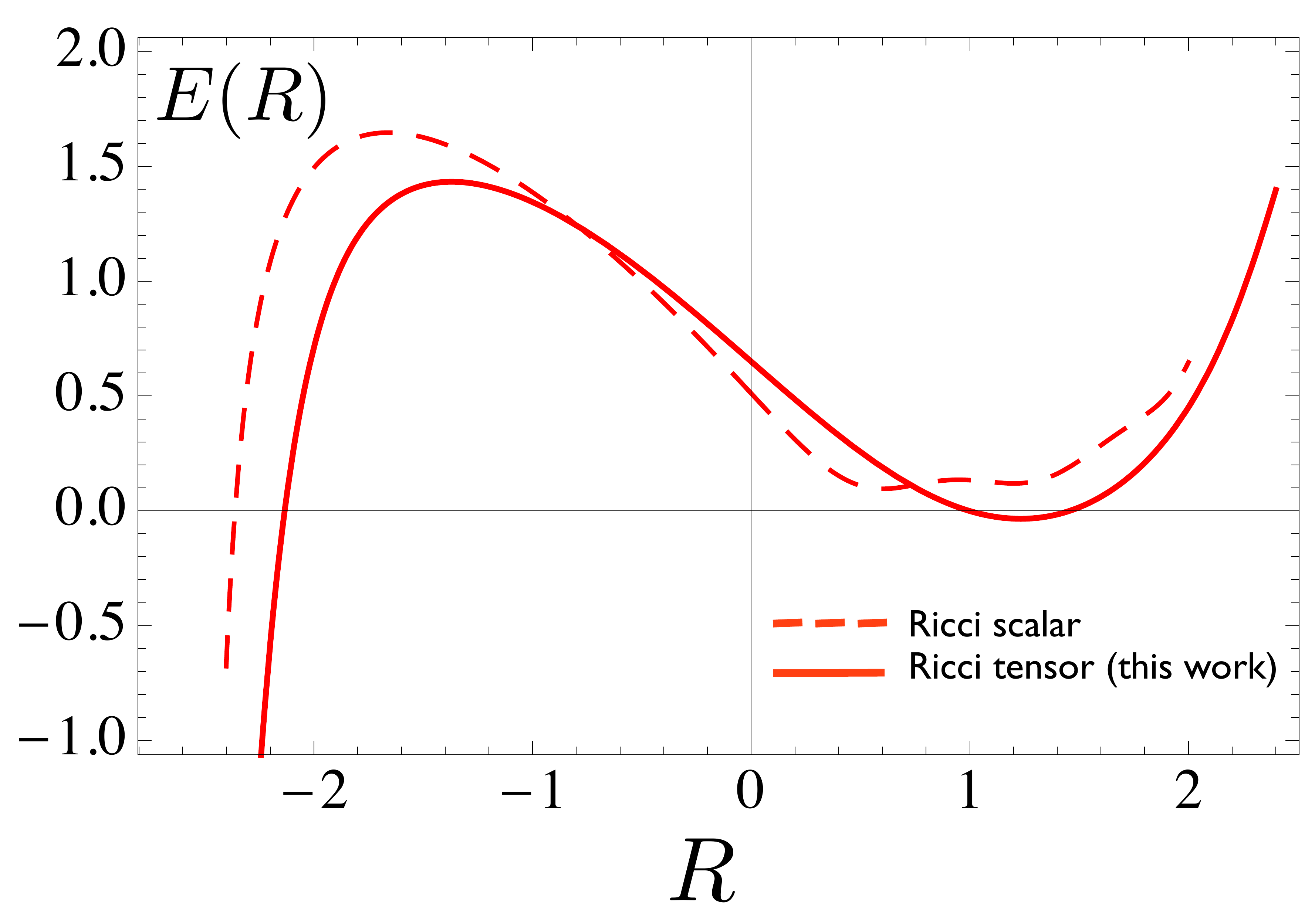}
\caption{\label{pCompareE} Comparison of the equations of motions of the present theory (full line) with results from $f(R)$ quantum gravity (dashed line) from \cite{Falls:2016wsa}, as functions of the scalar Ricci curvature. Both studies use the exact same set of regularisation and gauge fixings. We observe that the fluctuations of the Ricci tensor ultimately generate de Sitter solutions $R_{\rm dS}>0$ which otherwise would be absent for small curvature  (see text).}
\end{center}
\end{figure*}

\item[f)] 
Recursive structure.  In the $f(R)$ case, it is straightforward to 
set up an iterative recursive fixed point solution for all polynomial couplings, as done in \cite{Falls:2013bv,Falls:2014tra}. In principle, the same philosophy could be used here as well, except that roots arise at the leading order  (see Sec.~\ref{FPdata}). The proliferation of nested roots,  after iteration, makes a computer-algebraic solution of the recursive relations  cumbersome. 
We have then limited ourselves to approximations up to the $21^{\rm st}$ order \cite{Nikolakopoulos:Thesis,King:MSc}. Although the approximation order is much lower than the  35 orders achieved in the corresponding $f(R)$ study \cite{Falls:2013bv,Falls:2014tra}, this is more than compensated by the substantially enhanced rate of convergence, see points a), b), c). 
\end{itemize}

In conclusion, we have established that the dynamics of the Ricci tensor has a strong impact onto fixed point solutions. Most noticeably, the UV fixed point is stabilised and convergence is faster. Moreover, de Sitter solutions now arise very naturally within the domain of validity. Clearly, $f(R)$ models for quantum gravity and the model put forward here are both ``ad hoc'' inasmuch as a full quantum theory of gravity is expected to contain additional invariants not account for here. It is encouraging that the broad features are not affected by these choices. For future work, it will be interesting to extend these investigations 
for actions involving functions of Riemann and Ricci tensor invariants, and to explore the extend to which the UV fixed point depends on it.

\section{\bf Summary}
\label{sec:ConclusionsRM}

We have put forward a systematic study of the asymptotic safety conjecture for gravity using actions beyond Ricci scalars. The main novelty is that our models are sensitive to the dynamics of more complicated tensor  invariants, while still benefitting from the  simplicity of $f(R)$ models for gravity. 
We have illustrated our approach for actions of the form \eq{AnsatzRM}, \eq{ansatzeaa} involving functions of the squared Ricci tensor $R_{\mu\nu}R^{\mu\nu}$. The results are quite promising: the theory displays a stable interacting UV fixed point which, in high-order polynomial approximations, 
shows fast convergence with the order of approximation (Tab.~\ref{FixedPointsScaled}). The radius of convergence 
comes out maximal
(Tab.~\ref{tCompare}), and much larger than the one found in $f(R)$-type models of gravity. Moreover, for small curvature and within the radius of convergence, the UV fixed point is identified non-perturbatively without resorting to a polynomial approximation (Fig.~\ref{Numfeff}). 

The theory has three relevant parameters related to  the cosmological constant, the Ricci scalar and the square of the Ricci tensor. Higher order invariants become increasingly irrelevant, and scaling exponents converge very fast (Tab.~\ref{tconverge}), much faster than in theories with Ricci scalars only. It is also found that the theory becomes near-Gaussian with increasing mass dimension of curvature invariants (Fig.~\ref{pBootstrap}, \ref{pRicci21}). Clearly, higher order curvature invariants stabilise the fixed point, and the theory becomes ``as Gaussian as it gets'' \cite{Falls:2013bv,Falls:2014tra}.  
Our findings seem to confirm  once more that quantum scale invariance of gravity can be tested self-consistently by means of a bootstrap, despite of the fact that no small parameter has been identified. Also, the expansion in curvature invariants according to canonical mass dimension is viable. On the other hand, we find that the precise nature of higher order curvature  invariants might not be  centrally important for a fixed point to arise.

We also investigated implications of our model for quantum cosmology and the existence of inflationary solutions in the early universe. Most notably, we find that the UV fixed point of quantum gravity admits stable de Sitter solutions (Tab.~\ref{tdeSitter}). Here, the Ricci tensor fluctuations play a decisive role: without them, such as in the $f(R)$ approximation  \cite{Falls:2016wsa}, de Sitter solutions do not arise (Fig.~\ref{pCompareE}). We conclude that the existence of de Sitter solutions is strongly sensitive to the dynamics of metric fluctuations in the deep UV.

Our setup can be  developed in several directions. It is straightforward to extend it towards more general theories 
and to explore the impact of Riemann or Weyl tensor invariants, also using general Einstein spaces for the projection.
In the same spirit, the role of matter fields on the gravitational fixed point can  be explored \cite{Schroeder:Thesis,King:MSc,Christiansen:2017cxa}. 
Further directions for improvement include alternative routes for the projection \cite{Falls:2016msz}, ideas from the resummed $\epsilon$-expansion \cite{Falls:2017cze}, and the role of the background field 
\cite{Litim:1998nf,Freire:2000bq}. 
It would be particularly interesting to clarify how findings persist in settings that distinguish between background and fluctuation fields 
 (see \cite{Christiansen:2017bsy, Knorr:2017fus} for recent progress), 
or which are manifestly background independent (see \cite{Falls:2017nnu} for  related advances in gauge theories). Ultimately, the bootstrap search strategy can be used with either of these to test, or even falsify, the asymptotic safety conjecture for gravity.
These topics are left for future work.

\section*{\bf Acknowledgements}

Parts of this work were supported by the Science Technology and
Facilities Council (STFC) under grant number [ST/G000573/1], and by the A.S.~Onassis
Public Benefit Foundation under grant number [F-ZG066].

\appendix
\addtocontents{toc}{\protect\setcounter{tocdepth}{0}} 

\renewcommand{\thesection}{{\bf \Alph{section}}}

\section{\bf Variations of  invariants}
\label{app:Variations}

The first and second variations for various terms appearing in \eqref{QuadraticPartFR} evaluated on the sphere  can be read off from Tab.~\ref{VariationsTable}.
\begin{center}
\begin{table*}[b]
\centering
\setlength{\extrarowheight}{6pt}
\normalsize
\begin{tabular}{G  c}
\toprule
\rowcolor{Yellow}\multicolumn{2}{c}{\bf first variations}\\
\midrule
 $\delta(\sqrt{g})$ & $\frac{1}{2}\sqrt{g}h$ 
 \\[2mm] \midrule
 $\delta(R)$ & $-\frac{R}{d}h+\nabla^{\mu}\nabla^{\nu}h_{\mu\nu}-\nabla^{2}h$
 \\[2mm] \midrule
$\delta(R_{\mu\nu}R^{\mu\nu})$ & $-2\frac{R^2}{d^2}h-2\frac{R}{d}\nabla^2h+2\frac{R}{d}\nabla^{\mu}\nabla^{\nu}h_{\mu\nu}$
\\ [2mm] 
\toprule
\rowcolor{Yellow}\multicolumn{2}{c}{\bf second variations}\\
\midrule
$\delta^{(2)}(\sqrt{g})$ & $\frac{1}{2}\sqrt{g}\left(\frac{1}{2}hh-h_{\mu\nu}h^{\mu\nu}\right)$
\\[2mm] & \\[-7mm]  
\midrule
 & $h_{\mu\nu}\left[\frac{1}{2}\nabla^2+\frac{d-2}{d(d-1)}R\right]h^{\mu\nu}$ \\[2mm] &\\[-7mm] 
$\delta^{(2)}(R)$&$+h\left[\frac{1}{2}\nabla^2+\frac{R}{d(d-1)}\right]h$
\\[2mm] &\\[-7mm] 
&$+(\nabla_{\rho}h^{\rho\nu})(\nabla^{\mu}h_{\mu\nu})$\\
\midrule
 & $h_{\mu\nu}\left[\frac{1}{2}\nabla^4+\frac{R}{d}\frac{d-3}{d-1}\nabla^2+2\frac{R^2}{d^2}\frac{d^2-3d+3}{(d-1)^2}\right]h^{\mu\nu}$\\
$\ \ \delta^{(2)}(R_{\mu\nu}R^{\mu\nu})\ \ $&$+h\left[\frac{1}{2}\nabla^4+\frac{R}{d}\frac{3d+1}{2(d-1)}\nabla^2+2\frac{R^2}{d^2}\frac{2d-3}{(d-1)^2}\right]h$\\
&$+(\nabla_{\mu}\nabla_{\nu}h^{\mu\nu})(-\nabla^2)h+(\nabla_{\rho}h^{\rho\nu})\left(\nabla^2+\frac{3R}{d}\right)(\nabla^{\mu}h_{\mu\nu})+(\nabla_{\mu}\nabla_{\nu}h^{\mu\nu})^2 $
\\
\bottomrule
\end{tabular}  
\caption{Summary of the first and second variations appearing in \eqref{QuadraticPartFR}, evaluated on the sphere.}\label{VariationsTable}
\end{table*} 
\end{center}

\section{\bf Heat kernel  formul\ae}
\label{app:HeatKernel}

In this appendix we summarise some technicalities of the heat kernel expansion. We examine how these are affected when we consider constrained fields. For a full overview of the heat kernel methods we refer to \cite{Avramidi:2000bm,Gilkey:1995mj}. Some of the information found here has also been discussed in \cite{Lauscher:2002sq,Codello:2008vh,Machado:2007ea}.

\subsection*{Constrained fields}
\label{sec:ConstrainedFields}

First, we need to know how the trace evaluation is modified due to the fact that we decompose our original fields. For example, any vector field $V^{\mu}$ can be decomposed in its transverse and longitudinal part as
\begin{equation}
V^{\mu}=V^{T\mu}+\nabla^{\mu}\eta\label{VectorDecomposition}
\end{equation}
with $\nabla_{\mu}V^{T\mu}=0$. Similarly, any symmetric tensor $h_{\mu\nu}$ is decomposed according to
\begin{equation}
h_{\mu\nu}=h^T_{\mu\nu}+\nabla_{\mu}\xi_{\nu}+\nabla_{\nu}\xi_{\mu}+\nabla_{\mu}\nabla_{\nu}\sigma-\frac{1}{d}g_{\mu\nu}\nabla^2\sigma+\frac{1}{d}g_{\mu\nu}h
\end{equation}
subject to the constraints 
\begin{equation}
g^{\mu\nu}h^T_{\mu\nu}=0, \qquad \nabla^{\mu}h^T_{\mu\nu}=0, \qquad \nabla^{\mu}\xi_{\mu}=0, \qquad h=g_{\mu\nu}h^{\mu\nu}
\end{equation}
so that $h^T_{\mu\nu}$ is the transverse-traceless part of $h_{\mu\nu}$, $\xi_{\mu}$ is a transverse vector, $\sigma$ is a scalar 
and $h$ is the trace part of $h_{\mu\nu}$. From now on we use the notation $(2T)$ for a transverse-traceless symmetric tensor, 
$(1T)$ for a transverse vector.

In order to see how this affects the calculation, we need to know how the coefficients $\mathbf{b}_n$ are modified when the operator 
is restricted to act on $(2T)$ tensors or on $(1T)$ vectors which can be related
to the spectrum of the unconstrained fields. 
We start with the transverse vectors and we note that the spectrum of every vector can be expressed as the union of the spectrum of a $(1T)$ vector and of the longitudinal mode $\nabla_{\mu}\eta$. The spectrum of $\nabla_{\mu}\eta$ can be related to that of the scalar field $\eta$ through commutation of covariant derivatives
\begin{equation}
-\nabla^2\nabla_{\mu}\eta=-\nabla_{\mu}\left(\nabla^2+\frac{R}{d}\right)\eta.
\end{equation}
We note however that for a constant scalar $\eta$, the vector $V^{\mu}$ receives no contribution from the longitudinal mode. So we have to subtract from the scalar trace the constant mode. Thus we write for the trace of a transverse vector
\begin{equation}
\textnormal{Tr}_{(1)}\left[e^{t\nabla^2}\right]=\textnormal{Tr}_{(1T)}\left[e^{t\nabla^2}\right]+\textnormal{Tr}_{(0)}\left[e^{t\left(\nabla^2+\frac{R}{d}\right)}\right]-e^{t\frac{R}{d}},
\end{equation}
where the last term corresponds to the zero mode of the scalar field. Thus we can relate the spectrum of the transverse vector to that of the unconstrained vector.

Now we turn our attention to the constraints of the transverse traceless tensors. For the symmetric tensors we use the 
commutation relations
\begin{equation}
-\nabla^2\left(\nabla_{\mu}\xi_{\nu}+\nabla_{\nu}\xi_{\mu}\right)=\nabla_{\mu}\left(-\nabla^2-\frac{d+1}{d(d-1)}R\right)\xi_{\nu}
+\nabla_{\nu}\left(-\nabla^2-\frac{d+1}{d(d-1)}R\right)\xi_{\nu}
\end{equation}
 and
 \begin{equation}
 -\nabla^2\left(\nabla_{\mu}\nabla_{\nu}-\frac{1}{d}g_{\mu\nu}\nabla^2\right)\sigma=\left(\nabla_{\mu}\nabla_{\nu}
 -\frac{1}{d}g_{\mu\nu}\right)\left(-\nabla^2-\frac{2}{d-1}R\right)\sigma.
 \end{equation}
As in the case of the transverse vector there are modes that do not contribute to the trace. These modes are $(i)$ the 
$\frac{d(d+1)}{2}$ Killing vectors for which $\nabla_{\mu}\xi_{\nu}+\nabla_{\nu}\xi_{\mu}=0$ so that they do not contribute to 
$h_{\mu\nu}$; $(ii)$ a constant scalar $\sigma$ as in the case of vectors and $(iii)$ the $d+1$ scalars which correspond to the
second lowest eigenvalue of $-\nabla^2-\frac{2}{d-1}R$. Thus we have for the trace of a symmetric tensor
\begin{equation}
\begin{split}
\textnormal{Tr}_{(2)}\left[e^{t\nabla^2}\right]=&\textnormal{Tr}_{(2T)}\left[e^{t\nabla^2}\right]+\textnormal{Tr}_{(1T)}\left[e^{t\left(\nabla^2+\frac{d+1}{d(d-1)}R\right)}\right]+\textnormal{Tr}_{(0)}\left[e^{t\nabla^2}\right]+\textnormal{Tr}_{(0)}\left[e^{t\left(\nabla^2+\frac{2}{d-1}R\right)}\right]\\
&-e^{t\frac{2}{d-1}R}-(d+1)e^{t\frac{1}{d-1}R}-\frac{d(d+1)}{2}e^{t\frac{2}{d(d-1)}R}.
\end{split}
\end{equation}
Again we can relate the spectrum of the constrained transverse traceless tensor to that of  rank-$2$ tensors, vectors, and scalars.

In order to clarify how the contributions from the exponents play a role in our calculation we expand the exponential in 
powers of $R$ such as $\sum_{m=0}^{\infty}c_mR^m$. Taking into account that the volume of the sphere goes like $V\sim R^{-d/2}$ 
and that the heat kernel coefficients go like $\mathbf{b}_n\sim R^{n/2}$ we find that
\begin{equation}
\int d^dx\sqrt{g}\,\mathbf{b}_n\sim R^{\frac{n-d}{2}}.
\end{equation}
Since ultimately we are interested in comparing powers of $R$, the exponentials contribute when $2m=n-d$,
and the coefficients $\mathbf{b}_n$ receive contributions only when $n\geq d$. Another way to see where the excluded modes enter is from the expansion $e^{-tz}=1-tz+\frac{1}{2}t^2z^2+...$. In order for the parameter $t$ to be included in $Q_i[W]=\int_0^{\infty}dt\,t^{-i}\tilde W(t)$ we see directly from \eqref{GeneralQ&B} that the corresponding power $m$ of the expansion $\sum_{m=0}^{\infty}c_mR^m$ is such that $2m=n-d$.

\subsection*{Heat-kernel coefficients}
Here we summarise the trace of the heat kernel coefficients $\textnormal{tr}_s\mathbf{b}_n=b_n|_{s}$ for the fields that we will be interested in after taking into account their constraints. We write 0 for the scalars, 1 for the vectors and 2 for the tensors. For the scalars we have
\begin{eqnarray}
b_0|_{0} &=& 1\\
b_2|_{0} &=& \frac{1}{6} R \\
b_4|_{0} &=& \frac{\left(5 d^2-7 d+6\right) R^2}{360 (d-1) d} \\
b_6|_{0} &=& \frac{\left(35 d^4-112 d^3+187 d^2-110 d+96\right) R^3}{45360 (d-1)^2d^2} \\
b_8|_0&=& \frac{\left(175 d^6-945 d^5+2389 d^4-3111 d^3+3304 d^2-516 d+2160\right)
  R^4}{5443200 (d-1)^3 d^3} \,.
\end{eqnarray}
For the transverse vector fields we have
\begin{eqnarray}
b_0|_1& &= d-1\\
b_2|_1 &=& \frac{R (6 \delta _{2,d}+(d-3) (d+2))}{6 d}\\
b_4|_1 &=&\frac{R^2 \left(360 \delta _{2,d}+720 \delta _{4,d}+5 d^4-12 d^3-47 d^2-186 d+180\right)}{360 (d-1) d^2}
 \\
b_6|_1&=& R^3 \left(\frac{\delta _{2,d}}{8}+\frac{\delta
  _{4,d}}{96}\right) \\
   &+&\frac{\left(35 d^6-147 d^5-331 d^4-3825 d^3-676
  d^2+10992 d-7560\right) R^3}{45360 (d-1)^2 d^3} \nonumber \\
b_8|_1 &=&R^4 \left(\frac{\delta _{2,d}}{96}+\frac{\delta _{4,d}}{768}+\frac{\delta
  _{6,d}}{2700}+\frac{15 \delta _{8,d}}{175616}\right) \\
  &+& \frac{\left(175
  d^7-2345 d^6-8531 d^5-15911 d^4+16144 d^3+133924 d^2-206400
  d+75600\right) R^4}{75600 (d-1)^3 d^4} \nonumber
\end{eqnarray}
Finally for the transverse traceless tensor fields we have
\begin{eqnarray}
b_0|_{2} & =& \frac{1}{2} (d-2) (d+1) 
\end{eqnarray}
\begin{eqnarray} 
b_2|_{2} &=& \frac{(d+1) (d+2) R (3 \delta _{2,d}+d-5)}{12 (d-1)} 
\\
b_4|_{2} &=& \frac{(d+1) R^2 \left(1440 \delta _{2,d}+3240 \delta _{4,d}+5 d^4-22
  d^3-83 d^2-392 d-228\right)}{720 (d-1)^2 d} 
\\
b_6|_{2} &=& R^3 \left(\frac{3 \delta _{2,d}}{2}+\frac{5 \delta
  _{4,d}}{36}\right) \\
  &+&\frac{(d+1) \left(35 d^6-217 d^5-667 d^4-7951
  d^3-13564 d^2-10084 d-28032\right) R^3}{90720 (d-1)^3 d^2} \nonumber \\
  b_8|_{2} &=& R^4 \left(\frac{\delta _{2,d}}{2}+\frac{5 \delta
  _{4,d}}{288}+\frac{\delta _{6,d}}{175} +\frac{675 \delta
  _{8,d}}{175616}\right) \\
  &+& \frac{1}{4!}\frac{\left(175 d^{10}-945 d^9+464 d^8-150566
  d^7+478295 d^6-2028005 d^5 \right) R^4}{453600 (d-1)^4 d^4}  \nonumber \\
  &+& \frac{1}{4!}\frac{(-2945774 d^4-5191124 d^3-10359960
  d^2-7018560 d-1814400)R^4}{453600 (d-1)^4 d^4}. \nonumber
\end{eqnarray}
In Table \ref{Multiplicities} we summarise the eigenvalues $\Lambda_l$ of the Laplace operator $-\nabla^2$ on scalars, transverse vectors and transverse-traceless symmetric tensors and their multiplicities $D_l$.

\begin{center}
\begin{table*}[t]
\addtolength{\tabcolsep}{4pt}
\setlength{\extrarowheight}{8pt}
\centering
\begin{tabular}{G | c | G } 
\toprule
\rowcolor{Yellow}
\bf  fields & \bf  eigenvalue $\bm{\Lambda_l(d,s)}$ & \bf  degeneracy $\bm{D_l(d,s)}$ 
 \\[1ex] \midrule &&\\[-4mm]
$T^{lm}_{\mu\nu}$ with $l\geq2$ &  $\displaystyle\frac{l(l+d-1)-2}{d(d-1)}R$ & $\displaystyle\frac{(d+1)(d-2)(l+d)(l-1)(2l+d-1)(l+d-3)!}{2(d-1)!(l+1)!}$  
\\[4mm] \midrule &&\\[-4mm]
$T^{lm}_{\mu}$ with $l\geq1$ & $\displaystyle\frac{l(l+d-1)-1}{d(d-1)}R$ & $\displaystyle\frac{l(l+d-1)(2l+d-1)(l+d-3)!}{(d-2)!(l+1)!}$  
\\[4mm] \midrule &&\\[-4mm]
$T^{lm}$ with $l\geq0$ & $\displaystyle\frac{l(l+d-1)}{d(d-1)}R$ &  $\displaystyle\frac{(2l+d-1)(l+d-2)!}{l!(d-1)!}$   \\[4mm] 
\bottomrule
\end{tabular}  
\caption{Summary of the discrete eigenvalue spectrum $\Lambda_l$  
of the operator $-\nabla^2$ on scalars, transverse vectors and transverse-traceless symmetric tensors ($s=0,1,2$ respectively) 
and their multiplicities  $D_l$ for dimension $d$ labeled by the parameter $l$.
}\label{Multiplicities}
\end{table*} 
\end{center}

\section{\bf Gravitational renormalisation group equations}\label{AppFlow}
In this appendix we summarise our formula for the exact functional RG flow given in \eq{FlowEquationRM}, \eq{TheFlowEquation}. Concretely, the RG flow reads
\begin{equation}
\partial_t f+ 4 f-\R^2\,f'+\R\left(\partial_t z+2  z-\R^2  z' \right)=I[f,z](\R) \label{FRG}
\end{equation}
where the LHS are the terms originating from the classical scaling dimensions of the functions $f$ and $z$. The RHS encodes the contributions from fluctuations. It splits into several parts as
\newpage
\begin{equation}
\begin{split}
I[f,z](\R)=&I_0[f,z](\R)+\partial_tz\,I_1[f,z](\R)+\partial_tf'\,I_2[f,z](\R)+\partial_tz'\,I_3[f,z](\R)\\
&+\partial_tf''\,I_4[f,z](\R)+\partial_tz''\,I_5[f,z](\R)\ .\label{FRG-RHS}
\end{split}
\end{equation}
The flow terms appearing in \eqref{FRG-RHS} arise through the Wilsonian momentum cutoff $\partial_tR_k$, which we have chosen to depend on the background field. All the terms $I_0[f,z],...,I_5[f,z]$ arise from tracing over the fluctuations of the metric field for which we have adopted the transverse traceless decomposition. The term $I_0[f,z]$ also receives $f$ and $z$-independent contributions from the ghosts and from the Jacobians originating from the split of the metric fluctuations into tensor, vector and scalar parts. To indicate the origin of the various contributions in the expressions below, we use superscipts T, V, and S to refer to the transverse traceless tensorial, vectorial, and scalar origin respectively. Then we have for the various components $I_i[f,z](\R)$ 
\begin{eqnarray}
I_0[f,z]=&& c\left(\frac{P_c^V}{D_c^V}+\frac{P_c^S}{D_c^S}+\frac{P_0^{Tz0}z+P_0^{Tf1}f'+P_0^{Tz1}z'+P_0^{T2}(f''+\R z'')}{D^T} \right.\nonumber\\ 
&+& \left. \frac{P_0^{Sz0}z+P_0^{Sf1}f'+P_0^{Sz1}z'+P_0^{Sf2}f''+P_0^{Sz2}z''+P_0^{S3}(f^{(3)}+\R z^{(3)})}{D^S}\right)\label{I0}
\\ 
I_1[f,z]=&&c\left(\frac{P^{T}_1}{D^T}+\frac{P^S_1}{D^S}\right)
\\
I_2[f,z]=&&c\left(\frac{P^{T}_2}{D^T}+\frac{P^S_2}{D^S}\right)
\\
I_3[f,z]=&&c\left(\frac{P^{T}_3}{D^T}+\frac{P^S_3}{D^S}\right)\\
I_4[f,z]=&&c\frac{P^S_4}{D^S}\\
I_5[f,z]=&&c\frac{P^S_5}{D^S} \,.
\end{eqnarray}
The numerical coefficient $c = 1/(24 \pi)$ arises due to the volume element and our convention to factor out $16 \pi$ from the definition of $f$ and $z$. The various denominators and polynomials appearing in the above expressions are given by
\begin{eqnarray}
D^T [f,z]&=& 36 f + (24\R+36) z - (7\R^2-6\R+36)(f'+\R\,z')\\
D^S [f,z]&=& 8 f +12 z +(-2\R^2-8\R+24) f' +(2\R^3-32\R^2+60\R) z' \\ \nonumber
&&+\R^2 (\R-3)^2 (f''+\R\, z'')\\
D_c^V &=& 4-\R\\
D_c^S &=& 3-\R
\\
P_c^V&=&\frac{607}{15}\R^2-24\R-144\\
P_c^S&=&\frac{511}{30}\R^2 -12\R-36\\ 
P_0^{Tz0}&=&\frac{311}{63}\R^3-4\R^2-1080\R +2880
 \\
P_0^{Tf1}&=& -\frac{\R^3}{3} -116\R^2+1800\R-4320
\\
P_0^{Tz1}&=& \frac{371}{108}\R^5+\R^4+122\R^3+840\R^2-3240\R
\end{eqnarray}
\begin{eqnarray} 
P_0^{T2}&=&-\frac{731}{1512}\R^6+\frac{\R^5}{6}+29\R^4-300\R^3+540\R^2\\
P_0^{Sz0}&=&\frac{37}{189}\R^3+\frac{116}{15}\R^2+72\R+192
\\
P_0^{Sf1}&=&-\frac{116}{45}\R^3-\frac{248}{15}\R^2+96\R+576 
\\
P_0^{Sz1}&=&\frac{1111}{2268}\R^5-\frac{29}{15}\R^4-\frac{170}{3}\R^3+40\R^2+1080\R\\
P_0^{Sf2}&=&\frac{1333}{4536}\R^6+\frac{29}{9}\R^5+\frac{62}{15}\R^4-16\R^3+36\R^2
\\
P_0^{Sz2}&=& \frac{27991}{45360}\R^7+\frac{406}{45}\R^6+\frac{943}{30}\R^5-16\R^4-126\R^3\\
\label{P0S3}
P_0^{S3}&=&\frac{181}{3360}\R^8+\frac{29}{30}\R^7+\frac{91}{20}\R^6-27\R^4\\ 
P_1^T&=&\frac{311}{126}\R^3-\R^2-180\R+360\\
P_1^S&=&\frac{37}{378}\R^3+\frac{29}{15}\R^2+12\R+24\\
P_2^T&=&
\frac{731}{1512}\R^4-\frac{\R^3}{6}-29\R^2+300\R-540\\
P_2^S&=&-\frac{127}{1620}\R^4-\frac{58}{45}\R^3-\frac{62}{15}\R^2+16\R+72\\ 
P_3^T&=&
\frac{731}{1512}\R^5-\frac{\R^4}{6}-29\R^3+300\R^2-540\R\\
P_3^S&=&-\frac{1333}{4536}\R^5-\frac{232}{45}\R^4-\frac{67}{3}\R^3+16\R^2+180\R\\
P_4^S&=&-\frac{181}{3360}\R^6-\frac{29}{30}\R^5-\frac{91}{20}\R^4+27\R^2\\ P_5^S&=&-\frac{181}{3360}\R^7-\frac{29}{30}\R^6-\frac{91}{20}\R^5+27\R^3
\end{eqnarray}

\bibliographystyle{JHEP_withtitle}
  
  \bibliography{bib_Rmunu}

\providecommand{\href}[2]{#2}\begingroup\raggedright\begin{thebibliography}{100}

\bibitem{Wilson:1971bg}
K.~G. Wilson, {\it {Renormalization group and critical phenomena. 1.
  Renormalization group and the Kadanoff scaling picture}},  Phys.Rev. {\bf B4}
  (1971) 3174--3183.

\bibitem{Wilson:1971dh}
K.~G. Wilson, {\it {Renormalization group and critical phenomena. 2. Phase
  space cell analysis of critical behavior}},  Phys.Rev. {\bf B4} (1971)
  3184--3205.

\bibitem{Gross:1973id}
D.~J. Gross and F.~Wilczek, {\it {Ultraviolet Behavior of Nonabelian Gauge
  Theories}},  Phys.Rev.Lett. {\bf 30} (1973) 1343--1346.

\bibitem{Politzer:1973fx}
H.~D. Politzer, {\it {Reliable Perturbative Results for Strong Interactions?}},
   Phys.Rev.Lett. {\bf 30} (1973) 1346--1349.

\bibitem{Weinberg:1980gg}
S.~Weinberg, {\it {Ultraviolet divergences in quantum theories of
  gravitation}},  General Relativity: An Einstein centenary survey, Eds.
  Hawking, S.W., Israel, W; Cambridge University Press (1979) 790--831.

\bibitem{Bond:2016dvk}
A.~D. Bond and D.~F. Litim, {\it {Theorems for Asymptotic Safety of Gauge
  Theories}},  Eur. Phys. J. {\bf C77} (2017), no.~6 429,
  [\href{http://xxx.lanl.gov/abs/1608.00519}{{\tt arXiv:1608.00519}}].

\bibitem{Litim:2014uca}
D.~F. Litim and F.~Sannino, {\it {Asymptotic safety guaranteed}},  JHEP {\bf
  12} (2014) 178, [\href{http://xxx.lanl.gov/abs/1406.2337}{{\tt
  arXiv:1406.2337}}].

\bibitem{Bond:2017tbw}
A.~D. Bond, D.~F. Litim, G.~Medina~Vazquez, and T.~Steudtner, {\it {UV
  conformal window for asymptotic safety}},  Phys. Rev. {\bf D97} (2018), no.~3
  036019, [\href{http://xxx.lanl.gov/abs/1710.07615}{{\tt arXiv:1710.07615}}].

\bibitem{Bond:2017lnq}
A.~D. Bond and D.~F. Litim, {\it {More asymptotic safety guaranteed}},
  [\href{http://xxx.lanl.gov/abs/1707.04217}{{\tt arXiv:1707.04217}}].

\bibitem{Bond:2017suy}
A.~D. Bond and D.~F. Litim, {\it {Asymptotic safety guaranteed in
  supersymmetry}},  Phys. Rev. Lett. {\bf 119} (2017), no.~21 211601,
  [\href{http://xxx.lanl.gov/abs/1709.06953}{{\tt arXiv:1709.06953}}].

\bibitem{Bond:2017wut}
A.~D. Bond, G.~Hiller, K.~Kowalska, and D.~F. Litim, {\it {Directions for model
  building from asymptotic safety}},  JHEP {\bf 08} (2017) 004,
  [\href{http://xxx.lanl.gov/abs/1702.01727}{{\tt arXiv:1702.01727}}].

\bibitem{Niedermaier:2006ns}
M.~Niedermaier, {\it {The asymptotic safety scenario in quantum gravity: An
  introduction}},  Class. Quant. Grav. {\bf 24} (2007) R171--230,
  [\href{http://xxx.lanl.gov/abs/gr-qc/0610018}{{\tt gr-qc/0610018}}].

\bibitem{Litim:2006dx}
D.~F. Litim, {\it {On fixed points of quantum gravity}},  AIP Conf.Proc. {\bf
  841} (2006) 322--329, [\href{http://xxx.lanl.gov/abs/hep-th/0606044}{{\tt
  hep-th/0606044}}].

\bibitem{Niedermaier:2006wt}
M.~Niedermaier and M.~Reuter, {\it {The Asymptotic Safety Scenario in Quantum
  Gravity}},  Living Rev.Rel. {\bf 9} (2006) 5.

\bibitem{Percacci:2007sz}
R.~Percacci, {\it {Asymptotic Safety}},  In *Oriti, D. (ed.): Approaches to
  quantum gravity* 111-128 (2007)
  [\href{http://xxx.lanl.gov/abs/0709.3851}{{\tt arXiv:0709.3851}}].

\bibitem{Litim:2008tt}
D.~F. Litim, {\it {Fixed Points of Quantum Gravity and the Renormalisation
  Group}},  PoS {\bf QG-Ph} (2007) 024,
  [\href{http://xxx.lanl.gov/abs/0810.3675}{{\tt arXiv:0810.3675}}].

\bibitem{Litim:2011cp}
D.~F. Litim, {\it {Renormalisation group and the Planck scale}},
  Phil.Trans.Roy.Soc.Lond. {\bf A369} (2011) 2759--2778,
  [\href{http://xxx.lanl.gov/abs/1102.4624}{{\tt arXiv:1102.4624}}].

\bibitem{Reuter:2012id}
M.~Reuter and F.~Saueressig, {\it {Quantum Einstein Gravity}},  New J. Phys.
  {\bf 14} (2012) 055022, [\href{http://xxx.lanl.gov/abs/1202.2274}{{\tt
  arXiv:1202.2274}}].

\bibitem{Falls:2013bv}
K.~Falls, D.~Litim, K.~Nikolakopoulos, and C.~Rahmede, {\it {A bootstrap
  towards asymptotic safety}},  [\href{http://xxx.lanl.gov/abs/1301.4191}{{\tt
  arXiv:1301.4191}}].

\bibitem{Reuter:1996cp}
M.~Reuter, {\it {Nonperturbative Evolution Equation for Quantum Gravity}},
  Phys. Rev. {\bf D57} (1998) 971--985,
  [\href{http://xxx.lanl.gov/abs/hep-th/9605030}{{\tt hep-th/9605030}}].

\bibitem{Falkenberg:1996bq}
S.~Falkenberg and S.~D. Odintsov, {\it {Gauge dependence of the effective
  average action in Einstein gravity}},  Int.J.Mod.Phys. {\bf A13} (1998)
  607--623, [\href{http://xxx.lanl.gov/abs/hep-th/9612019}{{\tt
  hep-th/9612019}}].

\bibitem{Souma:1999at}
W.~Souma, {\it {Nontrivial ultraviolet fixed point in quantum gravity}},
  Prog.Theor.Phys. {\bf 102} (1999) 181--195,
  [\href{http://xxx.lanl.gov/abs/hep-th/9907027}{{\tt hep-th/9907027}}].

\bibitem{Lauscher:2001ya}
O.~Lauscher and M.~Reuter, {\it {Ultraviolet fixed point and generalized flow
  equation of quantum gravity}},  Phys. Rev. {\bf D65} (2002) 025013,
  [\href{http://xxx.lanl.gov/abs/hep-th/0108040}{{\tt hep-th/0108040}}].

\bibitem{Lauscher:2002sq}
O.~Lauscher and M.~Reuter, {\it {Flow equation of quantum Einstein gravity in a
  higher- derivative truncation}},  Phys. Rev. {\bf D66} (2002) 025026,
  [\href{http://xxx.lanl.gov/abs/hep-th/0205062}{{\tt hep-th/0205062}}].

\bibitem{Litim:2003vp}
D.~F. Litim, {\it {Fixed points of quantum gravity}},  Phys.Rev.Lett. {\bf 92}
  (2004) 201301, [\href{http://xxx.lanl.gov/abs/hep-th/0312114}{{\tt
  hep-th/0312114}}].

\bibitem{Bonanno:2004sy}
A.~Bonanno and M.~Reuter, {\it {Proper time flow equation for gravity}},  JHEP
  {\bf 02} (2005) 035, [\href{http://xxx.lanl.gov/abs/hep-th/0410191}{{\tt
  hep-th/0410191}}].

\bibitem{Fischer:2006fz}
P.~Fischer and D.~F. Litim, {\it {Fixed points of quantum gravity in extra
  dimensions}},  Phys.Lett. {\bf B638} (2006) 497--502,
  [\href{http://xxx.lanl.gov/abs/hep-th/0602203}{{\tt hep-th/0602203}}].

\bibitem{Fischer:2006at}
P.~Fischer and D.~F. Litim, {\it {Fixed points of quantum gravity in higher
  dimensions}},  AIP Conf.Proc. {\bf 861} (2006) 336--343,
  [\href{http://xxx.lanl.gov/abs/hep-th/0606135}{{\tt hep-th/0606135}}].

\bibitem{Codello:2006in}
A.~Codello and R.~Percacci, {\it {Fixed Points of Higher Derivative Gravity}},
  Phys. Rev. Lett. {\bf 97} (2006) 221301,
  [\href{http://xxx.lanl.gov/abs/hep-th/0607128}{{\tt hep-th/0607128}}].

\bibitem{Codello:2007bd}
A.~Codello, R.~Percacci, and C.~Rahmede, {\it {Ultraviolet properties of
  f(R)-gravity}},  Int. J. Mod. Phys. {\bf A23} (2008) 143--150,
  [\href{http://xxx.lanl.gov/abs/0705.1769}{{\tt arXiv:0705.1769}}].

\bibitem{Machado:2007ea}
P.~F. Machado and F.~Saueressig, {\it {On the renormalization group flow of
  f(R)-gravity}},  Phys. Rev. {\bf D77} (2008) 124045,
  [\href{http://xxx.lanl.gov/abs/0712.0445}{{\tt arXiv:0712.0445}}].

\bibitem{Codello:2008vh}
A.~Codello, R.~Percacci, and C.~Rahmede, {\it {Investigating the Ultraviolet
  Properties of Gravity with a Wilsonian Renormalization Group Equation}},
  Annals Phys. {\bf 324} (2009) 414--469,
  [\href{http://xxx.lanl.gov/abs/0805.2909}{{\tt arXiv:0805.2909}}].

\bibitem{Niedermaier:2009zz}
M.~R. Niedermaier, {\it {Gravitational Fixed Points from Perturbation Theory}},
   Phys. Rev. Lett. {\bf 103} (2009) 101303.

\bibitem{Benedetti:2009rx}
D.~Benedetti, P.~F. Machado, and F.~Saueressig, {\it {Asymptotic safety in
  higher-derivative gravity}},  Mod. Phys. Lett. {\bf A24} (2009) 2233--2241,
  [\href{http://xxx.lanl.gov/abs/0901.2984}{{\tt arXiv:0901.2984}}].

\bibitem{Eichhorn:2009ah}
A.~Eichhorn, H.~Gies, and M.~M. Scherer, {\it {Asymptotically free scalar
  curvature-ghost coupling in Quantum Einstein Gravity}},  Phys. Rev. {\bf D80}
  (2009) 104003, [\href{http://xxx.lanl.gov/abs/0907.1828}{{\tt
  arXiv:0907.1828}}].

\bibitem{Manrique:2009uh}
E.~Manrique and M.~Reuter, {\it {Bimetric Truncations for Quantum Einstein
  Gravity and Asymptotic Safety}},  Annals Phys. {\bf 325} (2010) 785--815,
  [\href{http://xxx.lanl.gov/abs/0907.2617}{{\tt arXiv:0907.2617}}].

\bibitem{Niedermaier:2010zz}
M.~Niedermaier, {\it {Gravitational fixed points and asymptotic safety from
  perturbation theory}},  Nucl.Phys. {\bf B833} (2010) 226--270.

\bibitem{Eichhorn:2010tb}
A.~Eichhorn and H.~Gies, {\it {Ghost anomalous dimension in asymptotically safe
  quantum gravity}},  Phys. Rev. {\bf D81} (2010) 104010,
  [\href{http://xxx.lanl.gov/abs/1001.5033}{{\tt arXiv:1001.5033}}].

\bibitem{Groh:2010ta}
K.~Groh and F.~Saueressig, {\it {Ghost wave-function renormalization in
  Asymptotically Safe Quantum Gravity}},  J. Phys. {\bf A43} (2010) 365403,
  [\href{http://xxx.lanl.gov/abs/1001.5032}{{\tt arXiv:1001.5032}}].

\bibitem{Manrique:2010am}
E.~Manrique, M.~Reuter, and F.~Saueressig, {\it {Bimetric Renormalization Group
  Flows in Quantum Einstein Gravity}},  Annals Phys. {\bf 326} (2011) 463--485,
  [\href{http://xxx.lanl.gov/abs/1006.0099}{{\tt arXiv:1006.0099}}].

\bibitem{Niedermaier:2011zz}
M.~Niedermaier, {\it {Can a nontrivial gravitational fixed point be identified
  in perturbation theory?}},  PoS {\bf CLAQG08} (2011) 005.

\bibitem{Manrique:2011jc}
E.~Manrique, S.~Rechenberger, and F.~Saueressig, {\it {Asymptotically Safe
  Lorentzian Gravity}},  Phys.Rev.Lett. {\bf 106} (2011) 251302,
  [\href{http://xxx.lanl.gov/abs/1102.5012}{{\tt arXiv:1102.5012}}].

\bibitem{Litim:2012vz}
D.~Litim and A.~Satz, {\it {Limit cycles and quantum gravity}},
  [\href{http://xxx.lanl.gov/abs/1205.4218}{{\tt arXiv:1205.4218}}].

\bibitem{Benedetti:2012dx}
D.~Benedetti and F.~Caravelli, {\it {The Local potential approximation in
  quantum gravity}},  JHEP {\bf 06} (2012) 017,
  [\href{http://xxx.lanl.gov/abs/1204.3541}{{\tt arXiv:1204.3541}}]. [Erratum:
  JHEP10,157(2012)].

\bibitem{Donkin:2012ud}
I.~Donkin and J.~M. Pawlowski, {\it {The phase diagram of quantum gravity from
  diffeomorphism-invariant RG-flows}},
  [\href{http://xxx.lanl.gov/abs/1203.4207}{{\tt arXiv:1203.4207}}].

\bibitem{Christiansen:2012rx}
N.~Christiansen, D.~F. Litim, J.~M. Pawlowski, and A.~Rodigast, {\it {Fixed
  points and infrared completion of quantum gravity}},  Phys.Lett. {\bf B728}
  (2014) 114--117, [\href{http://xxx.lanl.gov/abs/1209.4038}{{\tt
  arXiv:1209.4038}}].

\bibitem{Dietz:2012ic}
J.~A. Dietz and T.~R. Morris, {\it {Asymptotic safety in the f(R)
  approximation}},  JHEP {\bf 01} (2013) 108,
  [\href{http://xxx.lanl.gov/abs/1211.0955}{{\tt arXiv:1211.0955}}].

\bibitem{Hindmarsh:2012rc}
M.~Hindmarsh and I.~D. Saltas, {\it {f(R) Gravity from the renormalisation
  group}},  Phys. Rev. {\bf D86} (2012) 064029,
  [\href{http://xxx.lanl.gov/abs/1203.3957}{{\tt arXiv:1203.3957}}].

\bibitem{Benedetti:2013jk}
D.~Benedetti, {\it {On the number of relevant operators in asymptotically safe
  gravity}},  Europhys. Lett. {\bf 102} (2013) 20007,
  [\href{http://xxx.lanl.gov/abs/1301.4422}{{\tt arXiv:1301.4422}}].

\bibitem{Dietz:2013sba}
J.~A. Dietz and T.~R. Morris, {\it {Redundant operators in the exact
  renormalisation group and in the f(R) approximation to asymptotic safety}},
  JHEP {\bf 07} (2013) 064, [\href{http://xxx.lanl.gov/abs/1306.1223}{{\tt
  arXiv:1306.1223}}].

\bibitem{Codello:2013fpa}
A.~Codello, G.~D'Odorico, and C.~Pagani, {\it {Consistent closure of
  renormalization group flow equations in quantum gravity}},  Phys. Rev. {\bf
  D89} (2014), no.~8 081701, [\href{http://xxx.lanl.gov/abs/1304.4777}{{\tt
  arXiv:1304.4777}}].

\bibitem{Ohta:2013uca}
N.~Ohta and R.~Percacci, {\it {Higher Derivative Gravity and Asymptotic Safety
  in Diverse Dimensions}},  Class. Quant. Grav. {\bf 31} (2014) 015024,
  [\href{http://xxx.lanl.gov/abs/1308.3398}{{\tt arXiv:1308.3398}}].

\bibitem{Christiansen:2014raa}
N.~Christiansen, B.~Knorr, J.~M. Pawlowski, and A.~Rodigast, {\it {Global Flows
  in Quantum Gravity}},  Phys. Rev. {\bf D93} (2016) 044036,
  [\href{http://xxx.lanl.gov/abs/1403.1232}{{\tt arXiv:1403.1232}}].

\bibitem{Becker:2014qya}
D.~Becker and M.~Reuter, {\it {En route to Background Independence: Broken
  split-symmetry, and how to restore it with bi-metric average actions}},
  Annals Phys. {\bf 350} (2014) 225--301,
  [\href{http://xxx.lanl.gov/abs/1404.4537}{{\tt arXiv:1404.4537}}].

\bibitem{Falls:2014zba}
K.~Falls, {\it {Asymptotic safety and the cosmological constant}},  JHEP {\bf
  01} (2016) 069, [\href{http://xxx.lanl.gov/abs/1408.0276}{{\tt
  arXiv:1408.0276}}].

\bibitem{Padilla:2014yea}
A.~Padilla and I.~D. Saltas, {\it {A note on classical and quantum unimodular
  gravity}},  Eur. Phys. J. {\bf C75} (2015), no.~11 561,
  [\href{http://xxx.lanl.gov/abs/1409.3573}{{\tt arXiv:1409.3573}}].

\bibitem{Falls:2014tra}
K.~Falls, D.~F. Litim, K.~Nikolakopoulos, and C.~Rahmede, {\it {Further
  evidence for asymptotic safety of quantum gravity}},  Phys. Rev. {\bf D93}
  (2016), no.~10 104022, [\href{http://xxx.lanl.gov/abs/1410.4815}{{\tt
  arXiv:1410.4815}}].

\bibitem{Saltas:2014cta}
I.~D. Saltas, {\it {UV structure of quantum unimodular gravity}},  Phys. Rev.
  {\bf D90} (2014), no.~12 124052,
  [\href{http://xxx.lanl.gov/abs/1410.6163}{{\tt arXiv:1410.6163}}].

\bibitem{Falls:2015qga}
K.~Falls, {\it {Renormalization of Newton{\rq}s constant}},  Phys. Rev. {\bf
  D92} (2015), no.~12 124057, [\href{http://xxx.lanl.gov/abs/1501.05331}{{\tt
  arXiv:1501.05331}}].

\bibitem{Falls:2015cta}
K.~Falls, {\it {Critical scaling in quantum gravity from the renormalisation
  group}},  [\href{http://xxx.lanl.gov/abs/1503.06233}{{\tt
  arXiv:1503.06233}}].

\bibitem{Christiansen:2015rva}
N.~Christiansen, B.~Knorr, J.~Meibohm, J.~M. Pawlowski, and M.~Reichert, {\it
  {Local Quantum Gravity}},  Phys. Rev. {\bf D92} (2015), no.~12 121501,
  [\href{http://xxx.lanl.gov/abs/1506.07016}{{\tt arXiv:1506.07016}}].

\bibitem{Gies:2015tca}
H.~Gies, B.~Knorr, and S.~Lippoldt, {\it {Generalized Parametrization
  Dependence in Quantum Gravity}},  Phys. Rev. {\bf D92} (2015), no.~8 084020,
  [\href{http://xxx.lanl.gov/abs/1507.08859}{{\tt arXiv:1507.08859}}].

\bibitem{Benedetti:2015zsw}
D.~Benedetti, {\it {Essential nature of Newton{\rq}s constant in unimodular
  gravity}},  Gen. Rel. Grav. {\bf 48} (2016), no.~5 68,
  [\href{http://xxx.lanl.gov/abs/1511.06560}{{\tt arXiv:1511.06560}}].

\bibitem{Eichhorn:2015bna}
A.~Eichhorn, {\it {The Renormalization Group flow of unimodular f(R) gravity}},
   JHEP {\bf 04} (2015) 096, [\href{http://xxx.lanl.gov/abs/1501.05848}{{\tt
  arXiv:1501.05848}}].

\bibitem{Ohta:2015efa}
N.~Ohta, R.~Percacci, and G.~P. Vacca, {\it {Flow equation for $f(R)$ gravity
  and some of its exact solutions}},  Phys. Rev. {\bf D92} (2015), no.~6
  061501, [\href{http://xxx.lanl.gov/abs/1507.00968}{{\tt arXiv:1507.00968}}].

\bibitem{Ohta:2015fcu}
N.~Ohta, R.~Percacci, and G.~P. Vacca, {\it {Renormalization Group Equation and
  scaling solutions for f(R) gravity in exponential parametrization}},  Eur.
  Phys. J. {\bf C76} (2016), no.~2 46,
  [\href{http://xxx.lanl.gov/abs/1511.09393}{{\tt arXiv:1511.09393}}].

\bibitem{Gies:2016con}
H.~Gies, B.~Knorr, S.~Lippoldt, and F.~Saueressig, {\it {Gravitational Two-Loop
  Counterterm Is Asymptotically Safe}},  Phys. Rev. Lett. {\bf 116} (2016),
  no.~21 211302, [\href{http://xxx.lanl.gov/abs/1601.01800}{{\tt
  arXiv:1601.01800}}].

\bibitem{Falls:2016wsa}
K.~Falls, D.~F. Litim, K.~Nikolakopoulos, and C.~Rahmede, {\it {On de Sitter
  solutions in asymptotically safe $f(R)$ theories}},
  [\href{http://xxx.lanl.gov/abs/1607.04962}{{\tt arXiv:1607.04962}}].

\bibitem{Falls:2016msz}
K.~Falls and N.~Ohta, {\it {Renormalization Group Equation for $f(R)$ gravity
  on hyperbolic spaces}},  Phys. Rev. {\bf D94} (2016), no.~8 084005,
  [\href{http://xxx.lanl.gov/abs/1607.08460}{{\tt arXiv:1607.08460}}].

\bibitem{Christiansen:2016sjn}
N.~Christiansen, {\it {Four-Derivative Quantum Gravity Beyond Perturbation
  Theory}},  [\href{http://xxx.lanl.gov/abs/1612.06223}{{\tt
  arXiv:1612.06223}}].

\bibitem{Biemans:2016rvp}
J.~Biemans, A.~Platania, and F.~Saueressig, {\it {Quantum gravity on foliated
  spacetimes: Asymptotically safe and sound}},  Phys. Rev. {\bf D95} (2017),
  no.~8 086013, [\href{http://xxx.lanl.gov/abs/1609.04813}{{\tt
  arXiv:1609.04813}}].

\bibitem{Pagani:2016dof}
C.~Pagani and M.~Reuter, {\it {Composite Operators in Asymptotic Safety}},
  Phys. Rev. {\bf D95} (2017), no.~6 066002,
  [\href{http://xxx.lanl.gov/abs/1611.06522}{{\tt arXiv:1611.06522}}].

\bibitem{Denz:2016qks}
T.~Denz, J.~M. Pawlowski, and M.~Reichert, {\it {Towards apparent convergence
  in asymptotically safe quantum gravity}},
  [\href{http://xxx.lanl.gov/abs/1612.07315}{{\tt arXiv:1612.07315}}].

\bibitem{Falls:2017cze}
K.~Falls, {\it Physical renormalization schemes and asymptotic safety in
  quantum gravity},  Phys. Rev. D {\bf 96} (Dec, 2017) 126016,
  [\href{http://xxx.lanl.gov/abs/1702.03577}{{\tt arXiv:1702.03577}}].

\bibitem{Hamada:2017rvn}
Y.~Hamada and M.~Yamada, {\it {Asymptotic safety of higher derivative quantum
  gravity non-minimally coupled with a matter system}},  JHEP {\bf 08} (2017)
  070, [\href{http://xxx.lanl.gov/abs/1703.09033}{{\tt arXiv:1703.09033}}].

\bibitem{Houthoff:2017oam}
W.~B. Houthoff, A.~Kurov, and F.~Saueressig, {\it {Impact of topology in
  foliated Quantum Einstein Gravity}},  Eur. Phys. J. {\bf C77} (2017) 491,
  [\href{http://xxx.lanl.gov/abs/1705.01848}{{\tt arXiv:1705.01848}}].

\bibitem{Gonzalez-Martin:2017gza}
S.~Gonzalez-Martin, T.~R. Morris, and Z.~H. Slade, {\it {Asymptotic solutions
  in asymptotic safety}},  Phys. Rev. {\bf D95} (2017), no.~10 106010,
  [\href{http://xxx.lanl.gov/abs/1704.08873}{{\tt arXiv:1704.08873}}].

\bibitem{Knorr:2017fus}
B.~Knorr and S.~Lippoldt, {\it {Correlation functions on a curved background}},
   Phys. Rev. {\bf D96} (2017), no.~6 065020,
  [\href{http://xxx.lanl.gov/abs/1707.01397}{{\tt arXiv:1707.01397}}].

\bibitem{Knorr:2017mhu}
B.~Knorr, {\it {Infinite order quantum-gravitational correlations}},
  [\href{http://xxx.lanl.gov/abs/1710.07055}{{\tt arXiv:1710.07055}}].

\bibitem{Becker:2017tcx}
D.~Becker, C.~Ripken, and F.~Saueressig, {\it {On avoiding Ostrogradski
  instabilities within Asymptotic Safety}},
  [\href{http://xxx.lanl.gov/abs/1709.09098}{{\tt arXiv:1709.09098}}].

\bibitem{Christiansen:2017cxa}
N.~Christiansen, D.~F. Litim, J.~M. Pawlowski, and M.~Reichert, {\it {One force
  to rule them all: asymptotic safety of gravity with matter}},
  [\href{http://xxx.lanl.gov/abs/1710.04669}{{\tt arXiv:1710.04669}}].

\bibitem{Christiansen:2017bsy}
N.~Christiansen, K.~Falls, J.~M. Pawlowski, and M.~Reichert, {\it {Curvature
  dependence of quantum gravity}},
  [\href{http://xxx.lanl.gov/abs/1711.09259}{{\tt arXiv:1711.09259}}].

\bibitem{Wetterich:1992yh}
C.~Wetterich, {\it {Exact evolution equation for the effective potential}},
  Phys.Lett. {\bf B301} (1993) 90--94.

\bibitem{Morris:1993qb}
T.~R. Morris, {\it {The Exact renormalization group and approximate
  solutions}},  Int. J. Mod. Phys. {\bf A9} (1994) 2411--2450,
  [\href{http://xxx.lanl.gov/abs/hep-ph/9308265}{{\tt hep-ph/9308265}}].

\bibitem{Nikolakopoulos:Thesis}
K.~Nikolakopoulos, {\em {Quantum gravity and the renormalisation group --
  theoretical advances and applications}}.
\newblock PhD thesis, University of Sussex, 2013.

\bibitem{Dou:1997fg}
D.~Dou and R.~Percacci, {\it {The running gravitational couplings}},  Class.
  Quant. Grav. {\bf 15} (1998) 3449--3468,
  [\href{http://xxx.lanl.gov/abs/hep-th/9707239}{{\tt hep-th/9707239}}].

\bibitem{York:1973ia}
J.~J.~W. York, {\it {Conformatlly invariant orthogonal decomposition of
  symmetric tensors on Riemannian manifolds and the initial value problem of
  general relativity}},  J. Math. Phys. {\bf 14} (1973) 456--464.

\bibitem{Litim:1998nf}
D.~F. Litim and J.~M. Pawlowski, {\it {On gauge invariant Wilsonian flows}},
  [\href{http://xxx.lanl.gov/abs/hep-th/9901063}{{\tt hep-th/9901063}}].

\bibitem{Litim:2000ci}
D.~F. Litim, {\it {Optimization of the exact renormalization group}},
  Phys.Lett. {\bf B486} (2000) 92--99,
  [\href{http://xxx.lanl.gov/abs/hep-th/0005245}{{\tt hep-th/0005245}}].

\bibitem{Litim:2001up}
D.~F. Litim, {\it {Optimized renormalization group flows}},  Phys.Rev. {\bf
  D64} (2001) 105007, [\href{http://xxx.lanl.gov/abs/hep-th/0103195}{{\tt
  hep-th/0103195}}].

\bibitem{Litim:2001fd}
D.~F. Litim, {\it {Mind the gap}},  Int.J.Mod.Phys. {\bf A16} (2001)
  2081--2088, [\href{http://xxx.lanl.gov/abs/hep-th/0104221}{{\tt
  hep-th/0104221}}].

\bibitem{Litim:2006ag}
D.~F. Litim and J.~M. Pawlowski, {\it {Non-perturbative thermal flows and
  resummations}},  JHEP {\bf 0611} (2006) 026,
  [\href{http://xxx.lanl.gov/abs/hep-th/0609122}{{\tt hep-th/0609122}}].

\bibitem{Litim:2007jb}
D.~F. Litim, {\it {Towards functional flows for hierarchical models}},
  Phys.Rev. {\bf D76} (2007) 105001,
  [\href{http://xxx.lanl.gov/abs/0704.1514}{{\tt arXiv:0704.1514}}].

\bibitem{Litim:2010tt}
D.~F. Litim and D.~Zappala, {\it {Ising exponents from the functional
  renormalisation group}},  Phys.Rev. {\bf D83} (2011) 085009,
  [\href{http://xxx.lanl.gov/abs/1009.1948}{{\tt arXiv:1009.1948}}].

\bibitem{Avramidi:2000bm}
I.~Avramidi, {\it {Heat kernel and quantum gravity}},  Lect.Notes Phys. {\bf
  M64} (2000) 1--149.

\bibitem{Gilkey:1995mj}
P.~B. Gilkey, {\it {Invariance theory, the heat equation and the Atiyah-Singer
  index theorem}}, . Book published in Boca Raton by CRC Press, 1995.

\bibitem{Litim:1998qi}
D.~F. Litim and J.~M. Pawlowski, {\it {Flow equations for Yang-Mills theories
  in general axial gauges}},  Phys.Lett. {\bf B435} (1998) 181--188,
  [\href{http://xxx.lanl.gov/abs/hep-th/9802064}{{\tt hep-th/9802064}}].

\bibitem{King:MSc}
C.~King, {\it {Quantum gravity and matter}},  Master's thesis, University of
  Sussex, 2015.

\bibitem{Morris:1994ki}
T.~R. Morris, {\it {On truncations of the exact renormalization group}},
  Phys.Lett. {\bf B334} (1994) 355--362,
  [\href{http://xxx.lanl.gov/abs/hep-th/9405190}{{\tt hep-th/9405190}}].

\bibitem{Litim:2002cf}
D.~F. Litim, {\it {Critical exponents from optimized renormalization group
  flows}},  Nucl.Phys. {\bf B631} (2002) 128--158,
  [\href{http://xxx.lanl.gov/abs/hep-th/0203006}{{\tt hep-th/0203006}}].

\bibitem{Litim:2016hlb}
D.~F. Litim and E.~Marchais, {\it {Critical $O(N)$ models in the complex field
  plane}},  Phys. Rev. {\bf D95} (2017), no.~2 025026,
  [\href{http://xxx.lanl.gov/abs/1607.02030}{{\tt arXiv:1607.02030}}].

\bibitem{Juttner:2017cpr}
A.~Juettner, D.~F. Litim, and E.~Marchais, {\it {Global Wilson�Fisher fixed
  points}},  Nucl. Phys. {\bf B921} (2017) 769--795,
  [\href{http://xxx.lanl.gov/abs/1701.05168}{{\tt arXiv:1701.05168}}].

\bibitem{Marchais:2017jqc}
D.~F. Litim, E.~Marchais, and P.~Mati, {\it {Fixed points and the spontaneous
  breaking of scale invariance}},  Phys. Rev. {\bf D95} (2017), no.~12 125006,
  [\href{http://xxx.lanl.gov/abs/1702.05749}{{\tt arXiv:1702.05749}}].

\bibitem{Kawai:1989yh}
H.~Kawai and M.~Ninomiya, {\it {Renormalization Group and Quantum Gravity}},
  Nucl.Phys. {\bf B336} (1990) 115.

\bibitem{Ade:2013uln}
{\bf Planck Collaboration} Collaboration, P.~Ade { et~al.}, {\it {Planck 2013
  results. XXII. Constraints on inflation}},
  [\href{http://xxx.lanl.gov/abs/1303.5082}{{\tt arXiv:1303.5082}}].

\bibitem{Ade:2013zuv}
{\bf Planck Collaboration} Collaboration, P.~Ade { et~al.}, {\it {Planck 2013
  results. XVI. Cosmological parameters}},  Astron.Astrophys. {\bf 571} (2014)
  A16, [\href{http://xxx.lanl.gov/abs/1303.5076}{{\tt arXiv:1303.5076}}].

\bibitem{Perlmutter:1998np}
{\bf Supernova Cosmology Project} Collaboration, S.~Perlmutter { et~al.}, {\it
  {Measurements of Omega and Lambda from 42 High-Redshift Supernovae}},
  Astrophys. J. {\bf 517} (1999) 565--586,
  [\href{http://xxx.lanl.gov/abs/astro-ph/9812133}{{\tt astro-ph/9812133}}].

\bibitem{Shapiro:2000dz}
I.~L. Shapiro and J.~Sola, {\it {Scaling behavior of the cosmological constant:
  Interface between quantum field theory and cosmology}},  JHEP {\bf 02} (2002)
  006, [\href{http://xxx.lanl.gov/abs/hep-th/0012227}{{\tt hep-th/0012227}}].

\bibitem{Weinberg:2009wa}
S.~Weinberg, {\it {Asymptotically Safe Inflation}},  Phys. Rev. {\bf D81}
  (2010) 083535, [\href{http://xxx.lanl.gov/abs/0911.3165}{{\tt
  arXiv:0911.3165}}].

\bibitem{Reuter:2005kb}
M.~Reuter and F.~Saueressig, {\it {From big bang to asymptotic de Sitter:
  Complete cosmologies in a quantum gravity framework}},  JCAP {\bf 0509}
  (2005) 012, [\href{http://xxx.lanl.gov/abs/hep-th/0507167}{{\tt
  hep-th/0507167}}].

\bibitem{Bonanno:2001xi}
A.~Bonanno and M.~Reuter, {\it {Cosmology of the Planck era from a
  renormalization group for quantum gravity}},  Phys. Rev. {\bf D65} (2002)
  043508, [\href{http://xxx.lanl.gov/abs/hep-th/0106133}{{\tt
  hep-th/0106133}}].

\bibitem{Bonanno:2001hi}
A.~Bonanno and M.~Reuter, {\it {Cosmology with self-adjusting vacuum energy
  density from a renormalization group fixed point}},  Phys. Lett. {\bf B527}
  (2002) 9--17, [\href{http://xxx.lanl.gov/abs/astro-ph/0106468}{{\tt
  astro-ph/0106468}}].

\bibitem{Bonanno:2009nj}
A.~Bonanno, {\it {Astrophysical implications of the Asymptotic Safety Scenario
  in Quantum Gravity}},  PoS {\bf CLAQG08} (2011) 008,
  [\href{http://xxx.lanl.gov/abs/0911.2727}{{\tt arXiv:0911.2727}}].

\bibitem{Bonanno:2010bt}
A.~Bonanno, A.~Contillo, and R.~Percacci, {\it {Inflationary solutions in
  asymptotically safe f(R) theories}},  Class. Quant. Grav. {\bf 28} (2011)
  145026, [\href{http://xxx.lanl.gov/abs/1006.0192}{{\tt arXiv:1006.0192}}].

\bibitem{Koch:2010nn}
B.~Koch and I.~Ramirez, {\it {Exact renormalization group with optimal scale
  and its application to cosmology}},  Class. Quant. Grav. {\bf 28} (2011)
  055008, [\href{http://xxx.lanl.gov/abs/1010.2799}{{\tt arXiv:1010.2799}}].

\bibitem{Bonanno:2012jy}
A.~Bonanno, {\it {An effective action for asymptotically safe gravity}},  Phys.
  Rev. {\bf D85} (2012) 081503, [\href{http://xxx.lanl.gov/abs/1203.1962}{{\tt
  arXiv:1203.1962}}].

\bibitem{Frolov:2011ys}
A.~V. Frolov and J.-Q. Guo, {\it {Small Cosmological Constant from Running
  Gravitational Coupling}},  [\href{http://xxx.lanl.gov/abs/1101.4995}{{\tt
  arXiv:1101.4995}}].

\bibitem{Cai:2011kd}
Y.-F. Cai and D.~A. Easson, {\it {Asymptotically safe gravity as a
  scalar-tensor theory and its cosmological implications}},  Phys.Rev. {\bf
  D84} (2011) 103502, [\href{http://xxx.lanl.gov/abs/1107.5815}{{\tt
  arXiv:1107.5815}}].

\bibitem{Contillo:2011ag}
A.~Contillo, M.~Hindmarsh, and C.~Rahmede, {\it {Renormalisation group
  improvement of scalar field inflation}},  Phys.Rev. {\bf D85} (2012) 043501,
  [\href{http://xxx.lanl.gov/abs/1108.0422}{{\tt arXiv:1108.0422}}]. 17 pages/
  2 figures.

\bibitem{Hindmarsh:2011hx}
M.~Hindmarsh, D.~Litim, and C.~Rahmede, {\it {Asymptotically Safe Cosmology}},
  JCAP {\bf 1107} (2011) 019, [\href{http://xxx.lanl.gov/abs/1101.5401}{{\tt
  arXiv:1101.5401}}].

\bibitem{Ahn:2011qt}
C.~Ahn, C.~Kim, and E.~V. Linder, {\it {From Asymptotic Safety to Dark
  Energy}},  Phys. Lett. {\bf B704} (2011) 10--14,
  [\href{http://xxx.lanl.gov/abs/1106.1435}{{\tt arXiv:1106.1435}}].

\bibitem{Copeland:2013vva}
E.~J. Copeland, C.~Rahmede, and I.~D. Saltas, {\it {Asymptotically Safe
  Starobinsky Inflation}},  Phys. Rev. {\bf D91} (2015), no.~10 103530,
  [\href{http://xxx.lanl.gov/abs/1311.0881}{{\tt arXiv:1311.0881}}].

\bibitem{Kaya:2013bga}
A.~Kaya, {\it {Exact renormalization group flow in an expanding Universe and
  screening of the cosmological constant}},  Phys.Rev. {\bf D87} (2013), no.~12
  123501, [\href{http://xxx.lanl.gov/abs/1303.5459}{{\tt arXiv:1303.5459}}].

\bibitem{Saltas:2015vsc}
I.~D. Saltas, {\it {Higgs inflation and quantum gravity: An exact
  renormalisation group approach}},  JCAP {\bf 1602} (2016), no.~02 048,
  [\href{http://xxx.lanl.gov/abs/1512.06134}{{\tt arXiv:1512.06134}}].

\bibitem{Bonanno:2015fga}
A.~Bonanno and A.~Platania, {\it {Asymptotically safe inflation from quadratic
  gravity}},  Phys. Lett. {\bf B750} (2015) 638--642,
  [\href{http://xxx.lanl.gov/abs/1507.03375}{{\tt arXiv:1507.03375}}].

\bibitem{Kofinas:2016lcz}
G.~Kofinas and V.~Zarikas, {\it {Asymptotically Safe gravity and non-singular
  inflationary Big Bang with vacuum birth}},  Phys. Rev. {\bf D94} (2016),
  no.~10 103514, [\href{http://xxx.lanl.gov/abs/1605.02241}{{\tt
  arXiv:1605.02241}}].

\bibitem{Schroeder:Thesis}
J.~Schroeder, {\em {Aspects of quantum gravity and matter}}.
\newblock PhD thesis, University of Sussex, 2015.

\bibitem{Freire:2000bq}
F.~Freire, D.~F. Litim, and J.~M. Pawlowski, {\it {Gauge invariance and
  background field formalism in the exact renormalization group}},  Phys.Lett.
  {\bf B495} (2000) 256--262,
  [\href{http://xxx.lanl.gov/abs/hep-th/0009110}{{\tt hep-th/0009110}}].

\bibitem{Falls:2017nnu}
K.~Falls and T.~R. Morris, {\it {Conformal anomaly from gauge fields without
  gauge fixing}},  [\href{http://xxx.lanl.gov/abs/1712.05011}{{\tt
  arXiv:1712.05011}}].

\end{thebibliography}\endgroup

\end{document}